\documentclass[10pt,a4paper]{article}
\usepackage[dvips]{color}
\usepackage{epsfig}
\usepackage{amsmath}
\usepackage{graphicx}
\usepackage{authblk}
\usepackage{amssymb,amsmath}
\usepackage{amsmath}
\usepackage[belowskip=-5pt,aboveskip=0pt]{caption}
\begin{document}
\title{\bf Phenomenological models of Universe with varying $G$ and $\Lambda$}
\author{ { Martiros Khurshudyan$^{a}$ \thanks{Email:khurshudyan@yandex.ru}}\\
$^{a}${\small {\em Department of Theoretical Physics, Yerevan State
University, 1 Alex Manookian, 0025, Yerevan, Armenia}}\\
  } \maketitle
\begin{abstract}
In this article we will consider several phenomenological models for the Universe with varying $G$ and $\Lambda(t)$, where $G$ is the gravitational "constant" and $\Lambda(t)$ is a varying cosmological "constant". Two-component fluid model taken into account. An interaction of the phenomenological form between a barotropic fluid and a quintessence DE is supposed. Three different forms of $\Lambda(t)$ will be considered. The problem is analysed numerically and behavior of different cosmological parameters investigated graphically. Conclusion and discussions are given at the end of the work. In an Appendix an information concerning to the other cosmological parameters is presented.
\end{abstract}
\section*{\large{Introduction}}
To make a correspondence between GR and observations theoreticians proposed an existance of a dark energy, which takes the properties to be mysterious and with negative pressure and works against gravity. This properties for the dark energy originated from the facts that we do not know about real origin of it, about physics of it , as well as that current observations give a believe that the present phase in history of the Universe corresponds to the accelerated expansion. It is obvious that the problem related to the dark energy is a complicated from different reasons, because, for instance, it is entered to GR by hand. Second, the proposed models with their properties compared with the results of the observations usually overlap to each other and do not allow to extract the final answer. Probably the last one is the most problematic, because imagine, You want to fit a model with observations and you found that a free parameter of the model can accept values from relatively long interval, therefore outcomes and conclusions can not be unique. In theoretical cosmology among other problems, a puzzle known as the problem of a Dark Matter is also exist. Mathematically, DM was introduced to keep balance i.e to have energy conservation in Universe. A connection between DE and DM is also under the hot discussions, an interaction between DE and DM is assumed to be the fifth force in nature. Fundamental physics does not forbid the existance of the interaction between DE and DM, even recently were found that observations could support interaction. Evidence of the interaction between DE and DM has been found, for instance, by Bertolami et al. \cite{Bertolami}-\cite{Bertolami3} through studies on the Abell Cluster $A586$. On the other hand, by using optical, X-ray and weak lensing data from 33 relaxed galaxy clusters Abdalla et al. found the signature of interaction between DE and DM \cite{Abdalla}. Conclusions are that this coupling is small but indicates that DE might be decaying into DM. Moreover recently were shown that in both directions a flow can be realised. Recall, that DE and DM operate on different scales and responsible for different phenomenas in Universe. DE has negative pressure and is responsible for the acceleration, while DM is gravitationally attractive and allows the accumulation of matter that leads to the formation of large scale structures. Most of the hypotheses assum that DM and DE are physically unrelated and the similarity in their energy densities is a purely accidental. But interactions between DM and DE might alleviate the problem, keeping close values for their densities up to large redshifts as has been shown \cite{Ellis}-\cite{Chimento2}. Attempts to unify DE and DM is not a stopping process. For instance, using a static massive spherically symmetric scalar field coupled to gravity in the Schwarzschild-de Sitter (SdS) background, two different behaviors of unique matter in the distinct regions of spacetime at present era can be interpreted as a phase transition from dark matter to dark energy in the cosmic scales and construct a unified scenario \cite{Davood0}. Over the history of the research in Cosmology, different ideas and models concerning to the DE and DM were putted forward, we will mention some of the models with some limitations on the topic. Hereafter we would like to start with modified theories of Gravity and then to approach to the usual models existing in literature. As we have mentioned already DE was introduced by hand, but with modified theories of Gravity seems that we can have fundamental formulation for DE (but still mathematical). If we will modify GR we have possibility to obtain DE models, which seems natural way to have them in theory, but still model dependent.
For instance, in the framework of $f (T)$ theory \cite{Cai}-\cite{Davood11}, the modified
teleparallel action is given by
\begin{equation}
S=\frac{1}{16 \pi G} \int{d^{4}x \sqrt{-g}[f(T)+L_{m}]},
\end{equation}
where $L_{m}$ is the Lagrangian density of the matter inside the universe, $G$ is the gravitational constant and $g$ is the determinant of the metric tensor $g^{\mu \nu}$. Using flat Friedmann-Robertson-Walker (FRW) we can model DE as a fluid with
\begin{equation}
\rho_{T}=\frac{1}{2}(2Tf_{T}-f-T),
\end{equation}
and
\begin{equation}
P_{T}=-\frac{1}{2}[-8\dot{H}Tf_{TT}+(2T-4\dot{H})f_{T}-f+4\dot{H}-T],
\end{equation}
where $H$ is Hubble parameter, $f(T)$ is a function of $T$ and $T=-6H^{2}$. $F(R,T)$ theory is an other powerfull alternative model \cite{Davood12}, but it violates the first law of thermodynamic \cite{Davood13}. Other example could be the $f (G_{GB})$ \cite{S. Nojiri}-\cite{Garcia} action that describes Einstein’s gravity coupled
with perfect fluid plus a Gauss-Bonnet $(G_{GB})$ term given as
\begin{equation}
S=\int{d^{4}x\sqrt{-g}\left[\frac{1}{2k^{2}} R +f(G_{GB}) + L_{m}\right]},
\end{equation}
where $G_{GB}=R^{2}-4R_{\mu \nu}R^{\mu \nu}+R_{\mu \nu\lambda\alpha}R^{\mu \nu\lambda\alpha}$. $R$ represents the Ricci scalar curvature, $R_{\mu \nu}$ represents the Ricci curvature tensor and $R_{\mu \nu\lambda\alpha}$ represents the Riemann curvature tensor, $k^{2}= 8\pi G$, $g$ is the determinant of the metric tensor $g_{\mu \nu}$ and
$L_{m}$ is the Lagrangian of the matter present in the universe. The variation of the action $S$ with respect to $g_{\mu \nu}$ generates the field equations and in this case we have a right to associate
\begin{equation}
\rho_{GB}=G_{GB}f-f-24\dot{G}_{GB}H^{3}f^{\prime \prime},
\end{equation}
to a DE, where $^{.}$ represents time derivation, while $^{\prime}$ corresponds to the derivation of $f$ with respect to $G_{GB}$. We also would like to mention, that for $R+f(G_{GB})$ models there exist few exact solutions in cylindrical cases as wormholes \cite{Davood14}-\cite{Davood15}. We found approprite to mention also, that Horava-Lifshitz gravity also gives us a possibility to construct a non relativistic model for gravity \cite{Davood16}-\cite{Davood19}. Generally, an idea that darkness of the Universe can be associated with the fluid gives us possibility to be more confortable with the problem, because even if we will assume that DE is a scalar field (or tensor field), we have possibility to represent it as a fluid. This mean that with this we will have class of modified fluids, which means that we can consider fluids with EoS of  the general form i.e $F(\rho,P)$. Some of the examples of the saying are Chaplygin gas with EoS
\begin{equation}
P_{\small{CG}}=A\rho_{\small{CG}}-\frac{B}{\rho_{\small{CG}}^{\alpha}},
\end{equation}
where $A$,$B$ are constants. Van der Waals gas \cite{Capozziello Walls}-\cite{Martiros} 
\begin{equation}
P=\frac{\gamma \rho}{1-\beta\rho}-\alpha\rho^{2},
\end{equation}
and a gas \cite{Linder} of 
\begin{equation}
P=P_{0}+\alpha \rho+\beta\rho^{2},
\end{equation} 
are other examples. Very general dark fluid models with an inhomogeneous EoS were introduced in \cite{Nojiri}-\cite{Capozziello}. When the cosmological constant, $\Lambda$ is interpreted as the energy density of vacuum for an equation of state $p=-\rho$, if it is positive, then this assumption is not free from fundamental problems as we know. Actually, the cosmological constant problem is one of the most challenging questions in fundamental physics, as it is very hard to propose a mechanism that explains how the vacuum energy density can be lowered from its most natural value. Quintessence has been invoked \cite{Caldwell}-\cite{Wang1} as an advantageous alternative to the cosmological constant in order to explain the apparent accelerating expansion of the universe \cite{Perlmutter}-\cite{Zehavi}. Although quintessence models do not solve this problem, they may improve the related fine-tuning problem in the sense that they can explain a tiny value for the vacuum energy density with a scale comparable with the scales of high energy physics. Quintessence is described by a canonical scalar field $\phi$ minimally coupled to gravity. Compared to other scalar-field models such as phantoms and k-essence, quintessence is the simplest scalar-field scenario without having theoretical problems such as the appearance of ghosts and Laplacian instabilities. A slowly varying field along a potential can lead to the acceleration of the Universe. This mechanism is similar to slowroll inflation in the early Universe, but the difference is that non-relativistic matter cannot be ignored to discuss the dynamics of dark energy correctly. The pressure and the energy density of quintessence are given, respectively, by 
\begin{equation}
P_{Q}=\frac{\dot{\phi}^{2}}{2}-V(\phi),
\end{equation} 
and
\begin{equation}
\rho_{Q}=\frac{\dot{\phi}^{2}}{2}+V(\phi),
\end{equation}
where a dot represents a derivative with respect to $t$. We also can generalize it and for the energy density and pressure to write \cite{Davood20}
\begin{equation}\label{eq:GenP}
P_{Q}=\frac{\omega  } {2}\phi^{k}\dot{\phi}^{2}-V(\phi),
\end{equation} 
and
\begin{equation}\label{eq:Genrho}
\rho_{Q}=\frac{\omega  } {2}\phi^{k}\dot{\phi}^{2}+V(\phi).
\end{equation}
When $k=0$, Eqs (\ref{eq:GenP}) and (\ref{eq:Genrho}) transform to the canonical scalar field model with a rescaling of the field. Broadly speaking, we can classify dark energy models into two classes. The first one is based on a specific form of matter-such as quintessence \cite{Fujii}-\cite{Zlatev}, kessence \cite{Chiba}-\cite{Armendariz-Picon}, and the Chaplygin gas \cite{Kamenshchik}-\cite{Sadeghi}. The second one is based on the modification of gravity at large distances \cite{Sotiriou}-\cite{Odintsov3}. In both classes the dark energy equation of state dynamically changes in time, by which the models can be distinguished from the $\Lambda CDM$ model. Also, we would like to stress important question concerning to the possibilities allowing us to test various theories of gravity. It is obvious that it is not an easy task and requires very deep technological achievements. One of  the ways to solve the question can be hidden in the potential detection of gravitational waves, which is a powerfull tool in order to test various theories of gravity and discriminate among them \cite{Corda}-\cite{Corda1}.\\\\
In this article we would like to consider some phenomenological models of the Universe with varying $G$ and $\Lambda$. We assume that Universe contains a fluid which can be modeled as two-component fluid, for which total energy density and pressure are simply $\rho=\rho_{Q}+\rho_{b}$ and $P=P_{Q}+P_{b}$, where $Q$ index corresponds to the quintessence DE, while $b$ is for a barotropic fluid. To make system of equations of our consideration closed i.e make the number of unknowns and number of equations equal, we suppose that form $\Lambda$ is given as a function of other parameters of the model, at the same time we take that the potential is given $V(\phi)=e^{-\beta \phi}$ and is the same for all three models . Assumptions of the work are taken from nowhere, analysis of the results will give a hint which model is more realistic. For $\Lambda(t)$ we will discuss following forms
\begin{equation}\label{eq:Lambda1}
\Lambda(t)=\phi^{2}-\delta V(\phi),
\end{equation}
\begin{equation}\label{eq:Lambda2}
\Lambda(t)=H^{2}\phi^{-2}-\delta V(\phi),
\end{equation}
and
\begin{equation}\label{eq:Lambda3}
\Lambda(t)=q(\phi^{2}+\phi^{-2})-\delta V(\phi),
\end{equation}
where $\delta$ is a parameter of the model, $H$ is the Hubble parameter and $q$ is the deceleration parameter 
\begin{equation}
q=-1-\frac{\dot{H}}{H^{2}},
\end{equation}
and $q<0$ should be for the Universe expanding with an acceleration, because in that case we have $\ddot{a}>0$. For our conclusion concerning to the sign of $q$ we should recall the definition of the deceleration parameter within the scale factor $a$ given as
\begin{equation}
q=-\frac{\ddot{a}a}{\dot{a}^{2}}.
\end{equation}
An interaction between quintessence and barotropic fluid is taken to be
\begin{equation}\label{eq:int}
Q=3Hb\rho_{Q}+\gamma(\rho_{b}-\rho_{Q})\frac{\dot{\phi}}{\phi},
\end{equation}
where $b$ and $\gamma$ are positive constants.\\\\
Paper organized as follow: After introduction, field equations and some limited historical review on the art of variable $\Lambda$ and from where it started is given. In coming sections we will investigate 3 different models with different forms of $\Lambda$ and some graphical analysis of cosmological parameters are given. Finally, at the end of the work some conclusion is provided. Appendix is also attached to this work
\section*{\large{Field equations}}
Physics behind the varying $G$ can be different. Concerning to this reason we would like to refer our readers to the following two works \cite{Davood21}-\cite{Davood22} (and references therein) for general introduction into the topic. Field equations that govern our model with variable $G(t)$ and $\Lambda(t)$ (see for instance \cite{Abdussattar}) are,
\begin{equation}\label{s1}
R^{ij}-\frac{1}{2}Rg^{ij}=-8 \pi G(t) \left[ T^{ij} -
\frac{\Lambda(t)}{8 \pi G(t)}g^{ij} \right],
\end{equation}
where $G(t)$ and $\Lambda(t)$ are the functions of time. They are an approximation of a more general case of models with $G(t)$ and $\Lambda(t)$ \cite{Alfio}. By using the
following FRW metric for a flat Universe,
\begin{equation}\label{s2}
ds^2=-dt^2+a(t)^2\left(dr^{2}+r^{2}d\Omega^{2}\right),
\end{equation}
field equations can be reduced to the following Friedmann equations,
\begin{equation}\label{eq: Fridmman vlambda}
H^{2}=\frac{\dot{a}^{2}}{a^{2}}=\frac{8\pi G(t)\rho}{3}+\frac{\Lambda(t)}{3},
\end{equation}
and,
\begin{equation}\label{eq:fridman2}
\frac{\ddot{a}}{a}=-\frac{4\pi
G(t)}{3}(\rho+3P)+\frac{\Lambda(t)}{3},
\end{equation}
where $d\Omega^{2}=d\theta^{2}+\sin^{2}\theta d\phi^{2}$, and $a(t)$
represents the scale factor. The $\theta$ and $\phi$ parameters are
the usual azimuthal and polar angles of spherical coordinates, with
$0\leq\theta\leq\pi$ and $0\leq\phi<2\pi$. The coordinates ($t, r,
\theta, \phi$) are called co-moving coordinates.\\
Energy conservation $T^{;j}_{ij}=0$ reads as,
\begin{equation}\label{eq:conservation}
\dot{\rho}+3H(\rho+P)=0.
\end{equation}
Combination of Eqs. (\ref{eq: Fridmman vlambda}), (\ref{eq:fridman2}) and Eq. (\ref{eq:conservation}) gives the relationship between $\dot{G}(t)$ and $\dot{\Lambda}(t)$
\begin{equation}\label{eq:glambda}
\dot{G}=-\frac{\dot{\Lambda}}{8\pi\rho}.
\end{equation}
Ever since Dirac's proposition of a possible time variation of $G$,
a volume of works has been centered around the act of calculating
the amount of variation of the gravitational constant. For instance,
observation of spinning-down rate of pulsar $PSR J2019+2425$
provides the result,
\begin{equation}\label{eq:gvar1}
\left|\frac{\dot{G}}{G} \right|\leq (1.4-3.2) \times 10^{-11} yr^{-1}.
\end{equation}
Depending on the observations of pulsating white dwarf star $G 117-B
15A$, Benvenuto et al. \cite{Benvenuto} have set up the
astroseismological bound as,
\begin{equation}\label{eq:gvar2}
 -2.50 \times 10^{-10} \leq \left|\frac{\dot{G}}{G} \right|\leq 4 \times 10^{-10} yr^{-1}.
\end{equation}
For a review to "Large Number Hypothesis" (LNH) we refer our readers
to\cite{Saibal} and references therein. Spirit of Dirac's LNH gave us 
\begin{equation}\label{eq:vlambda}
\Lambda \propto t^{-2},
\end{equation}
for $\Lambda$. Other forms for $\Lambda(t)$ were considered over years based on phenomenological approach, some of examples are, for instance, $\Lambda \propto (\dot{a}/a)$, $\Lambda \propto \ddot{a}/a$ or $\Lambda \propto \rho$ to mention a few. Consideration of the two (many) component fluid models gives us possibility to perform some general manipulations for instance with $Q$ or $\Lambda$. For a two-fluid component universe recently we propose some phenomenological forms for $\Lambda(t)$:
\begin{equation}\label{eq:lambda1_old}
\Lambda(t)=\rho_{1}+\rho_{2}e^{-tH},
\end{equation}
\begin{equation}\label{eq:lambda2_old}
\Lambda(t)=H^{2}+(\rho_{1}+\rho_{2})e^{-tH},
\end{equation}
and
\begin{equation}\label{eq:lambda3_old}
\Lambda(t)=t^{-2}+(\rho_{1}-\rho_{2})e^{-tH},
\end{equation}
where $\rho_{1}$ and $\rho_{2}$ are energy densities of components, while $H$ is the Hubble parameter. For more details see Ref. \cite{Khurshudyan-Movsisyan}. In Ref. \cite{Khurshudyan-Hakobyan} we consider
\begin{equation}
\Lambda(t)=H^{2}+Aa^{-k},
\end{equation}
with $A$, $k$ as a constant and $a$ as a scale factor.

\section*{\large{Numerical results}}
In order to perform analysis of the models, first we will pay an attention to the mathematical part of the problem and then to explain obtained results. Well known fact is that the interaction between fluid components splits energy conservation equation and for each component we have
\begin{equation}
\dot{\rho}_{Q}+3H(\rho_{Q}+P_{Q})=-Q,
\end{equation}
and
\begin{equation}
\dot{\rho}_{b}+3H(\rho_{b}+P_{b})=Q.
\end{equation}
After some mathematics for quintessence model we will have
\begin{equation}\label{eq:engQ}
\ddot{\phi}+3H(\dot{\phi}(1+\frac{b}{2})+b\frac{e^{-\beta \phi}}{\dot{\phi}}) - \gamma \frac{\dot{\phi}}{\phi}(\frac{1}{2}\dot{\phi}+\frac{e^{-\beta \phi}}{\dot{\phi}})+\frac{dV}{d\phi}=-\gamma \frac{\rho_{b}}{\phi}.
\end{equation}
For a barotropic fluid with $P_{b}=\omega \rho_{b}$ we will have
\begin{equation}\label{eq:engb}
\dot{\rho}_{b}+ \left [ 3H(1+\omega)-\gamma \frac{\dot{\phi}}{\phi}  \right] \rho_{b}=(3Hb-\gamma \frac{\dot{\phi}}{\phi})(\frac{1}{2}\dot{\phi}^{2}+e^{-\beta \phi}),
\end{equation}
where the Hubble parameter can be obtained from Friedmann equation
\begin{equation}\label{eq:H}
H=\sqrt{\frac{8 \pi G}{3} (\rho_{Q}+\rho_{b})+\frac{\Lambda(t)}{3}}.
\end{equation}
Dynamics of $G(t)$ can be recovered from 
\begin{equation}\label{eq:G}
\frac{\dot{G}(t)}{G(t)}+\frac{\dot{\Lambda}(t)}{3H^{2}-\Lambda(t)}=0.
\end{equation}
Future analysis is based on the last 4 equations and the specific form of $\Lambda(t)$. Such combination will allow us to recover the scale factor $a$, therefore the Hubble parameter $H$, the deceleration parameter $q$, cosmological parameters like $r,s, j, k$ for statefinder diagnostics. Moreover, with numerical analysis we can recover behavior of $G$ and field $\phi$, therefore the behavior of potential of the field $V(\phi)$. 

\subsection*{\large{Model 1}}
\begin{figure}[h!]
 \begin{center}$
 \begin{array}{cccc}
\includegraphics[width=50 mm]{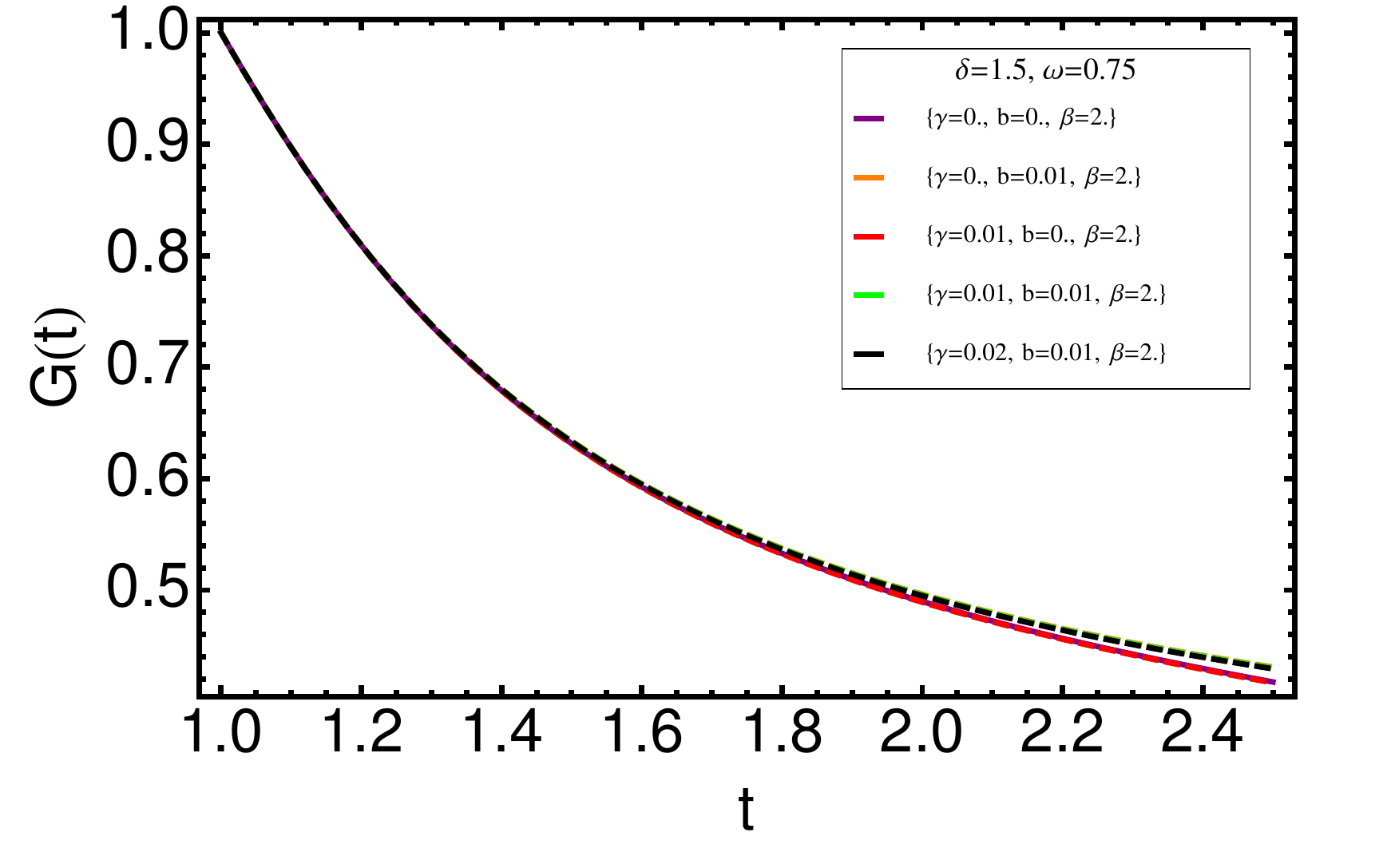} &
\includegraphics[width=50 mm]{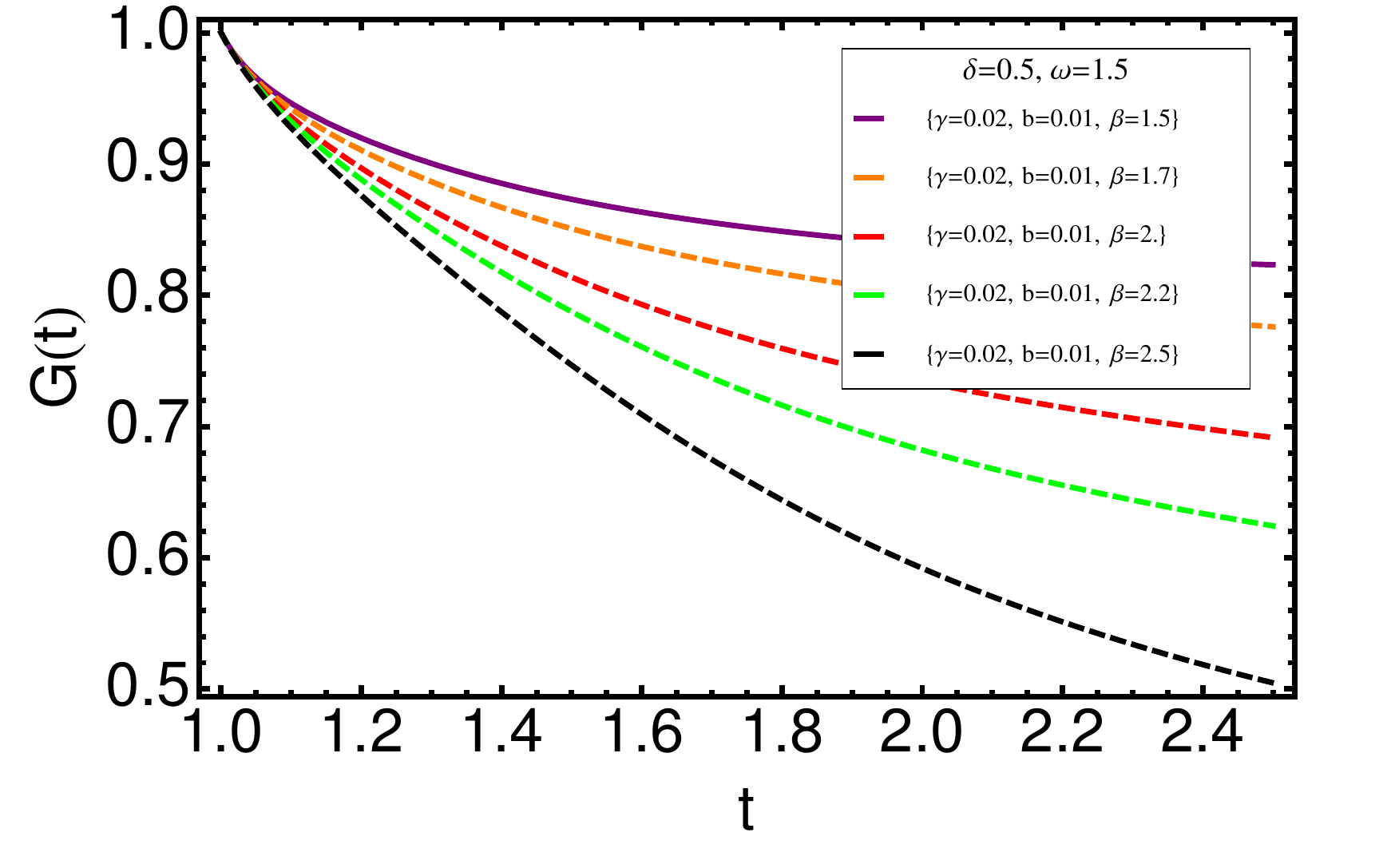}\\
\includegraphics[width=50 mm]{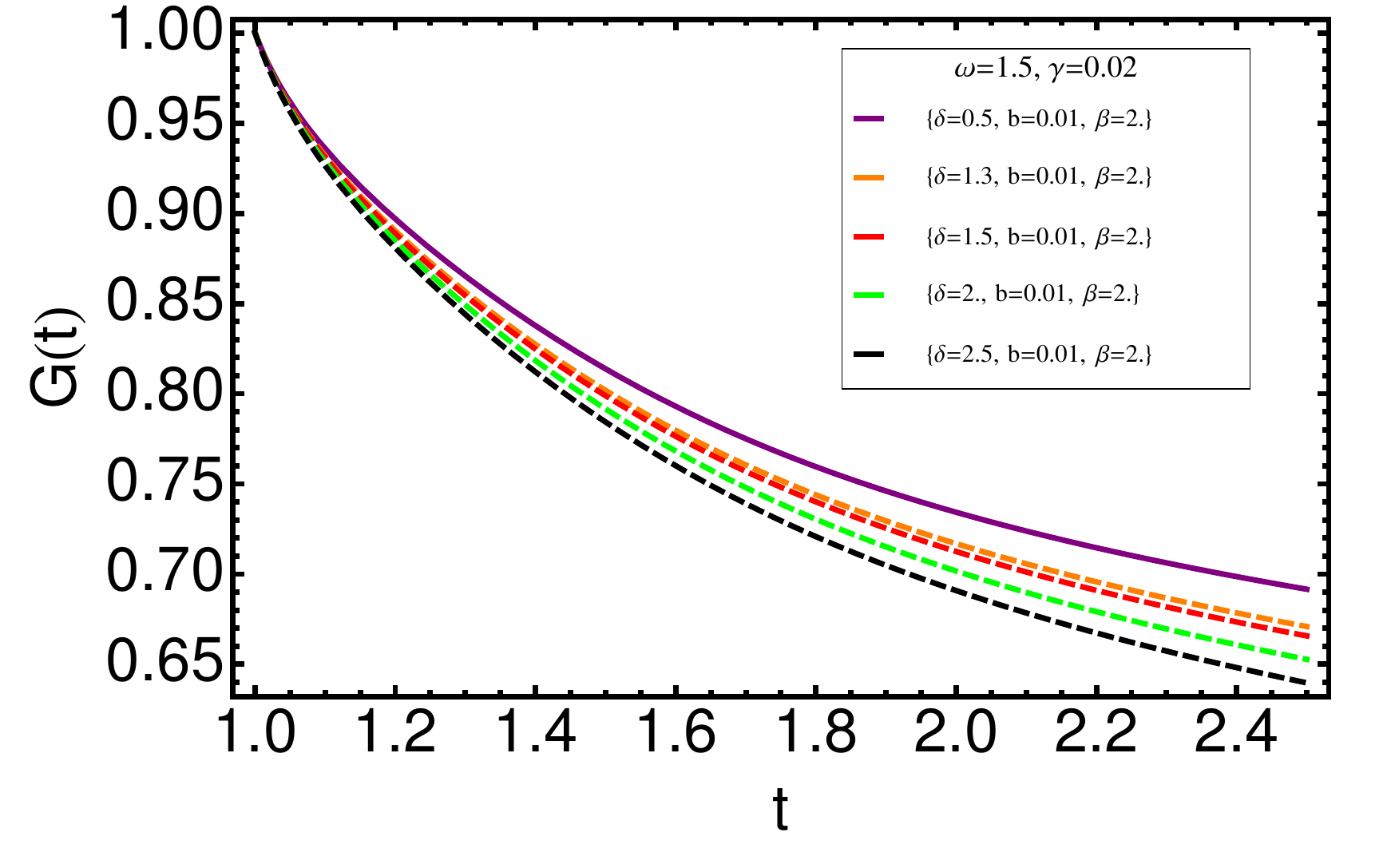}&
\includegraphics[width=50 mm]{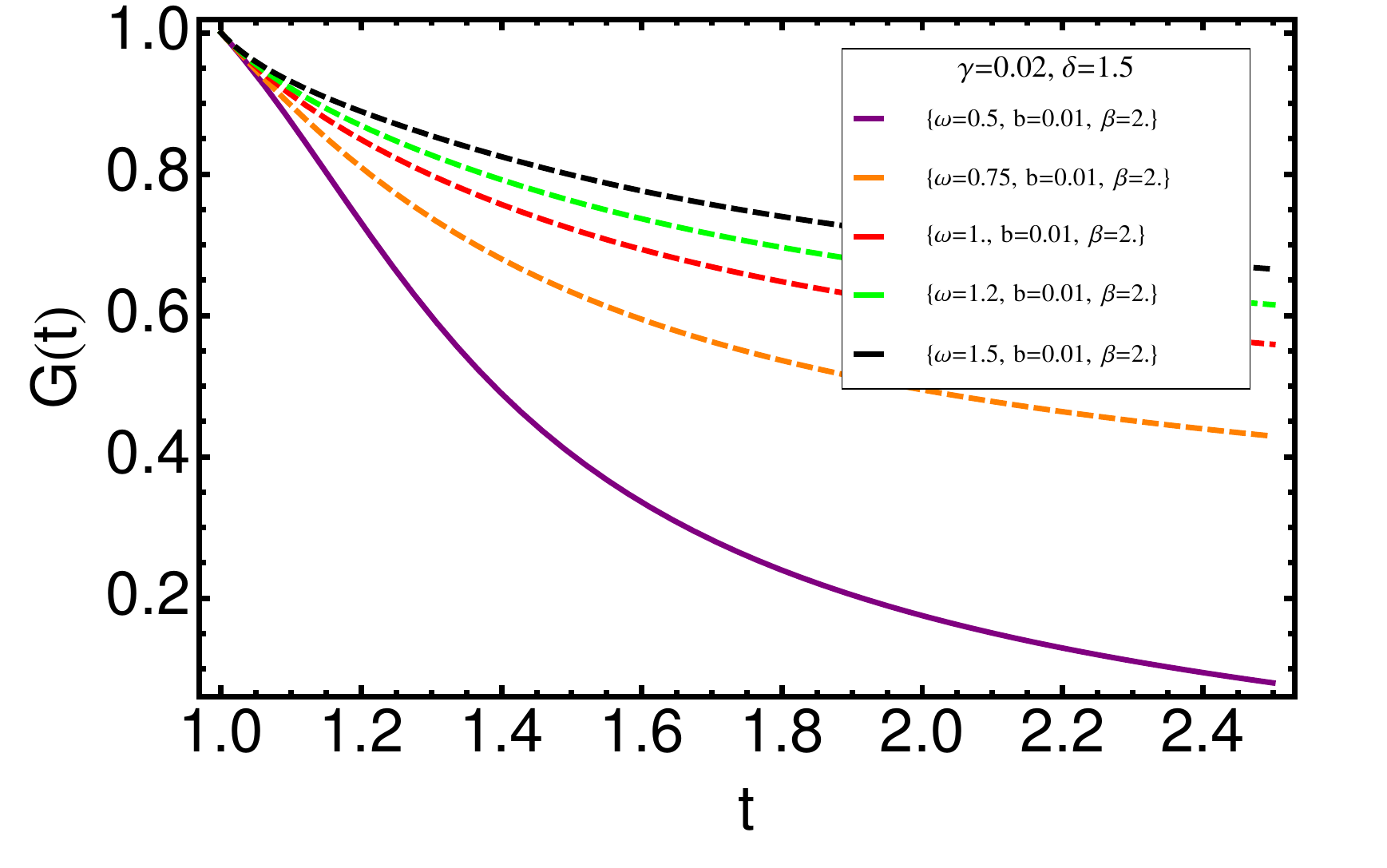}
 \end{array}$
 \end{center}
\caption{Behavior of $G$ against $t$ based on the solution of Eq. (\ref{eq:G1}). Model 1}
 \label{fig:1}
\end{figure}

\begin{figure}[h!]
 \begin{center}$
 \begin{array}{cccc}
\includegraphics[width=50 mm]{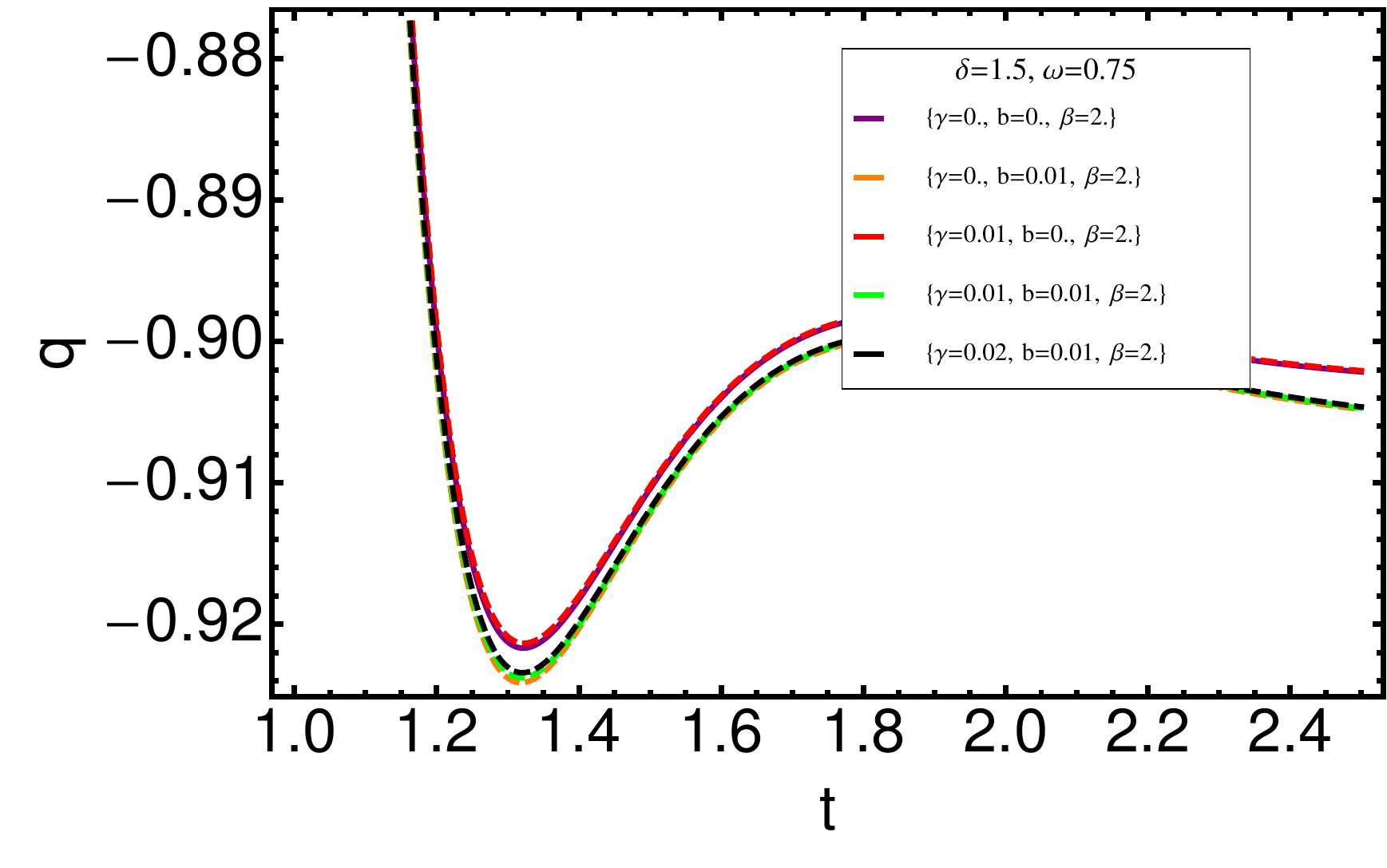} &
\includegraphics[width=50 mm]{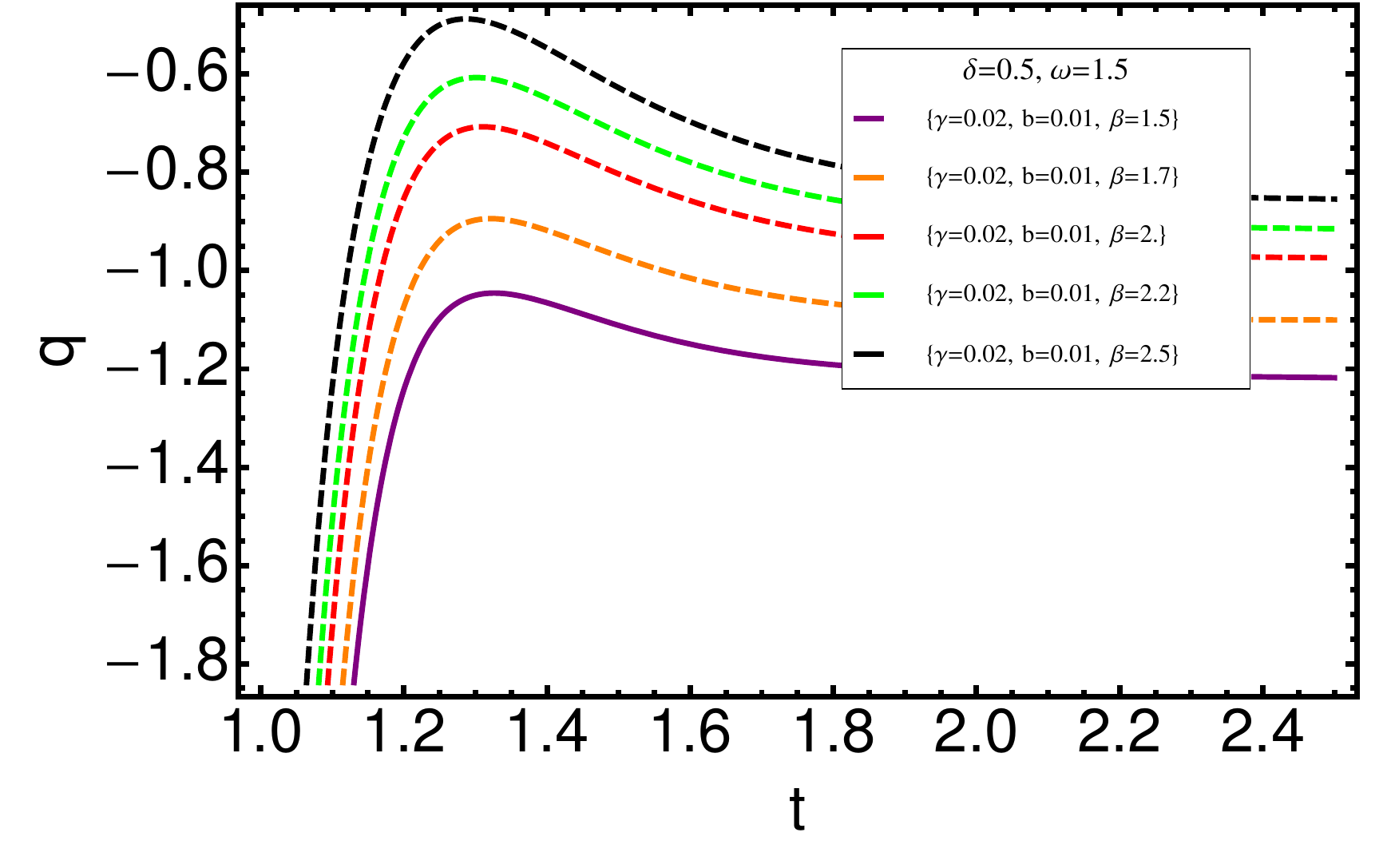}\\
\includegraphics[width=50 mm]{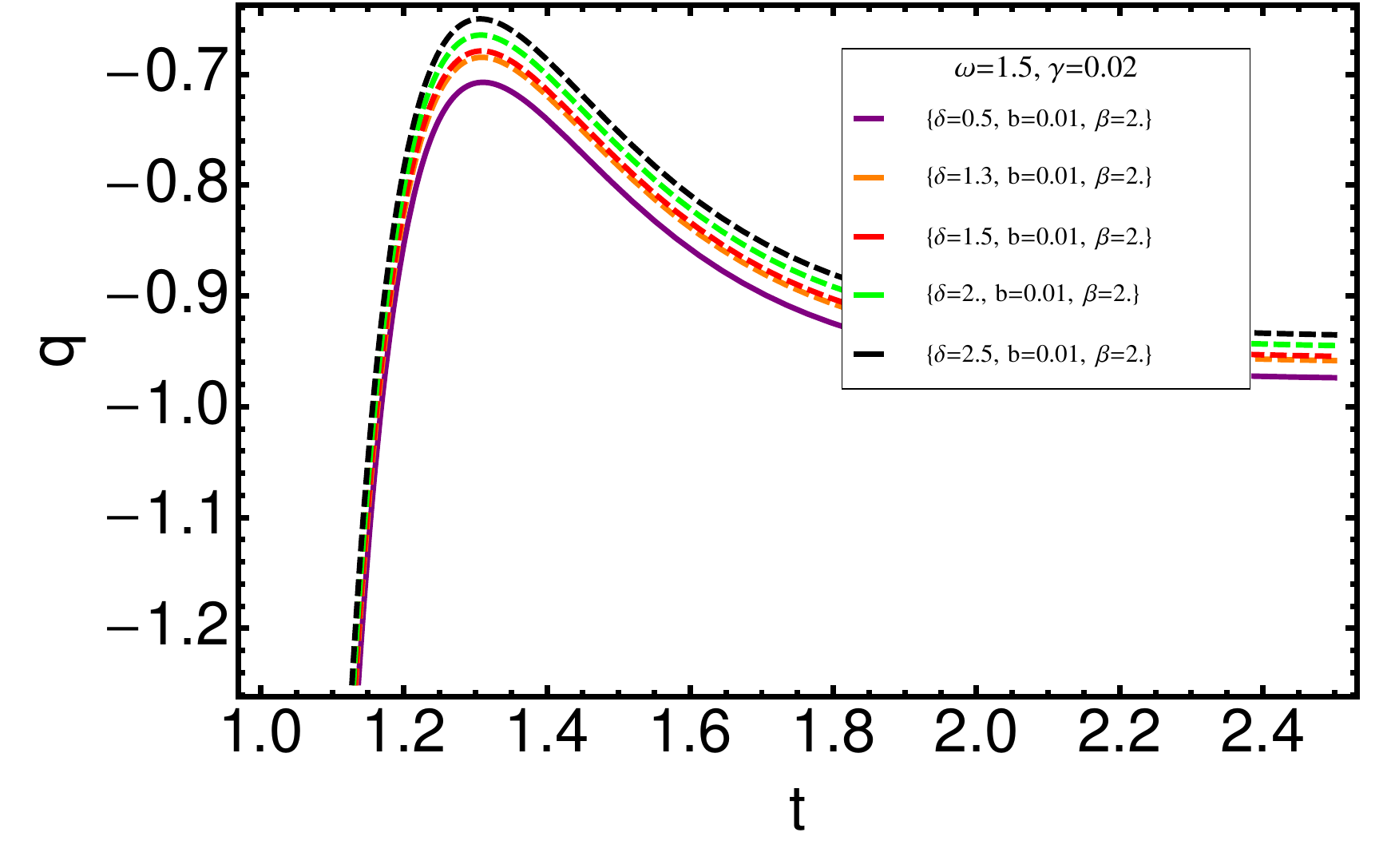}&
\includegraphics[width=50 mm]{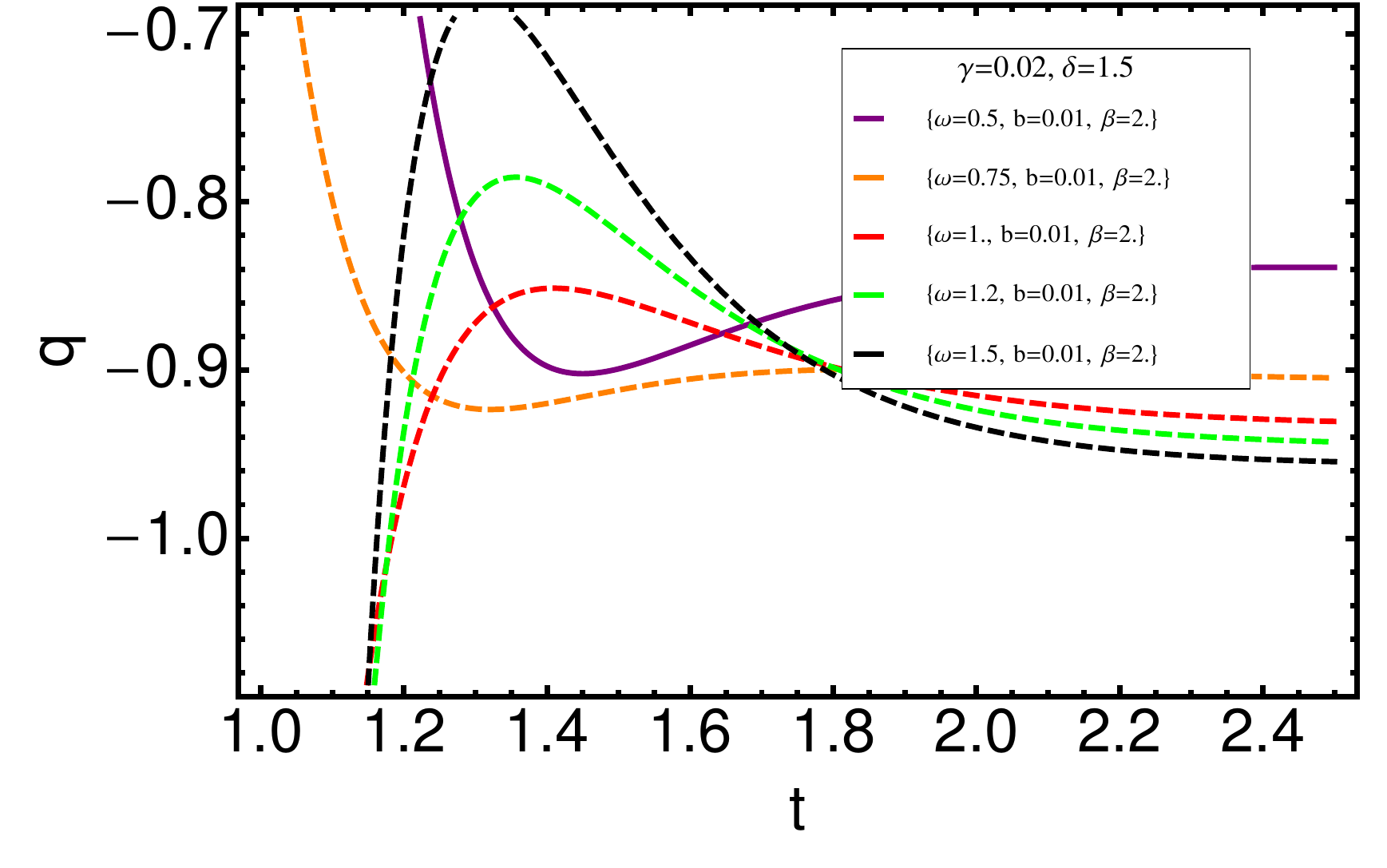}
 \end{array}$
 \end{center}
\caption{Behavior of $q$ against $t$ based on the solution of Eq. (\ref{eq:H1}). Model 1}
 \label{fig:2}
\end{figure}

\begin{figure}[h!]
 \begin{center}$
 \begin{array}{cccc}
\includegraphics[width=50 mm]{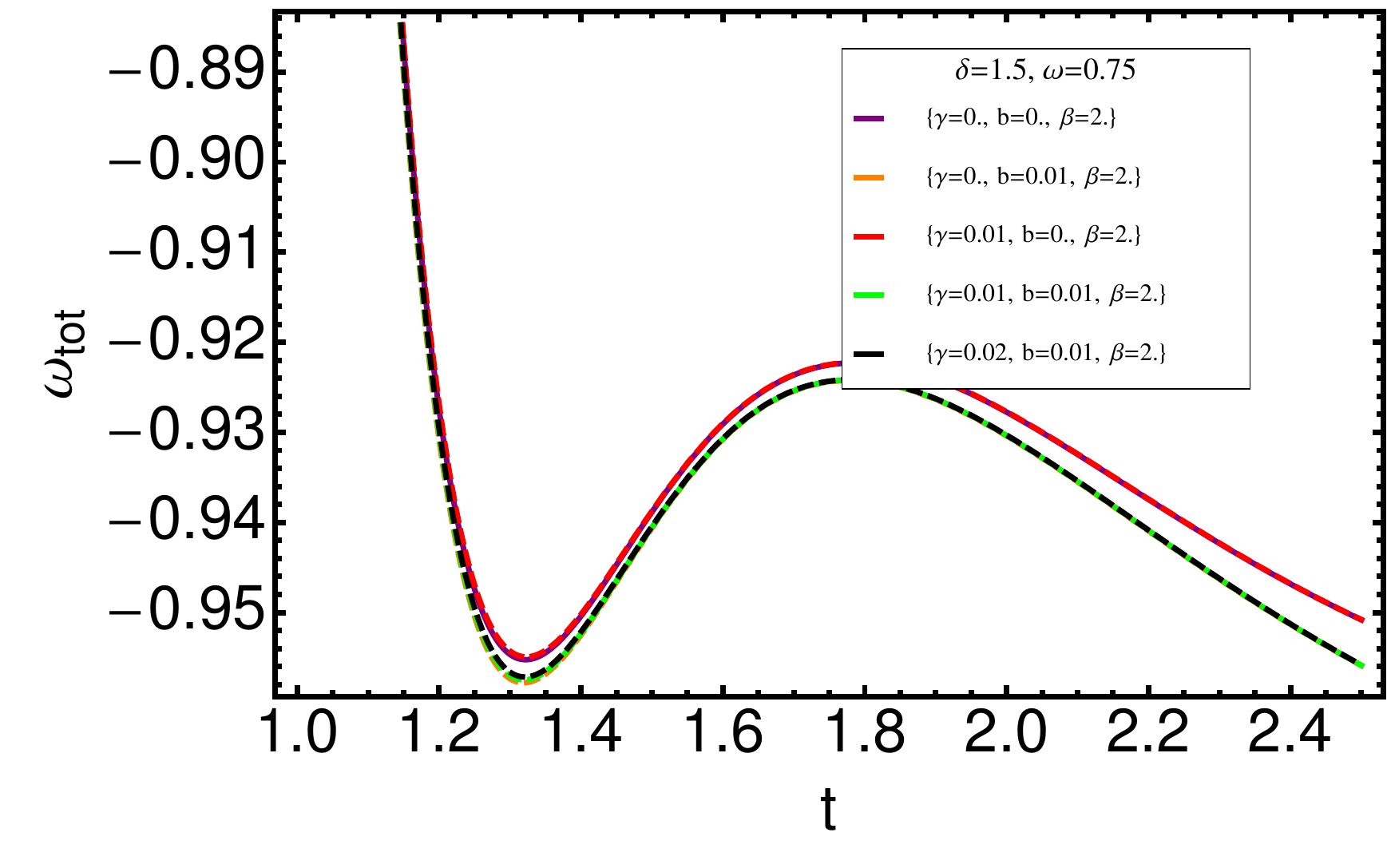} &
\includegraphics[width=50 mm]{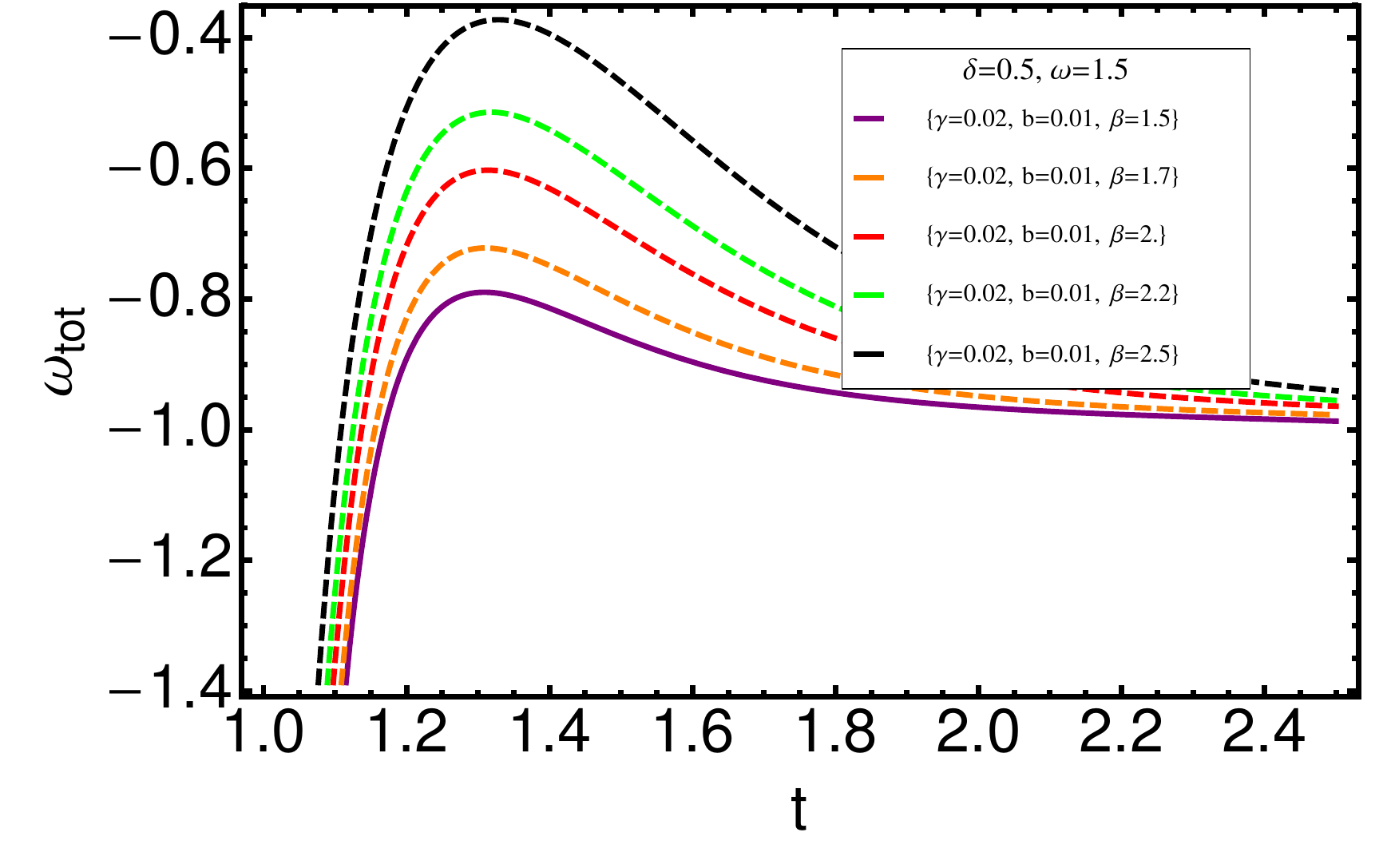}\\
\includegraphics[width=50 mm]{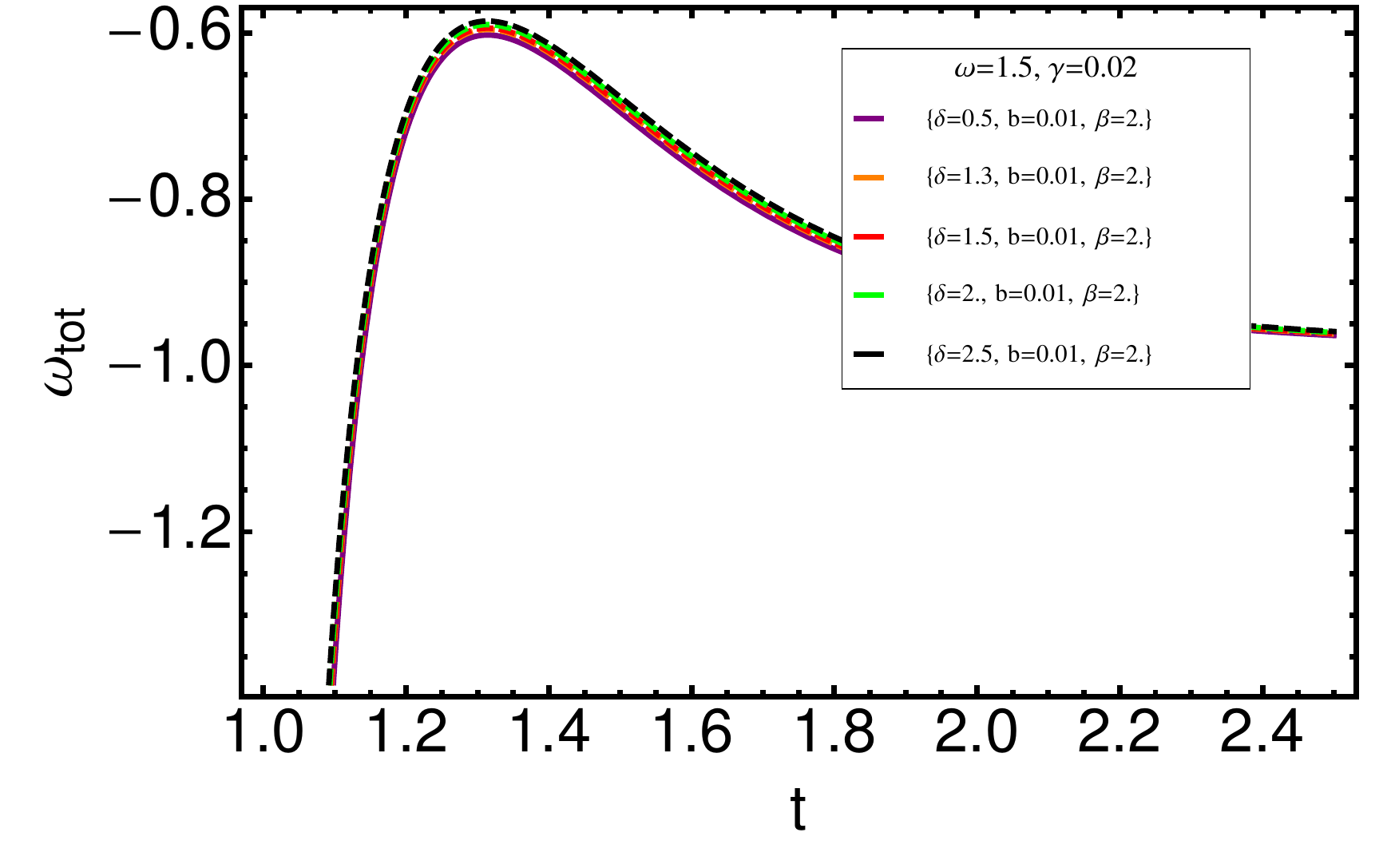}&
\includegraphics[width=50 mm]{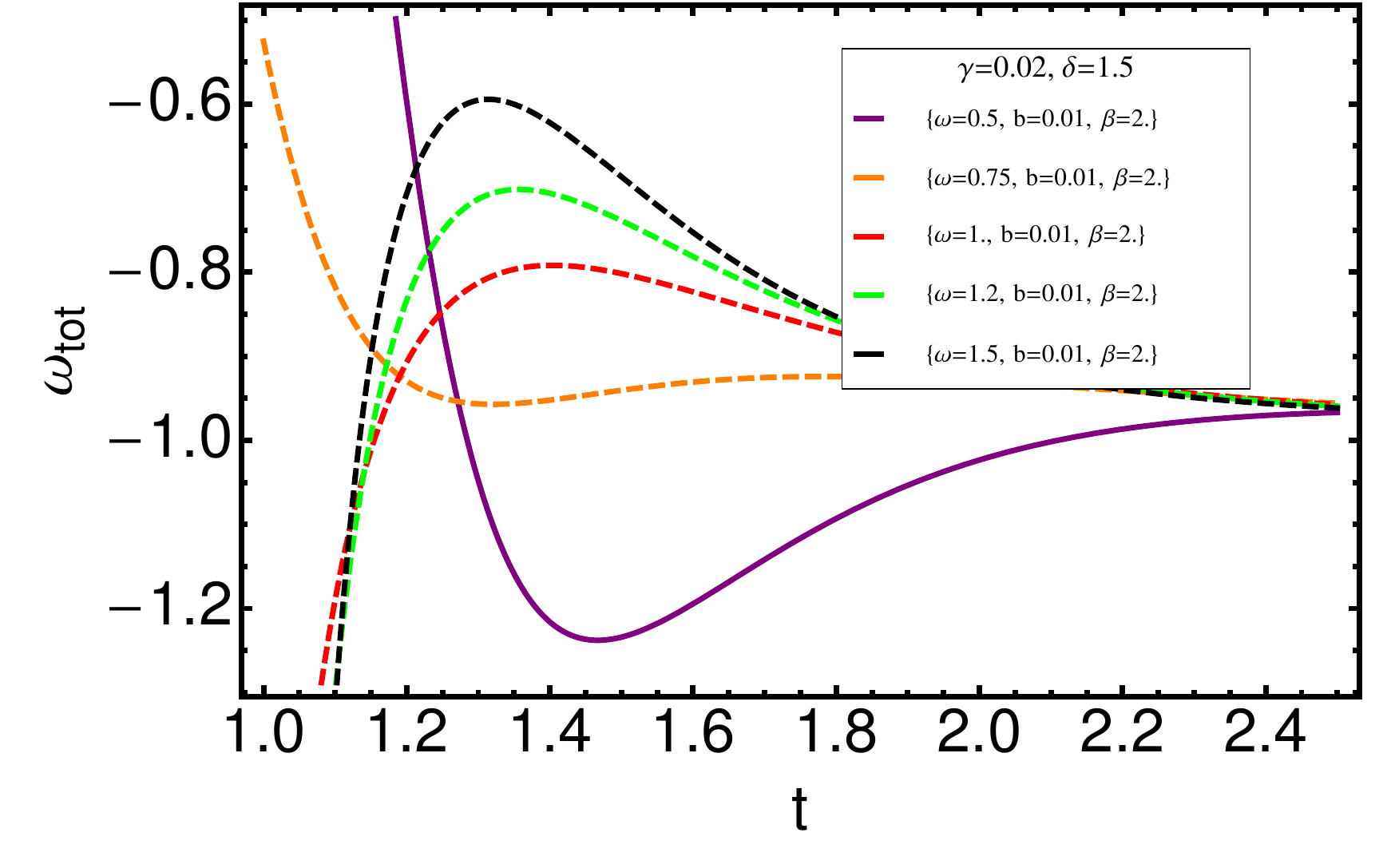}
 \end{array}$
 \end{center}
\caption{Behavior of $\omega_{tot}$ against $t$. Model 1}
 \label{fig:3}
\end{figure}

In this section we will start analyse model 1, for which we have taken $\Lambda$ to be
\begin{equation}\label{eq:mod1}
\Lambda(t)=\phi^{2}-\delta e^{-\beta \phi}.
\end{equation}
Then, for the dynamics of $G$ we will have
\begin{equation}\label{eq:G1}
\frac{\dot{G}(t)}{G(t)}+\frac{\dot{\phi} ( 2\phi + \beta \delta e^{-\beta \phi} ) }{3H^{2} -\phi^{2}+\delta e^{-\beta \phi}}=0,
\end{equation}
and for the Hubble parameter 
\begin{equation}\label{eq:H1}
H=\sqrt{\frac{8 \pi G}{3}(\rho_{b}+\frac{1}{2}\dot{\phi}^{2})+\frac{8 \pi G-\delta}{3}e^{-\beta \phi}+\frac{\phi^{2}}{3}}.
\end{equation}
Performing numerical analysis of the model we obtained graphical behavior of the several important cosmological parameters ($H$, $q$, $\phi$, $V$, $\omega_{tot}$, $\omega_{Q}$ etc) as a function of the parameters ($\omega$, $b$, $\gamma$, $\beta$, $\delta$) of the model. Graphical behavior of $G$ is given in Fig. \ref{fig:1}. We see that $G$ is a decreasing function over time. First plot of the Fig. \ref{fig:1} (top-left) presents behavior of $G$ as a function of interaction parameters $\gamma$ and $b$. The plot includes the case of the non interacting components (blue line with $b=0$ and $\gamma=0$), two different cases including only one component of interaction term $Q$ ($\gamma=0$, $\beta\neq 0$ and $b=0$ with $\gamma\neq0$). Green and black lines corresponds to the general case when $Q$ has a full form as proposed in this article. We see that an existence of the interaction between components in behavior of $G$ will be seen for later stages of evolution only. Second plot (top-right) shows behavior of $G$ as a function of $\beta$. It can be noticed, that within increasing $\beta$ we increase the rate of decreasing i.e with higher $\beta$ $G$ will decrease faster (black line). The third plot (bottom-left) is for study the behavior of $G$ as a function of $\delta$. We clearly see that for fixed values of $\gamma=0.02$, $b=0.01$, $\omega=1.5$ and $\beta=2$ with incresing $\delta$ we again will have faster decreasing of $G$. Last plot indicates, that with decreasing $\omega$ we will increase decreasing rate of $G$. Generally, we believe that  deceleration parameter must be $q>-1$. Behavior of the deceleration parameter $q$ can be found in Fig. \ref{fig:2}. For whole history of the Universe we have $q<0$. First plot of 
Fig. \ref{fig:2} (top-left) which represents behavior of $q$ as a function of interaction parameters $b$ and $\gamma$, when $\delta=1.5$, $\beta=2$ and $\omega=0.75$ (as an example). We observe that $q$ is decreasing-increasing-decreasing function. We see that the model including interaction differs from the non-interacting model. We also conclude that the models with $Q=3Hb\rho_{Q}$ and $Q=\gamma(\rho_{b}-\rho_{Q})\frac{\dot{\phi}}{\phi}$ give identical behavior for $q$. The top-right plot shows that $q$ is an increasing function for early stages of evolution,  then it will decrease and for later stages of evolution of the Universe will become a constant. This behavior is obtained for the $\omega=1.5$, for different values of $\beta$. Our analysis shows that for $\omega>1$ $\beta>2$ should take a place in order to obtain $q>-1$. Bottom-left  and bottom-right plots give a behavior of deceleration parameter as a function of $\delta$ and $\omega$ respectively. At the end of this section we will analyse behavior of $\omega_{tot}=\frac{P_{Q}+P_{b}}{\rho_{Q}+\rho_{b}}$ (Fig. \ref{fig:3}). For $\omega<1$, $\omega_{tot}$ is a decreasing-increasing-decreasing function and always $\omega_{tot}>-1$ (see the top-left plot, which presents the behavior of $\omega_{tot}$ as a function of $b$ and $\gamma$). With $\omega>1$ and varying $\beta$ we have increasing-decreasing $\omega_{tot}$ for early stages of evolution, while for later stages of evolution $\omega_{tot}=-1$ (top-right plot of Fig. \ref{fig:3}). The bottom-right plot of Fig. \ref{fig:3} represents behavior of $\omega_{tot}$ for two cases $\omega<1$ and $\omega>1$ (upper limit on $\omega$ is $2$, while lower limit is $0$) and $\omega=1$. For all cases for later stages $\omega_{tot}=-1$, while for early stages of evolution, we have completely opposite and different behavior.
\subsection*{\large{Model 2}}
\begin{figure}[h!]
 \begin{center}$
 \begin{array}{cccc}
\includegraphics[width=50 mm]{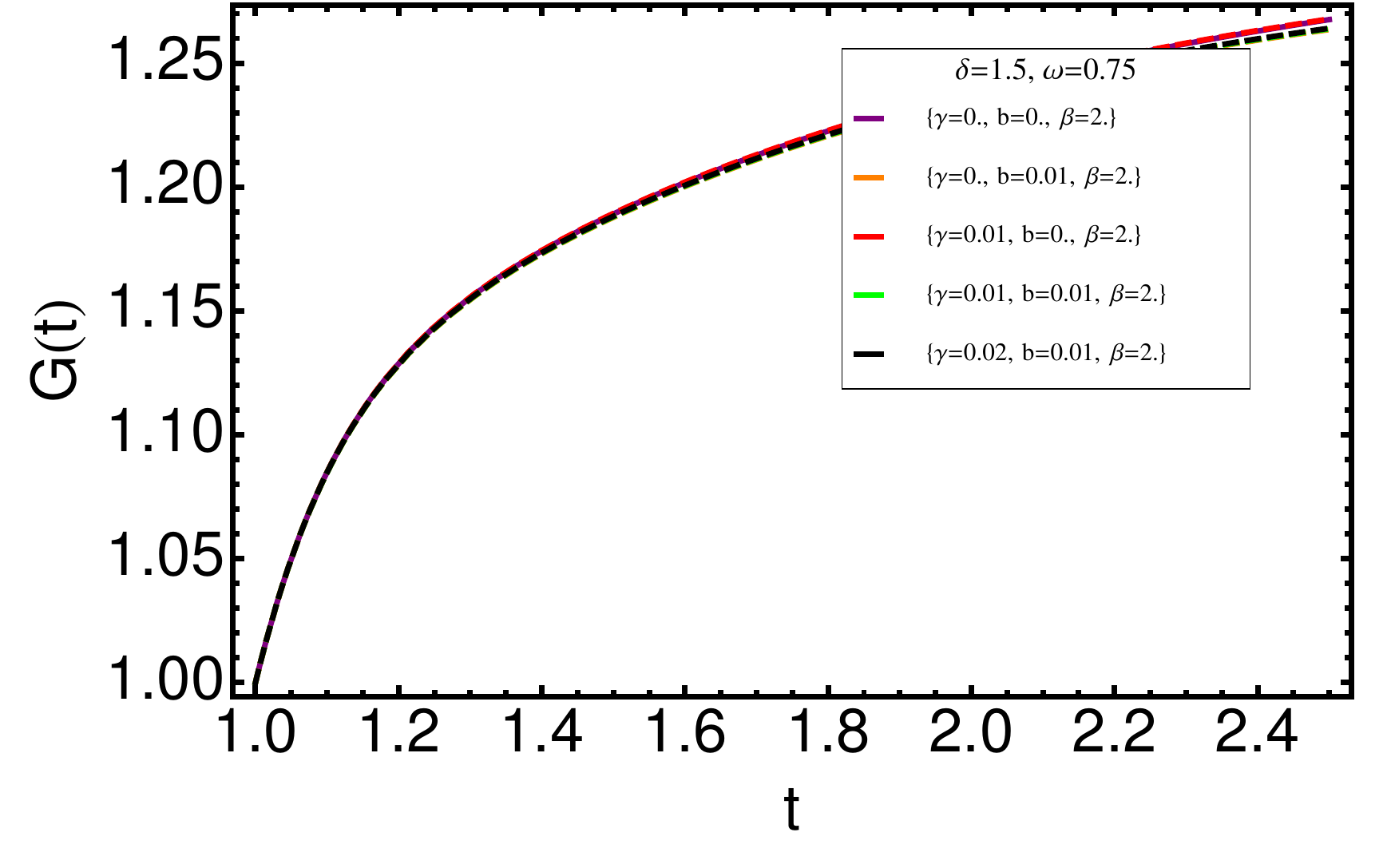} &
\includegraphics[width=50 mm]{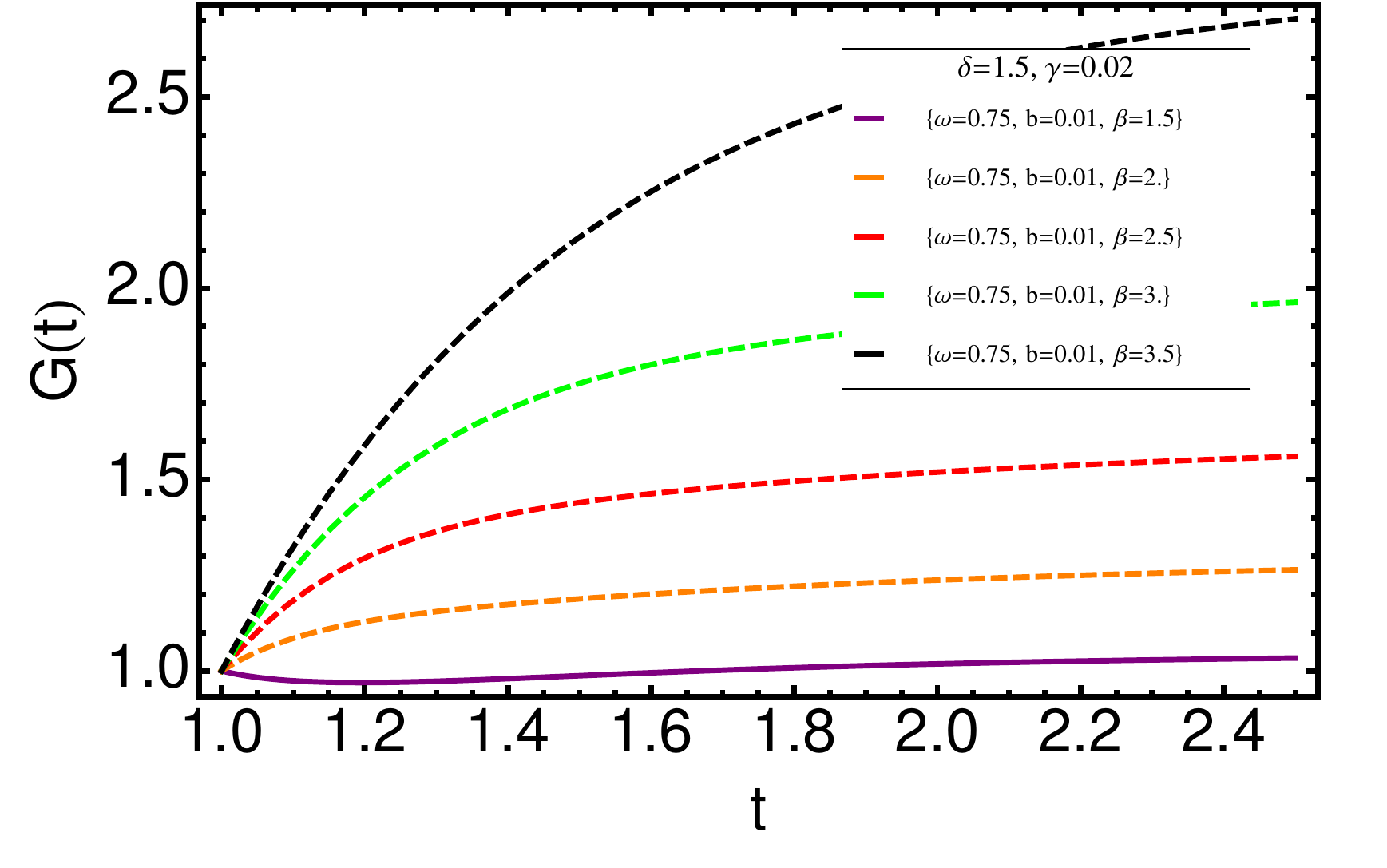}\\
\includegraphics[width=50 mm]{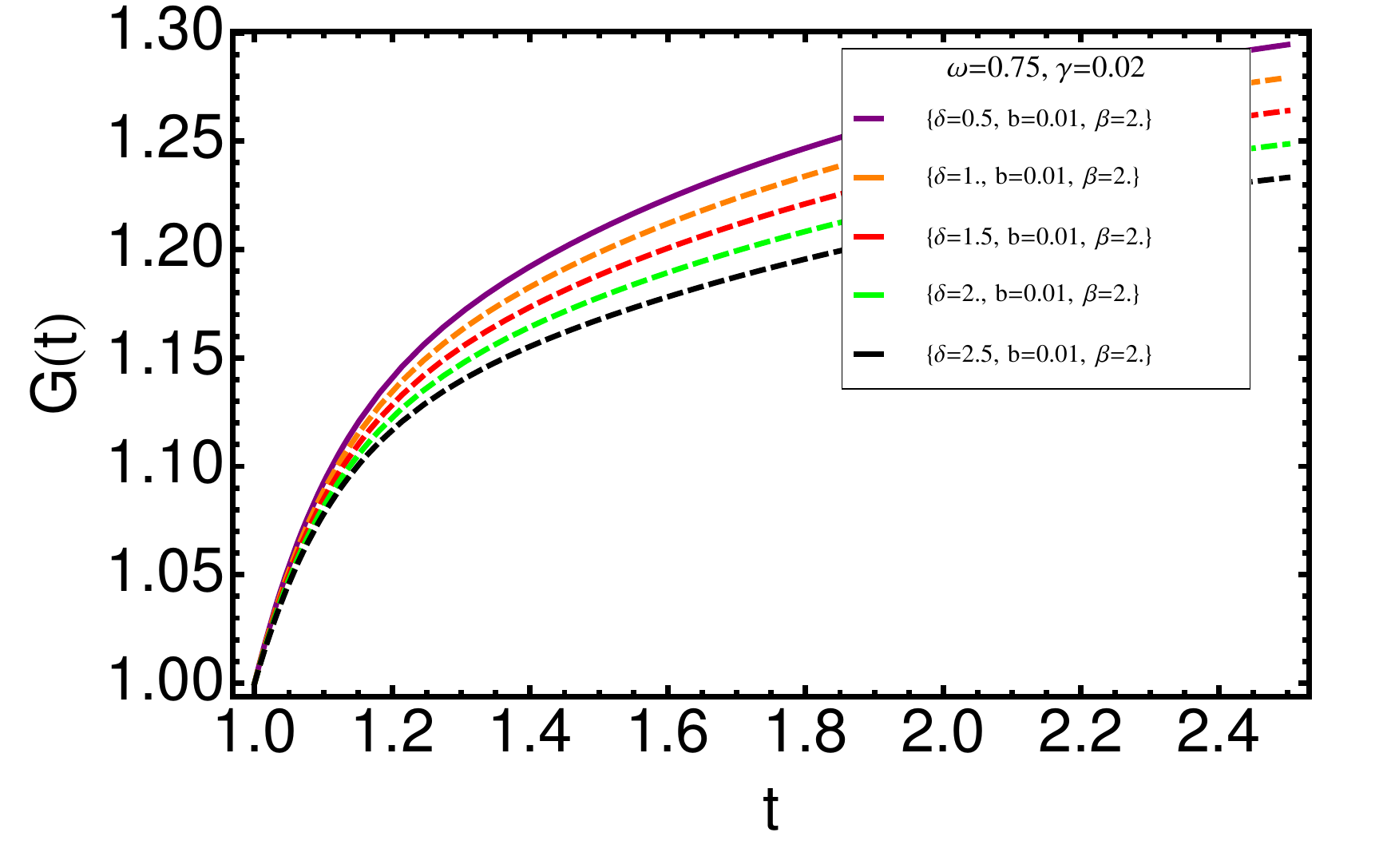}&
\includegraphics[width=50 mm]{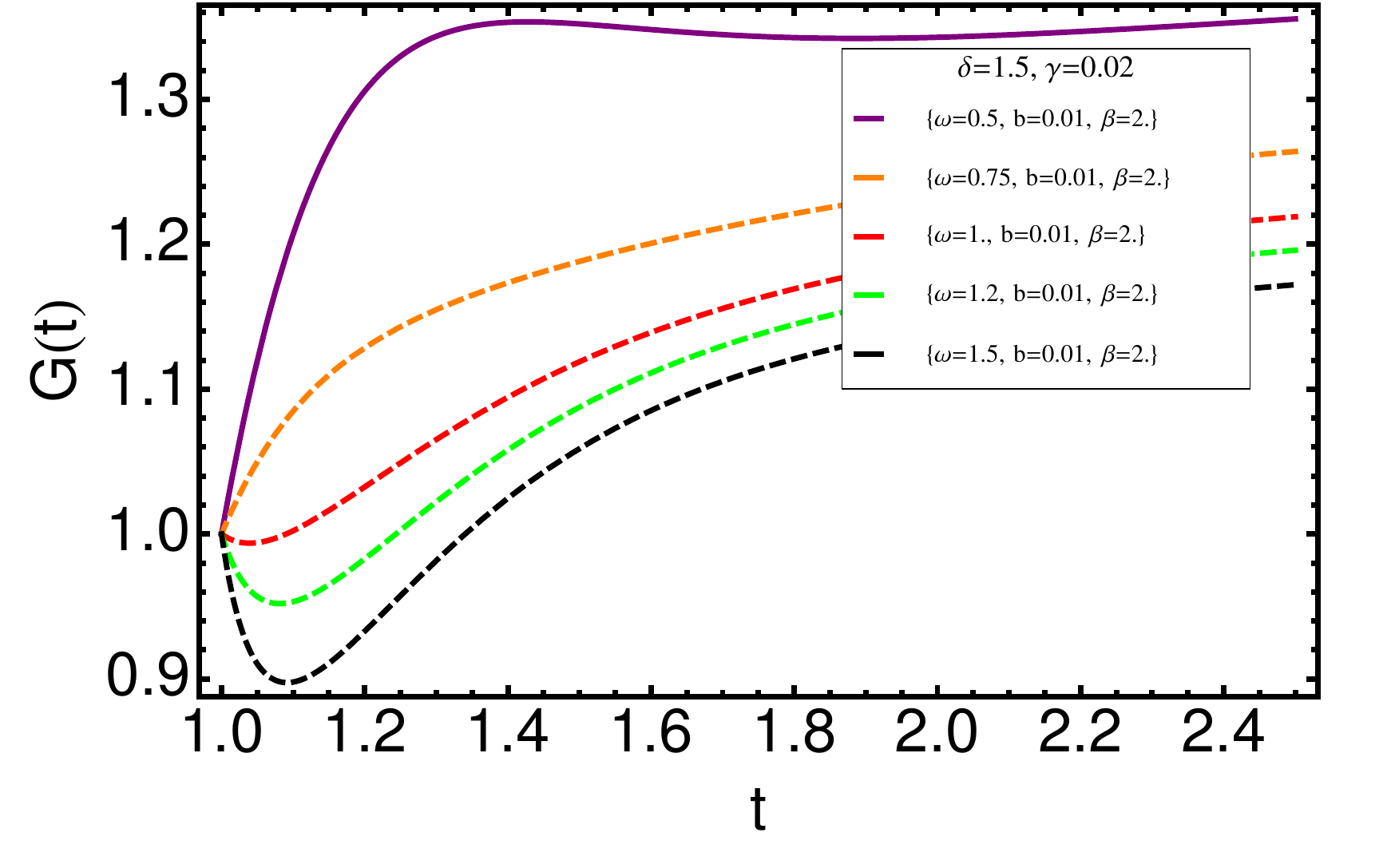}
 \end{array}$
 \end{center}
\caption{Behavior of $G$ against $t$ based on the solution of Eq. (\ref{eq:G2}). Model 2}
 \label{fig:4}
\end{figure}

\begin{figure}[h!]
 \begin{center}$
 \begin{array}{cccc}
\includegraphics[width=50 mm]{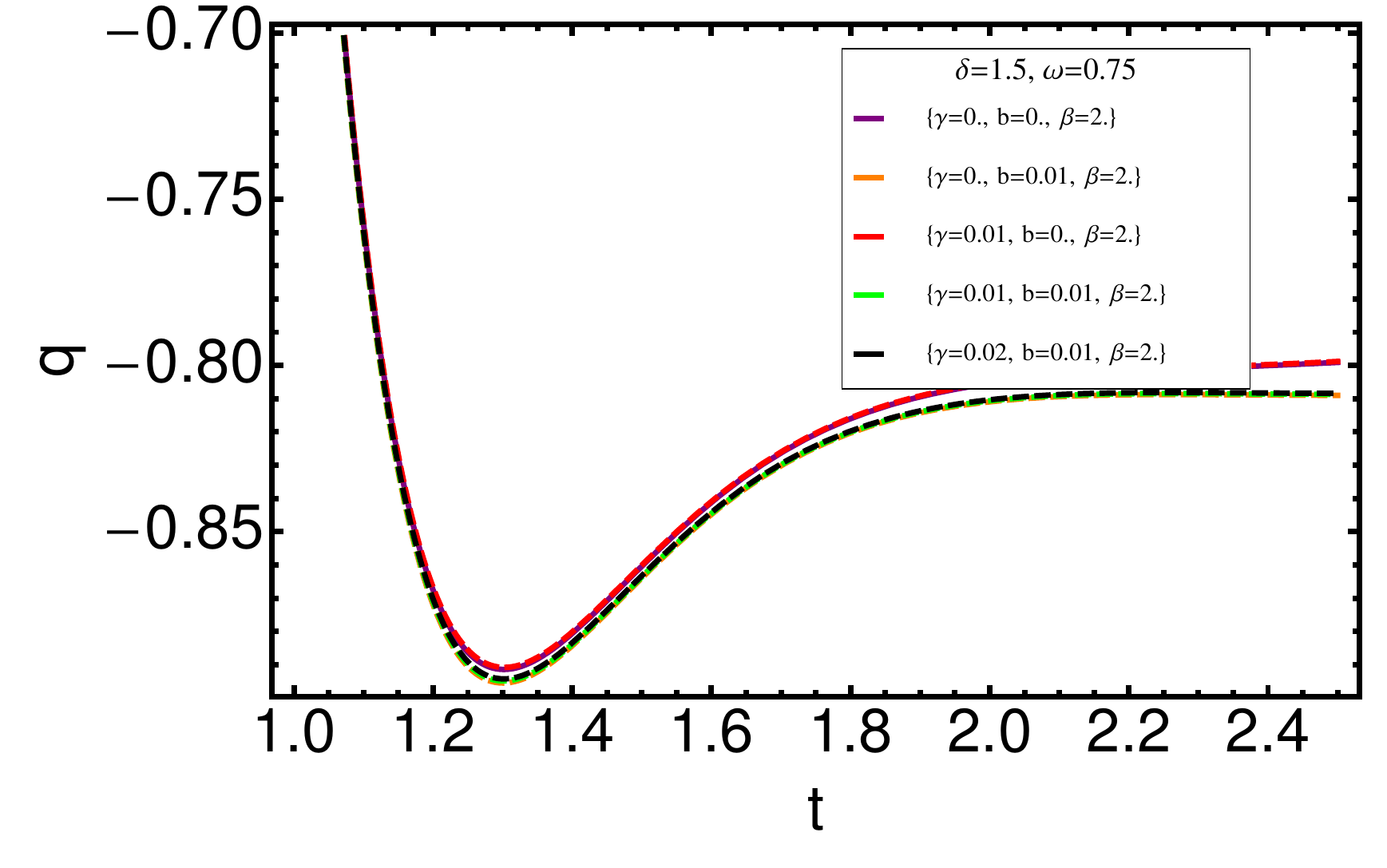} &
\includegraphics[width=50 mm]{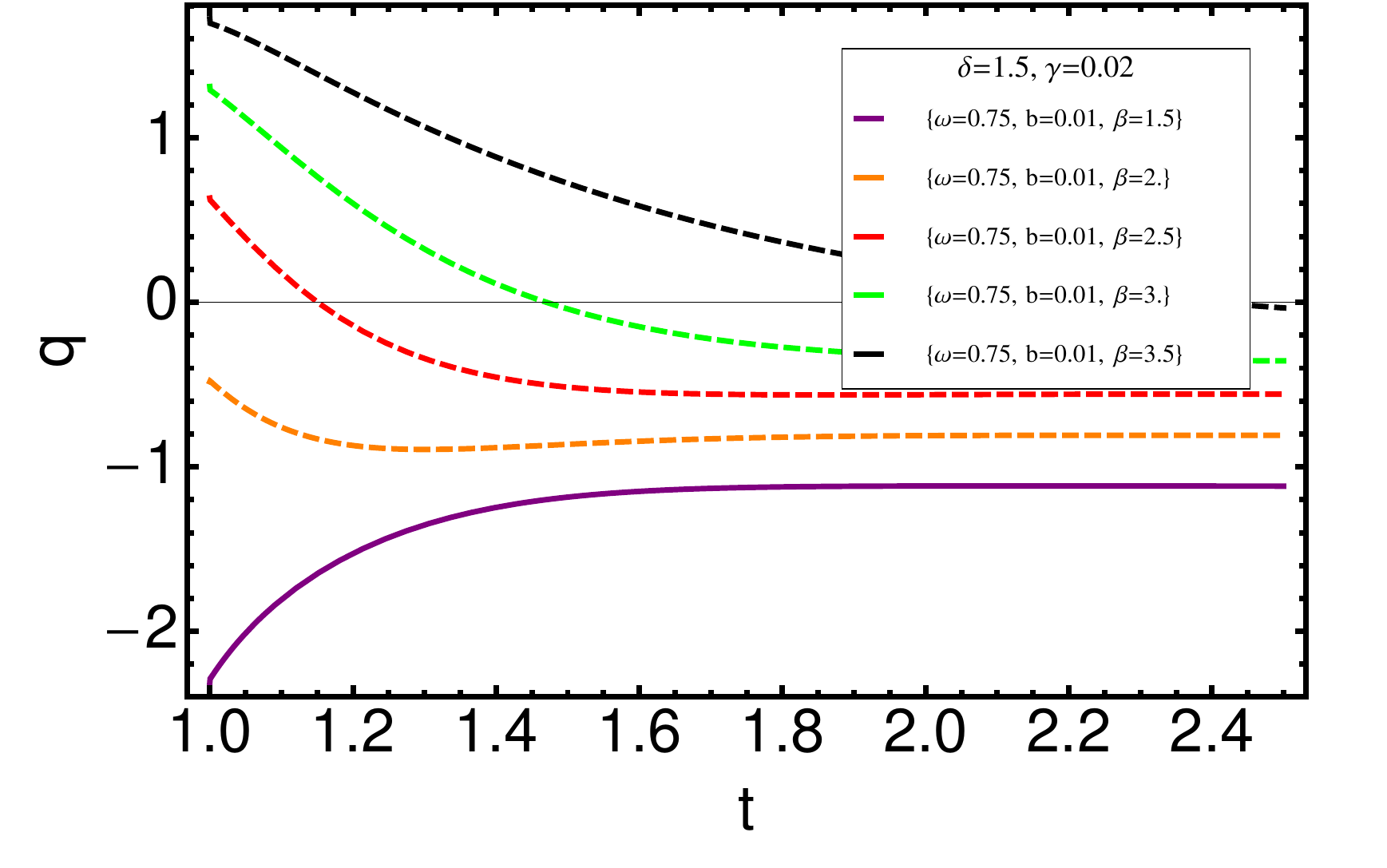}\\
\includegraphics[width=50 mm]{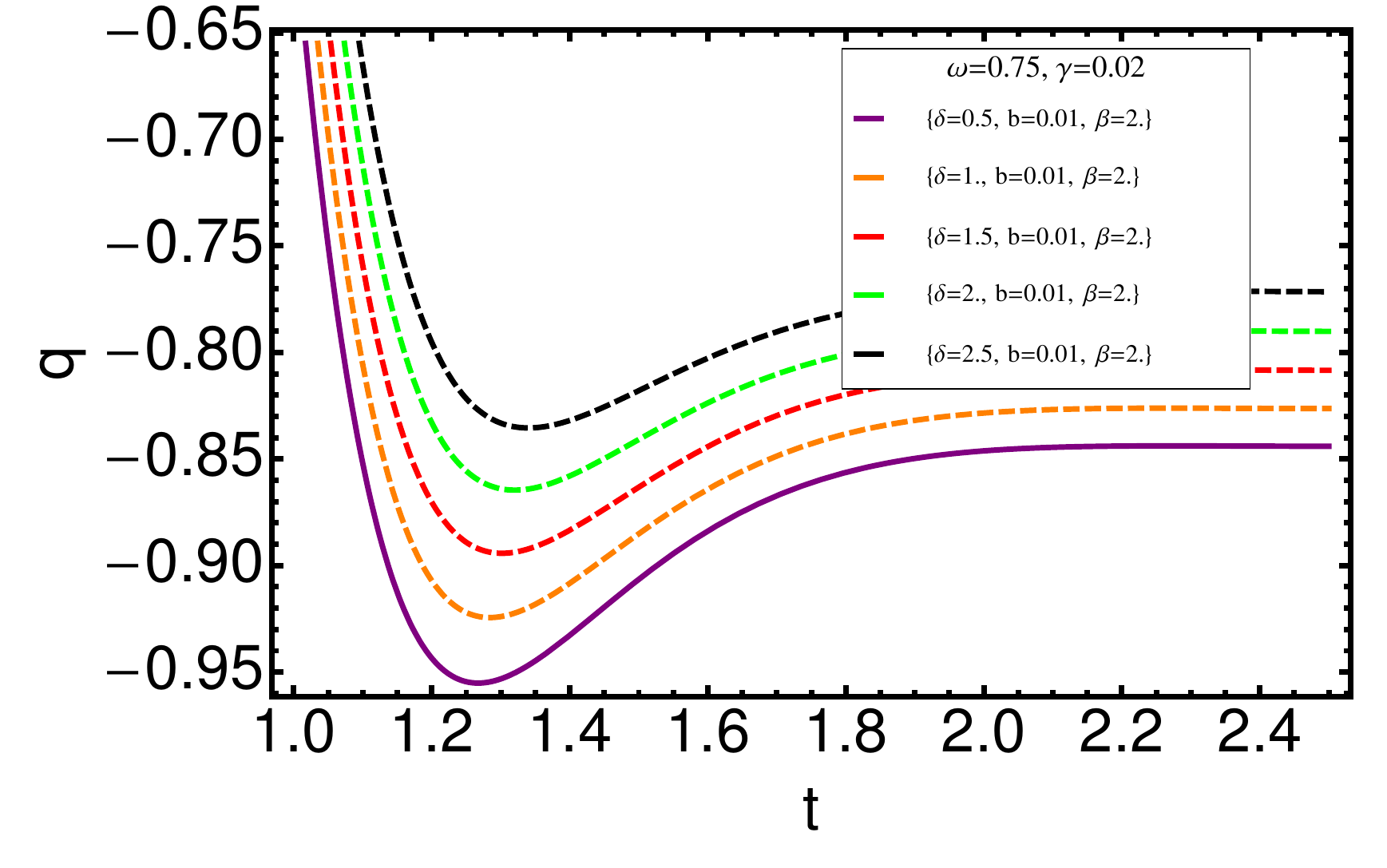}&
\includegraphics[width=50 mm]{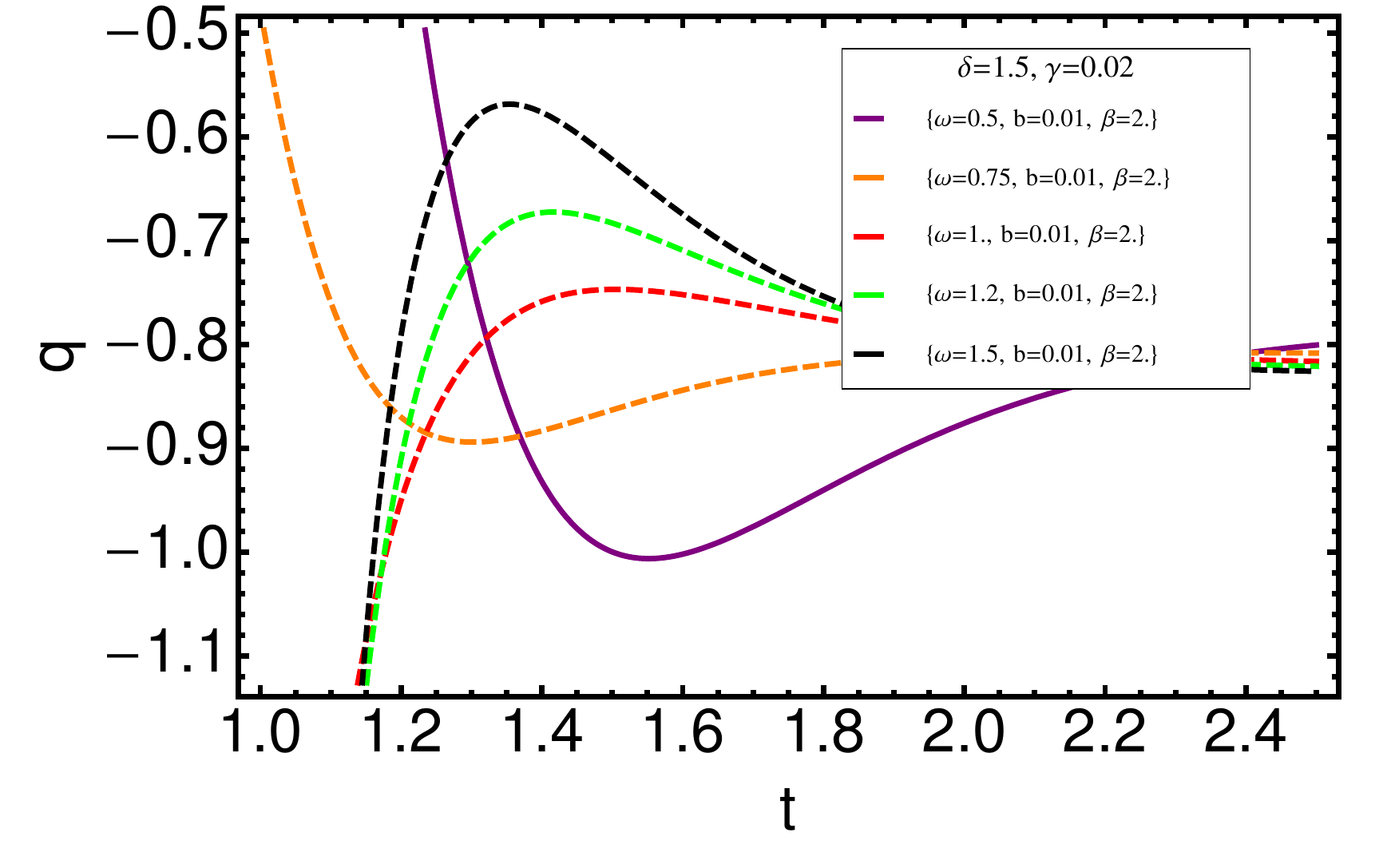}
 \end{array}$
 \end{center}
\caption{Behavior of $q$ against $t$ based on the solution of Eq. (\ref{eq:H2}). Model 2}
 \label{fig:5}
\end{figure}

\begin{figure}[h!]
 \begin{center}$
 \begin{array}{cccc}
\includegraphics[width=50 mm]{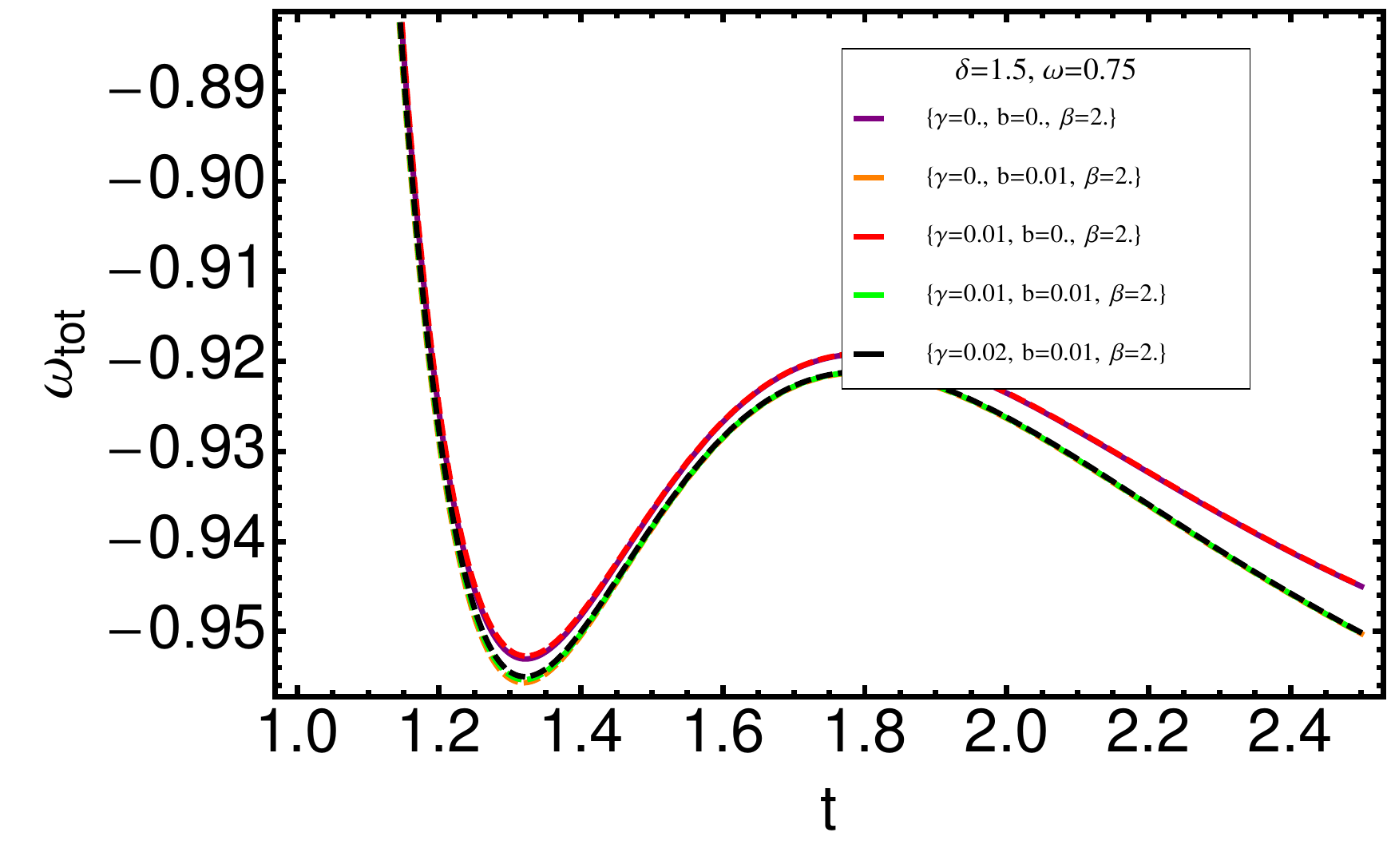} &
\includegraphics[width=50 mm]{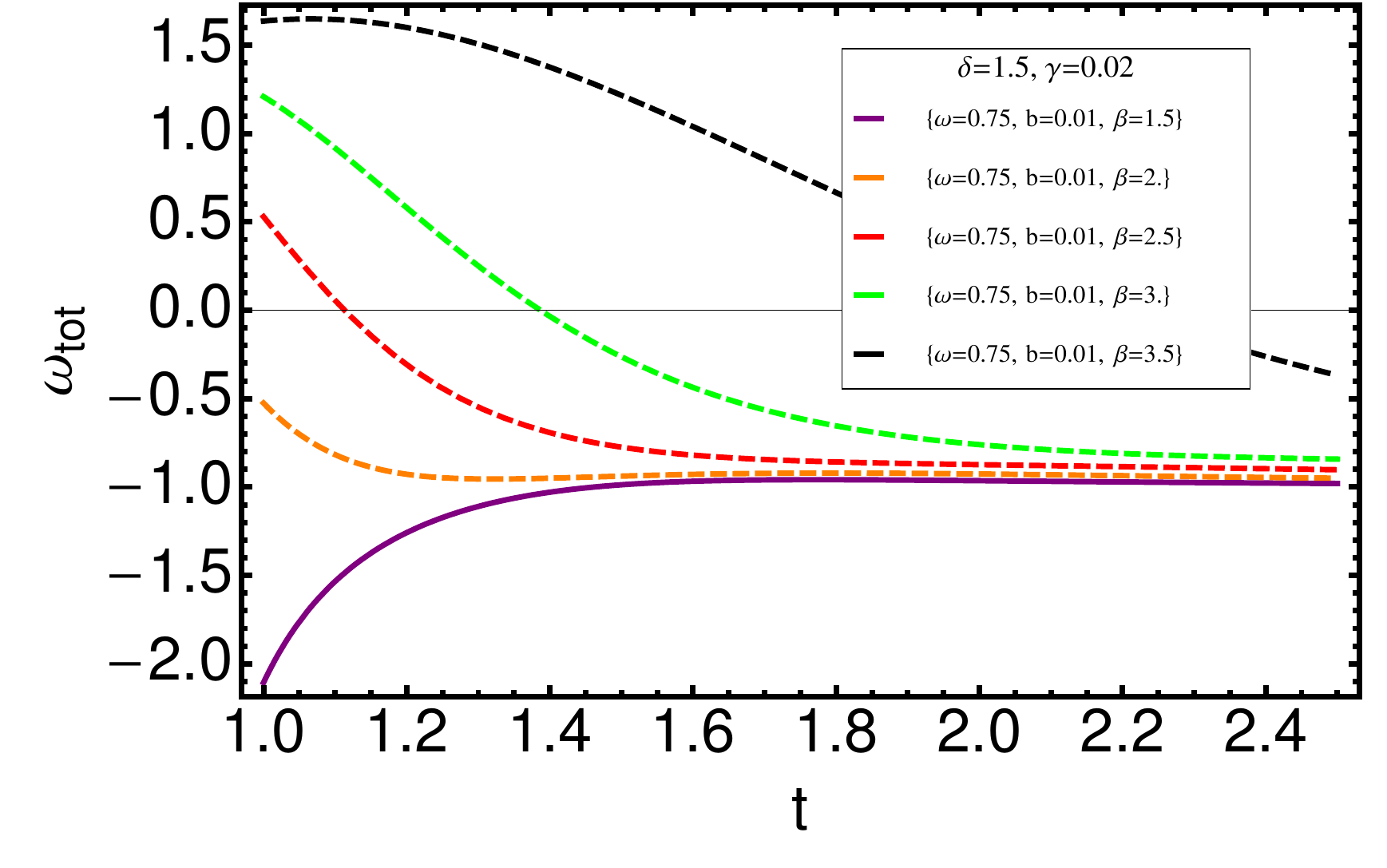}\\
\includegraphics[width=50 mm]{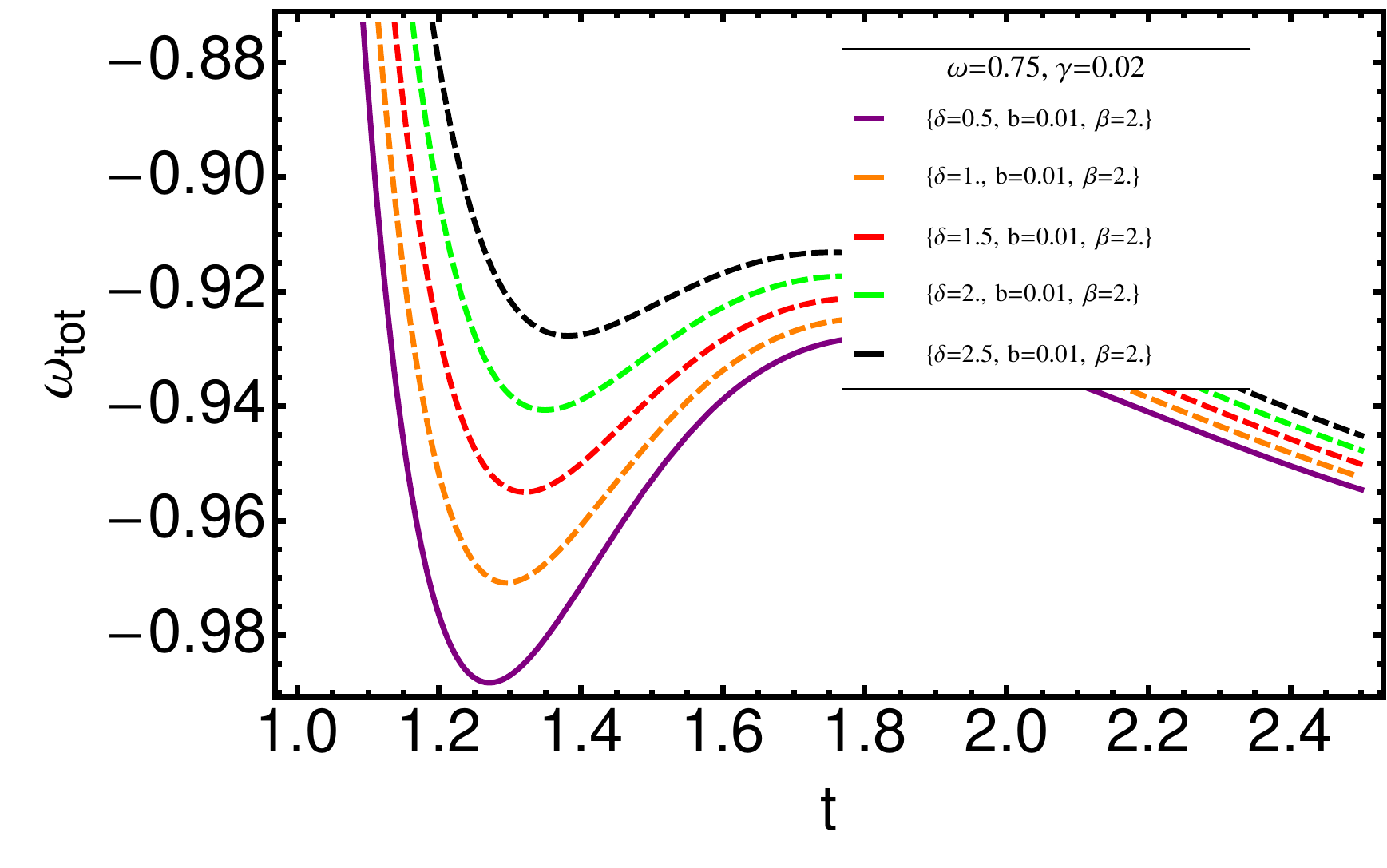}&
\includegraphics[width=50 mm]{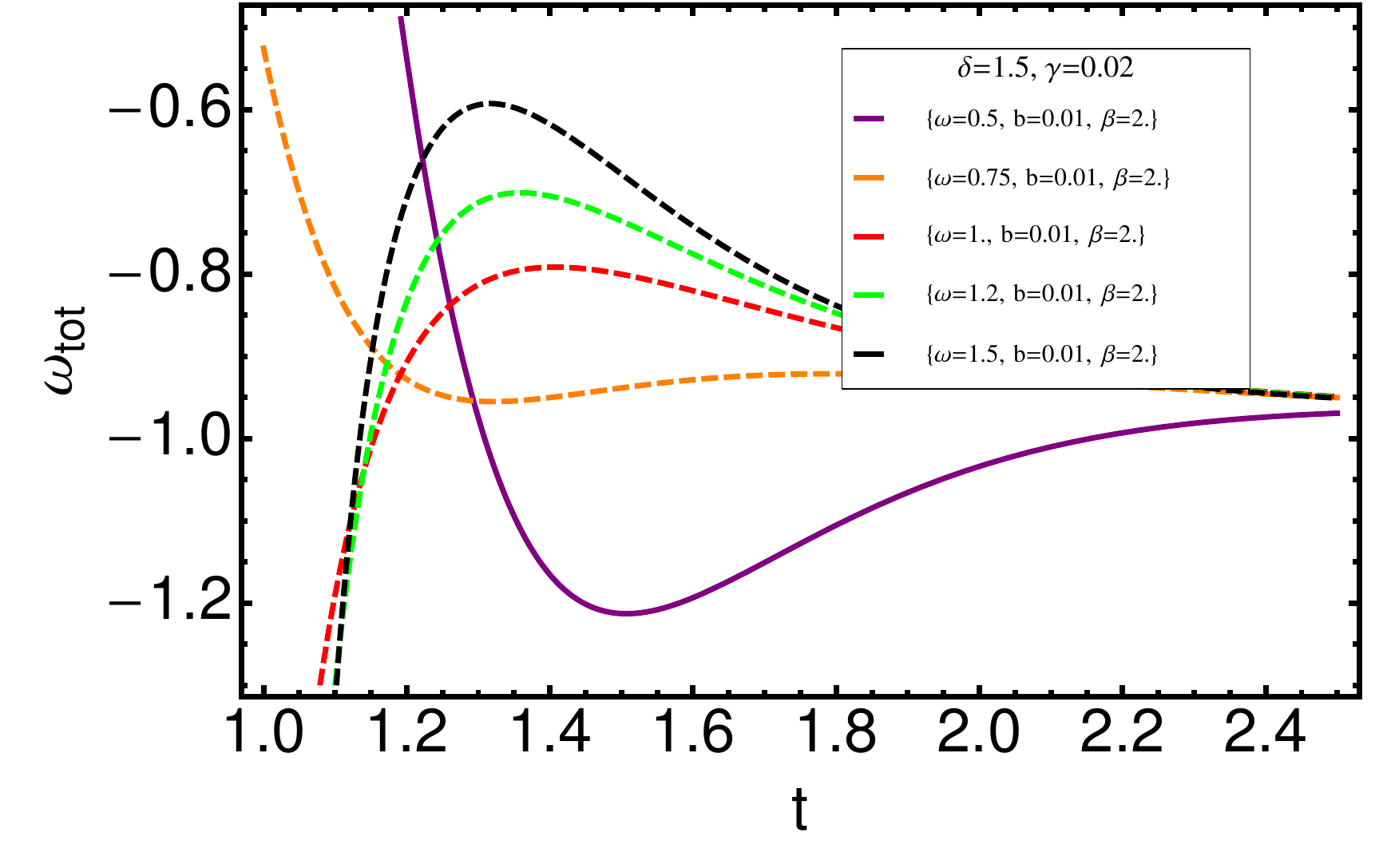}
 \end{array}$
 \end{center}
\caption{Behavior of $\omega_{tot}$ against $t$. Model 2}
 \label{fig:6}
\end{figure}

In this section we will investigate our second model which differs from the first model by the form of $\Lambda(t)$
\begin{equation}\label{eq:M2}
\Lambda(t)=H^{2}\phi^{-2}-\delta V(\phi).
\end{equation}
For the dynamics of $G$ with the Eq. (\ref{eq:M2}) we will have
\begin{equation}\label{eq:G2}
\frac{\dot{G}(t)}{G(t)}+\frac{\dot{\Lambda}(t) }{3H^{2} -H^{2}\phi^{-2}+\delta e^{-\beta \phi}}=0,
\end{equation}
where
\begin{equation}
\dot{\Lambda}(t)=-2\frac{H^{3}}{\phi^{2}}+\beta \delta \dot{\phi}e^{-\beta \phi}-2H^{2}\frac{\dot{\phi}}{\phi^{3}}+\frac{2H}{\phi^{2}}\frac{\ddot{a}}{a},
\end{equation} 
and 
\begin{equation}
\frac{\ddot{a}}{a}=-\frac{4\pi G}{3}( ( 1+3\omega)\rho_{b} + 2\dot{\phi}^{2}+4e^{-\beta \phi} )+\frac{H^{2}\phi^{-2}-\delta e^{-\beta \phi}}{3}.
\end{equation}
For the Hubble parameter we will have
\begin{equation}\label{eq:H2}
H=\sqrt{\frac{8 \pi G}{3-\phi^{-2}}(\rho_{b}+\frac{1}{2}\dot{\phi}^{2})+\frac{8 \pi G-\delta}{3-\phi^{-2}}e^{-\beta \phi}}.
\end{equation}
We will start from the analysis of behavior of $G$ in Fig. \ref{fig:4}. We observe that with $0<\omega<1$ we have increasing $G$. In behavior of $G$ interaction will start play a role only at later stages of evolution, this can be seen form top-left plot of Fig. \ref{fig:4}. We have considered models including full interaction $Q$, $Q=0$, $Q=3Hb\rho_{Q}$  and $Q=\gamma(\rho_{b}-\rho_{Q})\frac{\dot{\phi}}{\phi}$. Behavior of the $G$ as a function of $\beta$ is on the top-right of the Fig. \ref{fig:4}, where we see that with increasing $\beta$ we will increase value of $G$ for a fixed time. With increasing $\delta$ we will decrease value of $G$ (bottom-left plot). Last plot of Fig. \ref{fig:5} presents graphical behavior of $G$ as a function of $\omega$. We see that with increasing $\omega$ and $\omega\geq 1$ we will have different behavior. $G$ will decrease for early stages and then will start increase becoming constant for later stages of evolution. The graphical behavior of the deceleration parameter $q$ for this model gives us believe that with the carefull choosing the values of the parameters we can reproduce results comparable with observational data. As in the case of the first model, here we can declare that the model is able to reproduce deceleration parameter greater $-1$: $q>-1$. In Fig. \ref{fig:5} we have 4 different cases, which provide behavior of $q$ as a function of model parameters. We see that when $\delta=1.5$, $\omega=0.75$, $b=0.01$, $\gamma=0.02$ and $\beta=1.5$ (blue line) $q$ will increase, then becomes a constant, however $q<-1$. Therefore, we conclude that $\beta>1.5$ should be. With increasing $\beta$ we see that we can obtain Universe where $q<0$ for whole evolution, however $\beta>2$ we see that transition from Universe with $q>0$ to the Universe where $q<0$ can be observed. The bottom-left plot describes behavior $q$ from behavior of $\delta$. The bottom-right plot dedicated to present graphical behavior of $q$ as a function of $\omega$. When $0<\omega<1$ we see that $q$ is decreasing-increasing function, which for later stages will become a constant and $q>-1$. When we arrive to $\omega\geq 1$we see changes in behavior of $q$. It is increasing-decreasing function, which is a constant for later stages of evolution. Moreover, considered two regimes for $\omega$ for later stages of evolution will give results with small differences.  Behavior of $\omega_{tot}$ shows us that for later stages $\omega_{tot}$ will become $-1$. Behavior of $q$ and $\omega_{tot}$ (Fig. \ref{fig:6}) found to be consistent with general believe.

\subsection*{\large{Model 3}}
\begin{figure}[h!]
 \begin{center}$
 \begin{array}{cccc}
\includegraphics[width=50 mm]{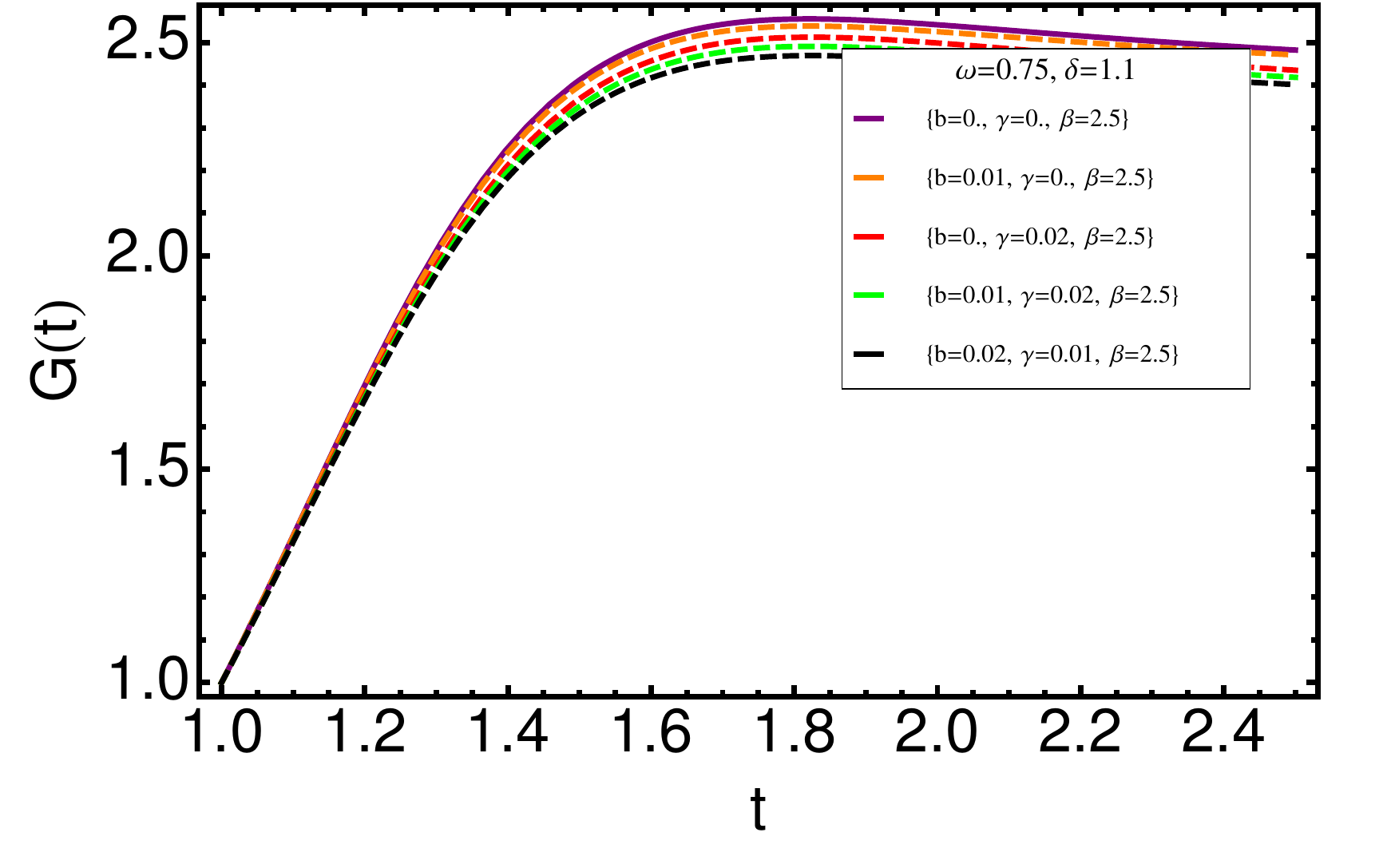} &
\includegraphics[width=50 mm]{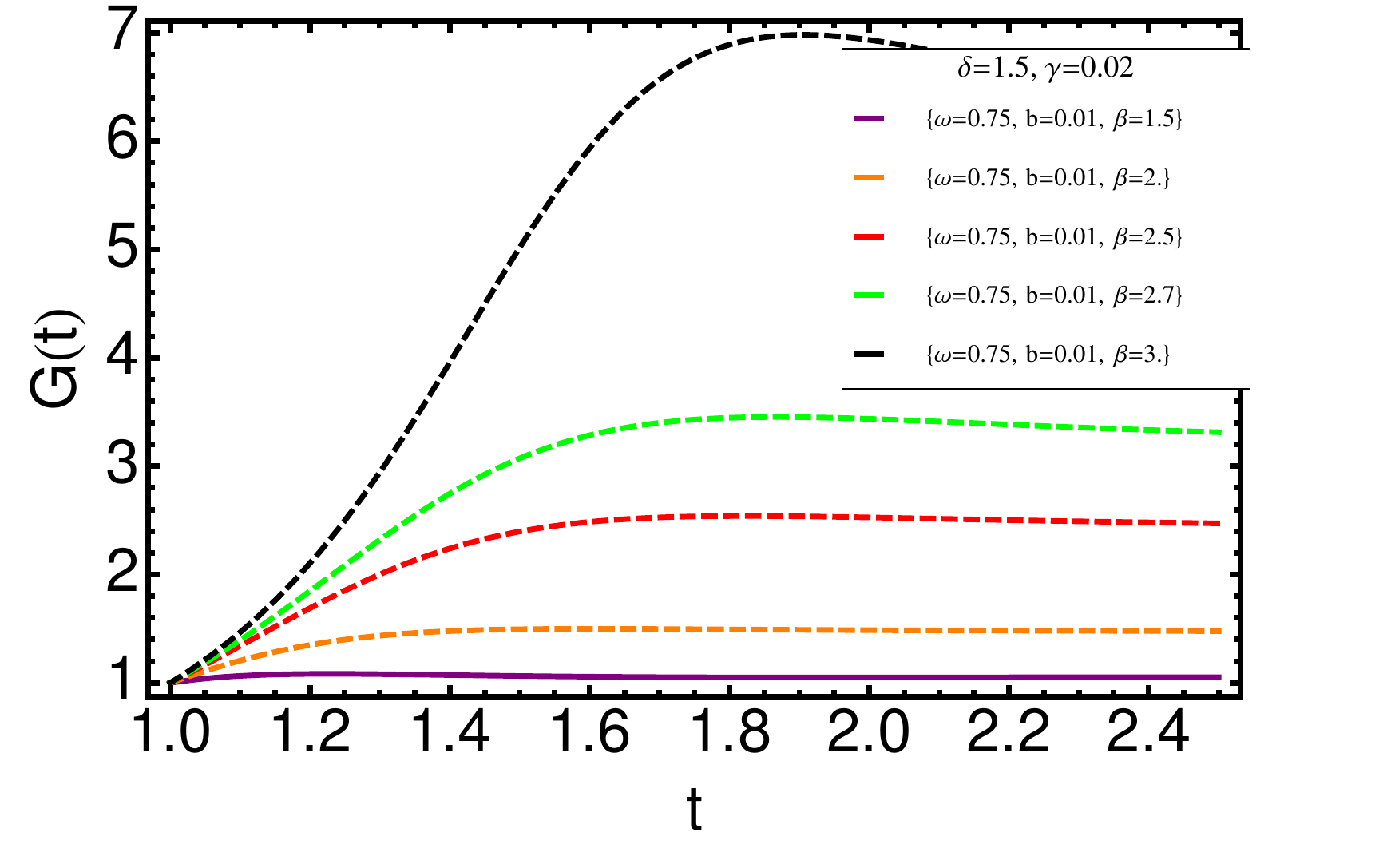}\\
\includegraphics[width=50 mm]{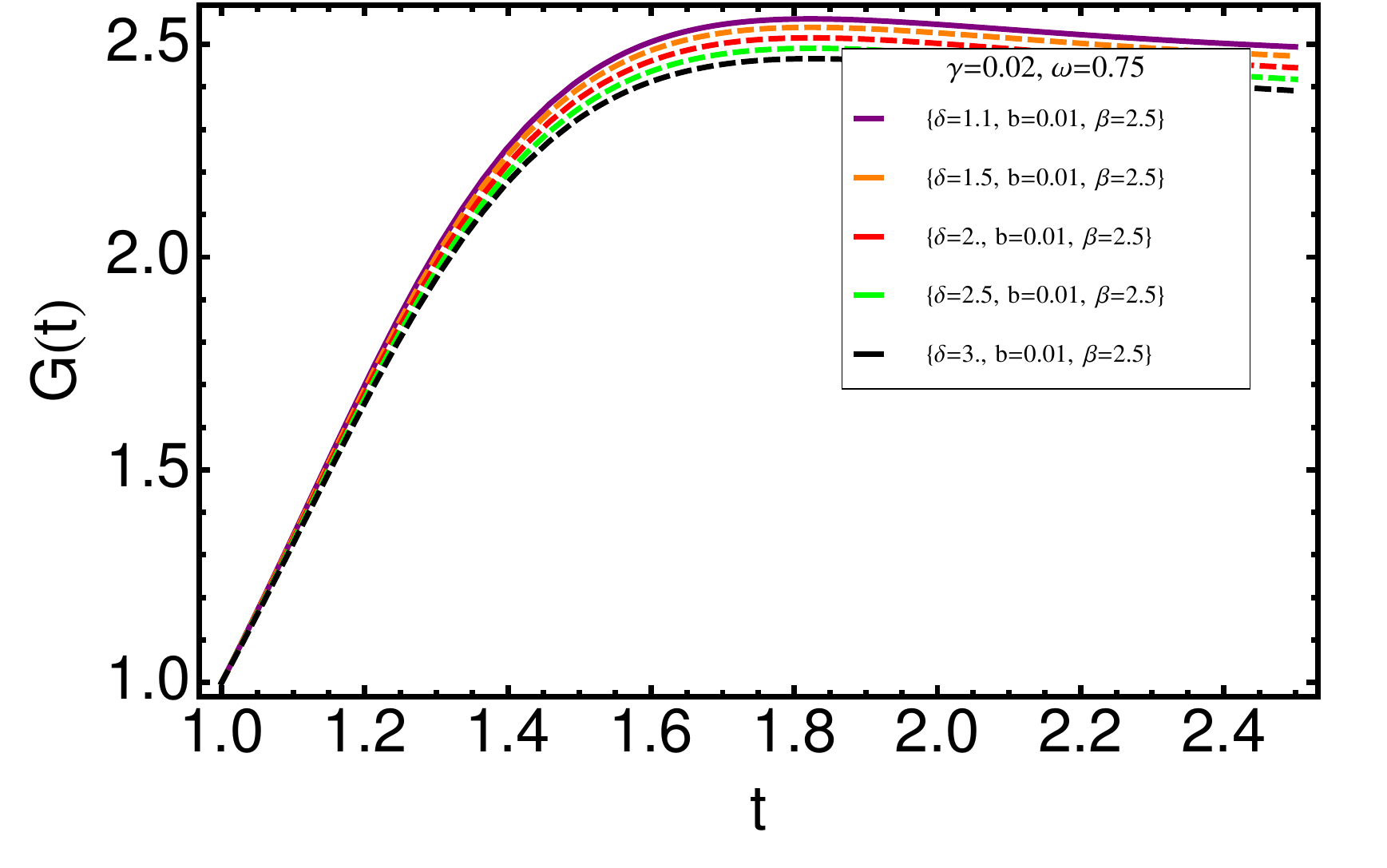}&
\includegraphics[width=50 mm]{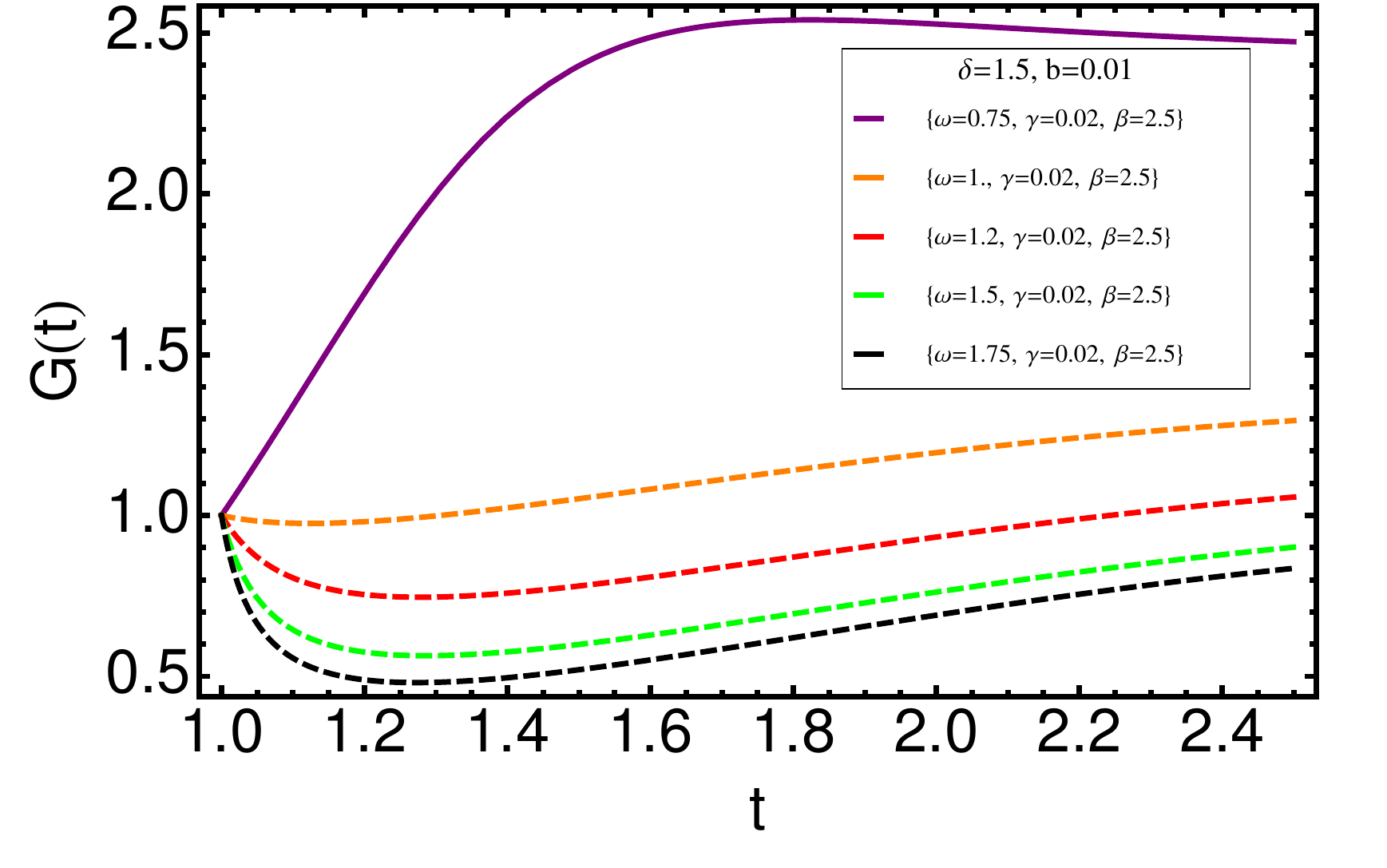}
 \end{array}$
 \end{center}
\caption{Behavior of $G$ against $t$  based on the solution of Eq. (\ref{eq:G3}). Model 3}
 \label{fig:7}
\end{figure}

\begin{figure}[h!]
 \begin{center}$
 \begin{array}{cccc}
\includegraphics[width=50 mm]{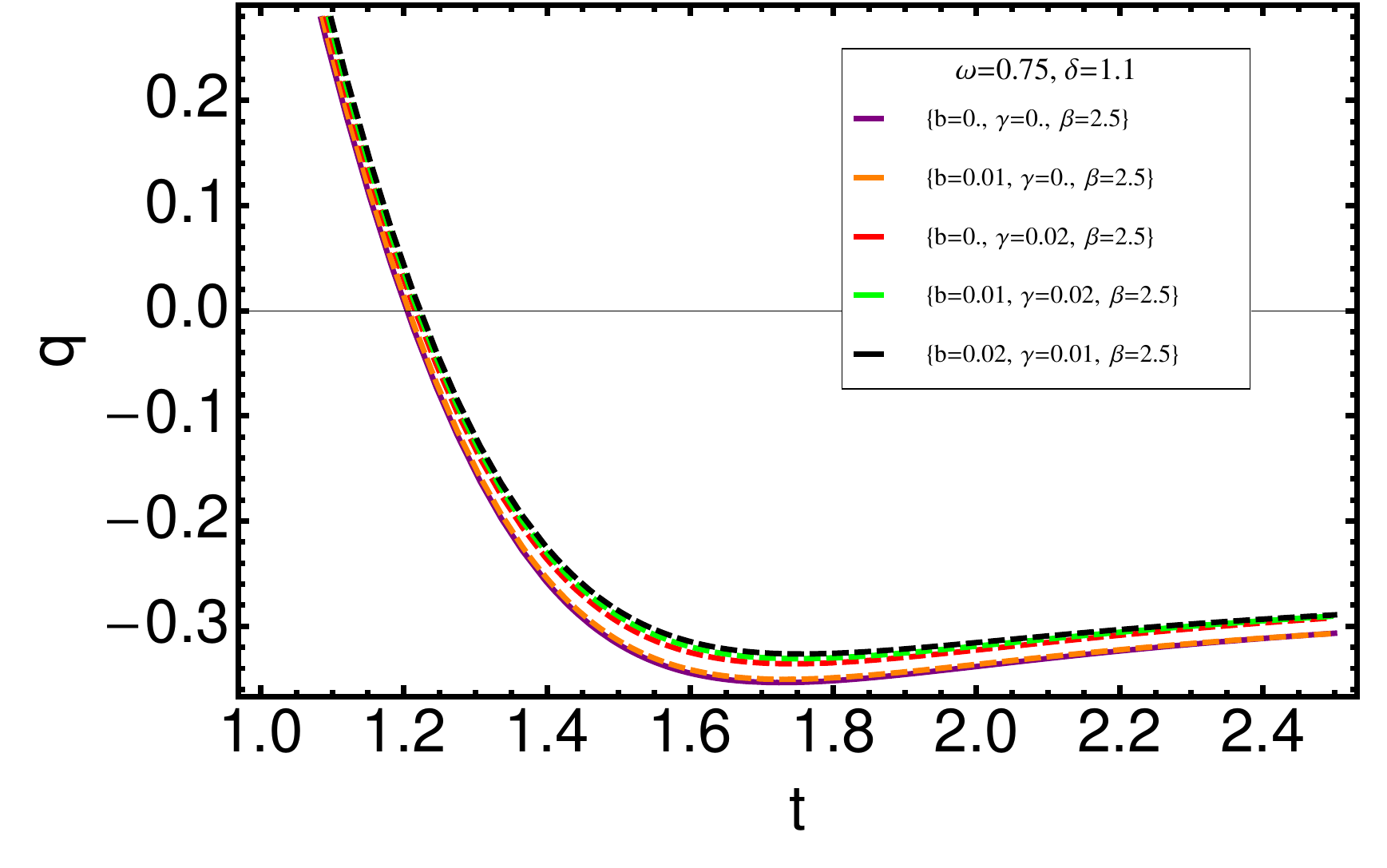} &
\includegraphics[width=50 mm]{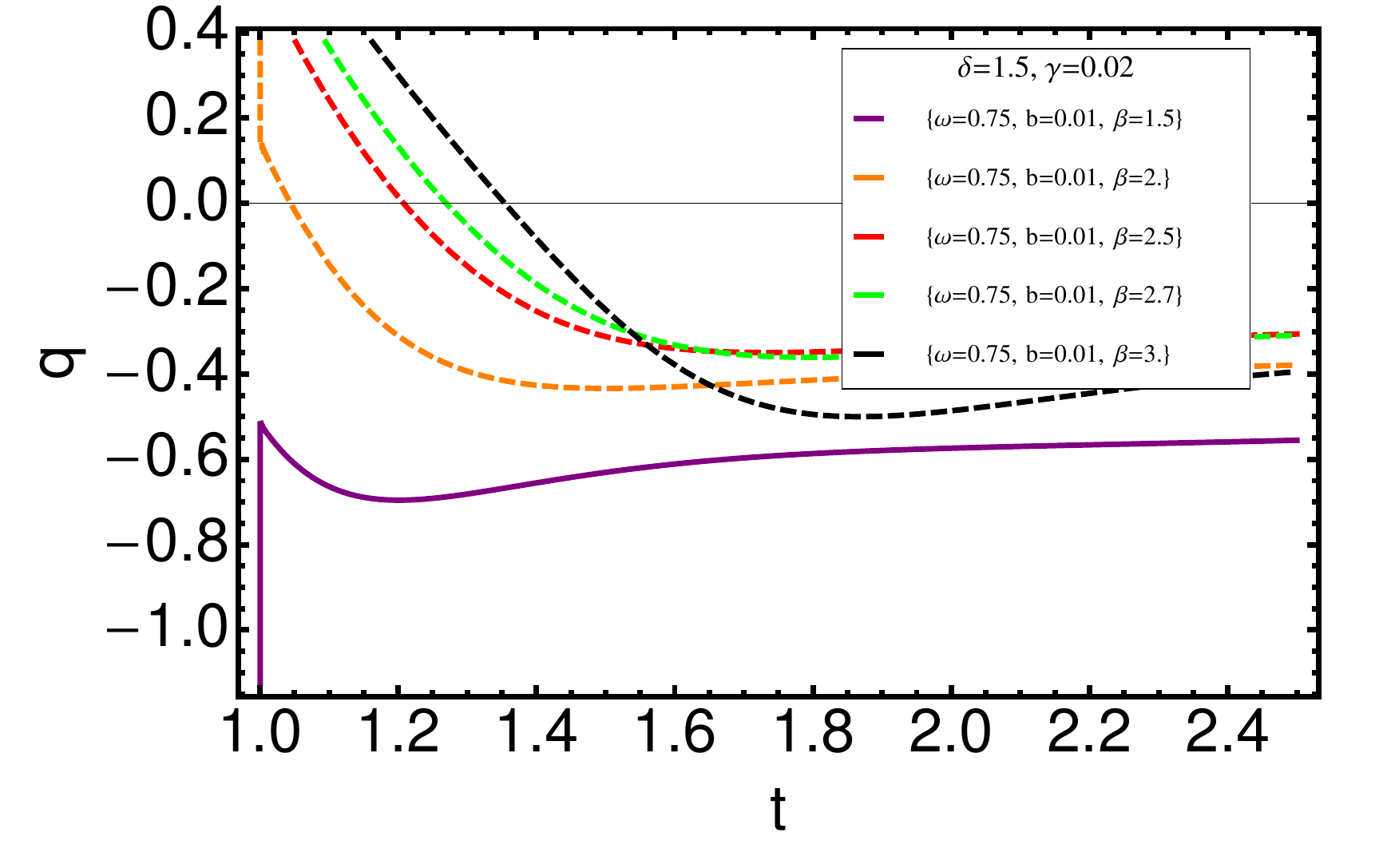}\\
\includegraphics[width=50 mm]{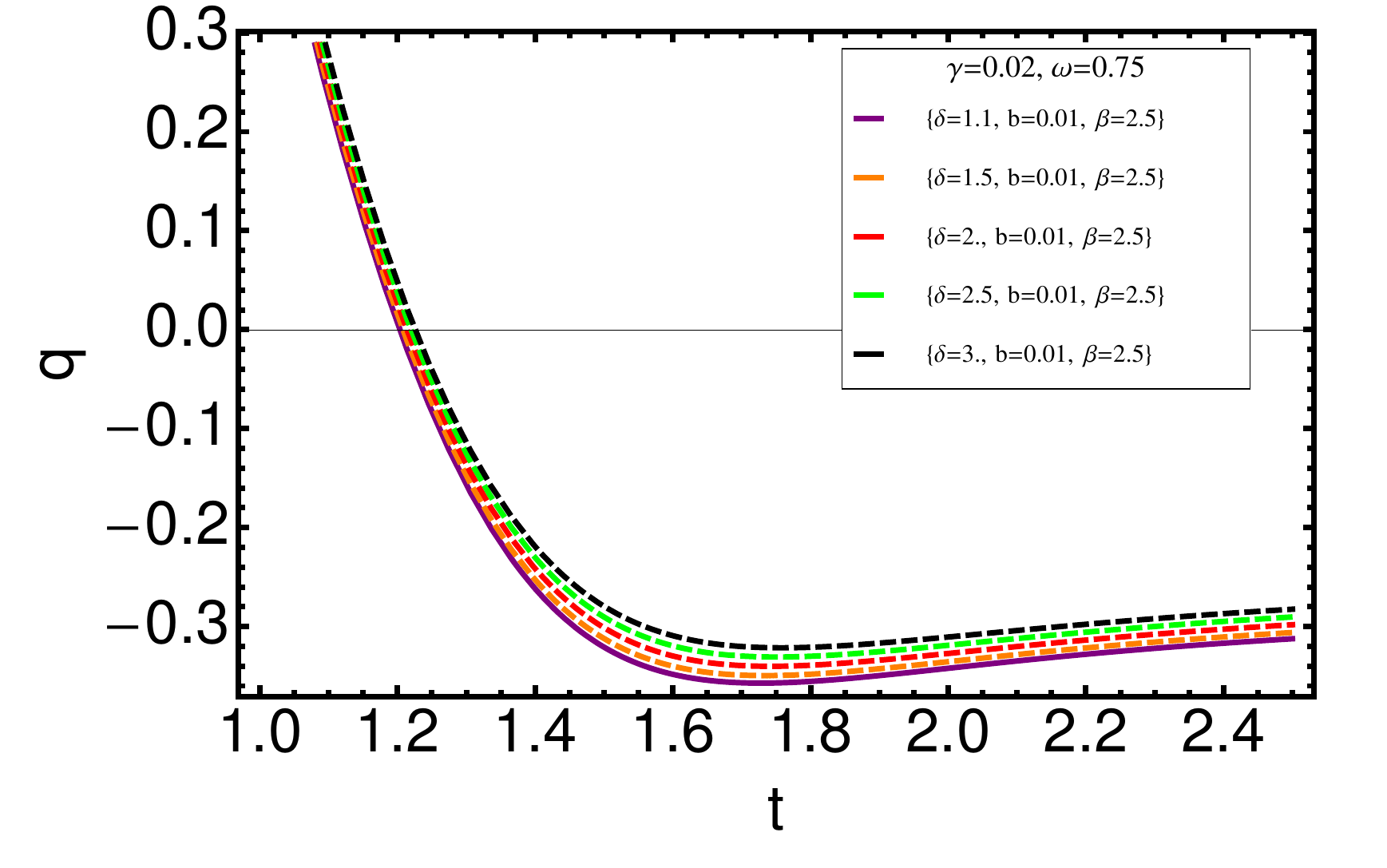}&
\includegraphics[width=50 mm]{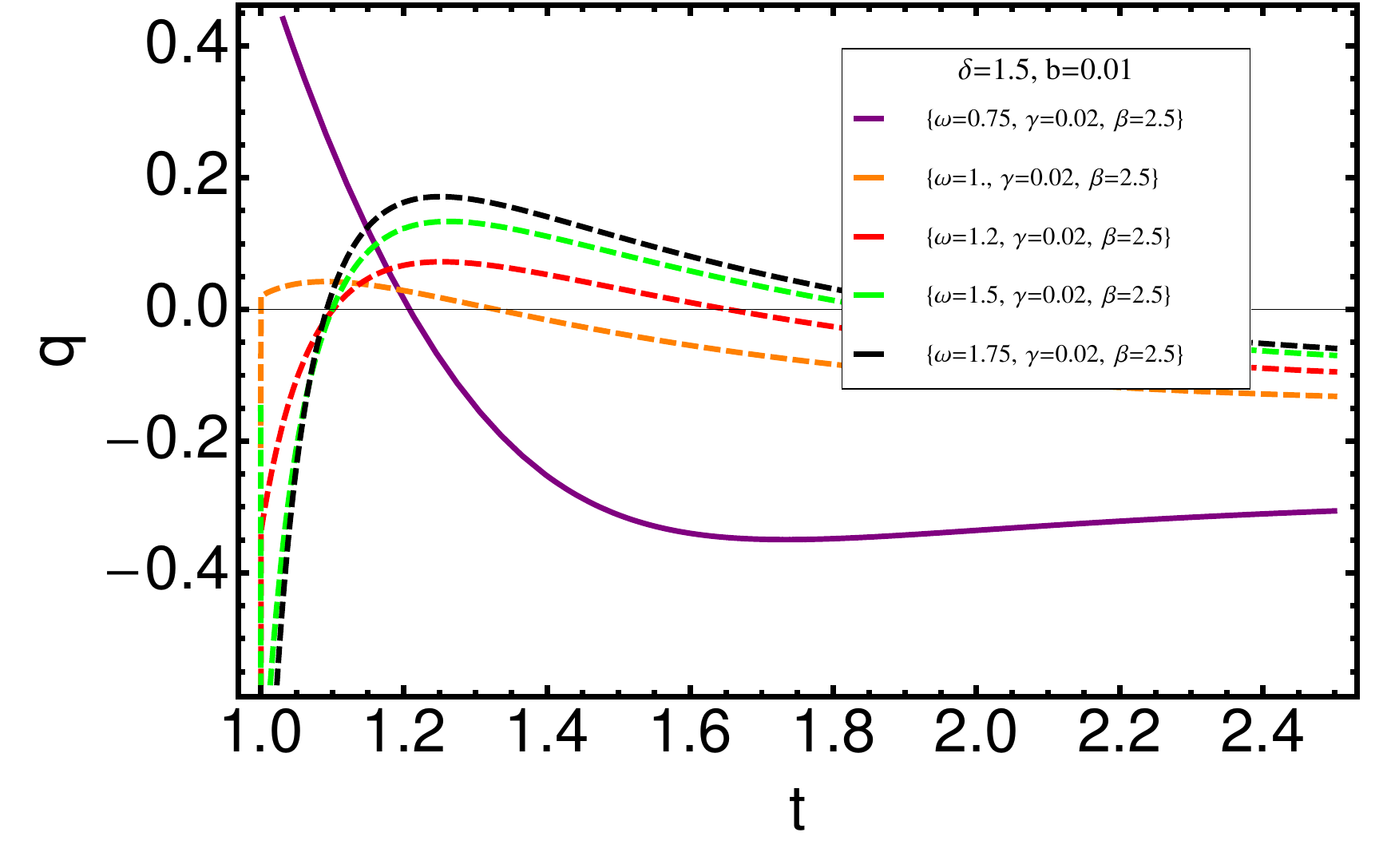}
 \end{array}$
 \end{center}
\caption{Behavior of $q$ against $t$  based on the solution of Eq. (\ref{eq:H3}). Model 3}
 \label{fig:8}
\end{figure}

\begin{figure}[h!]
 \begin{center}$
 \begin{array}{cccc}
\includegraphics[width=50 mm]{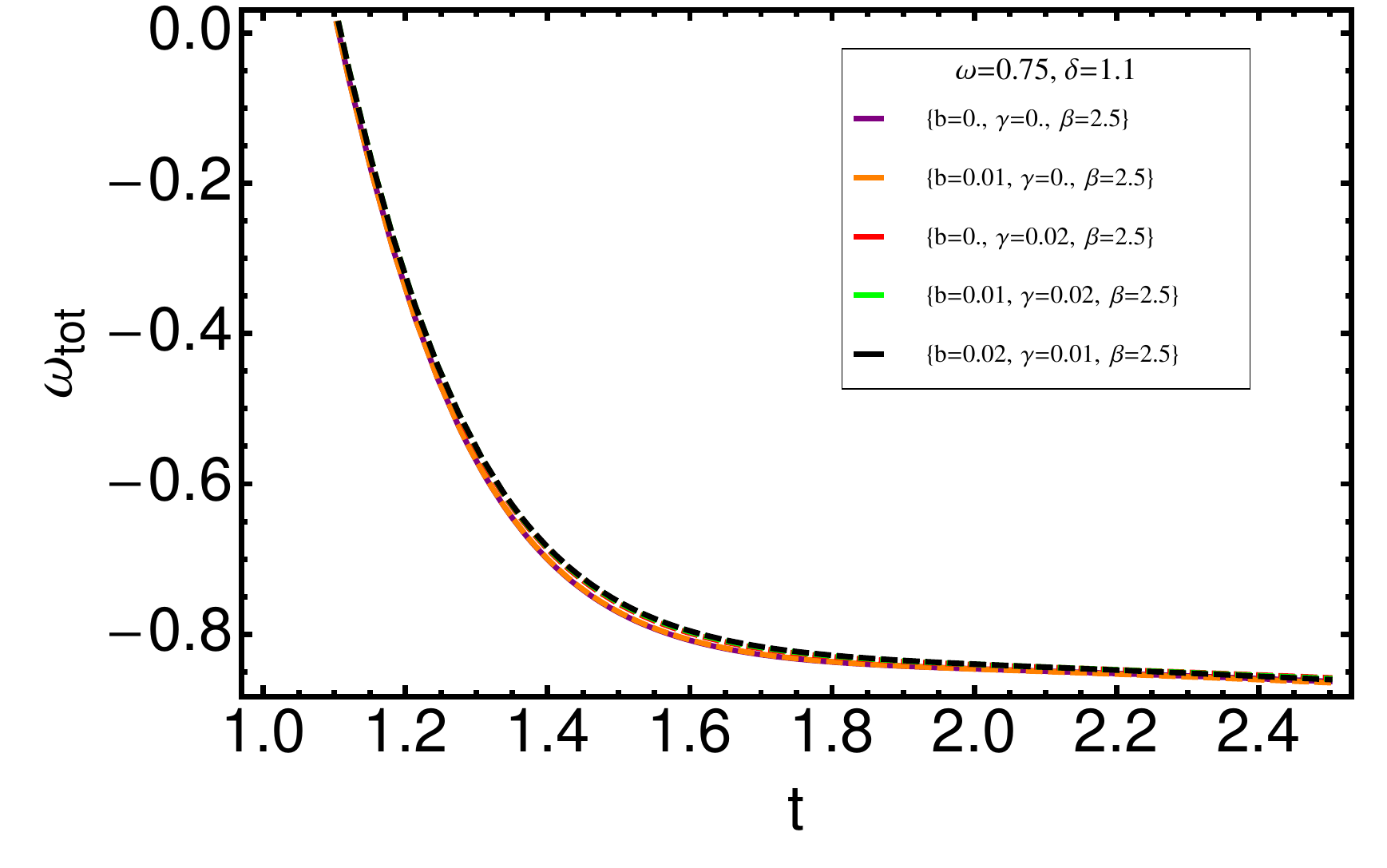} &
\includegraphics[width=50 mm]{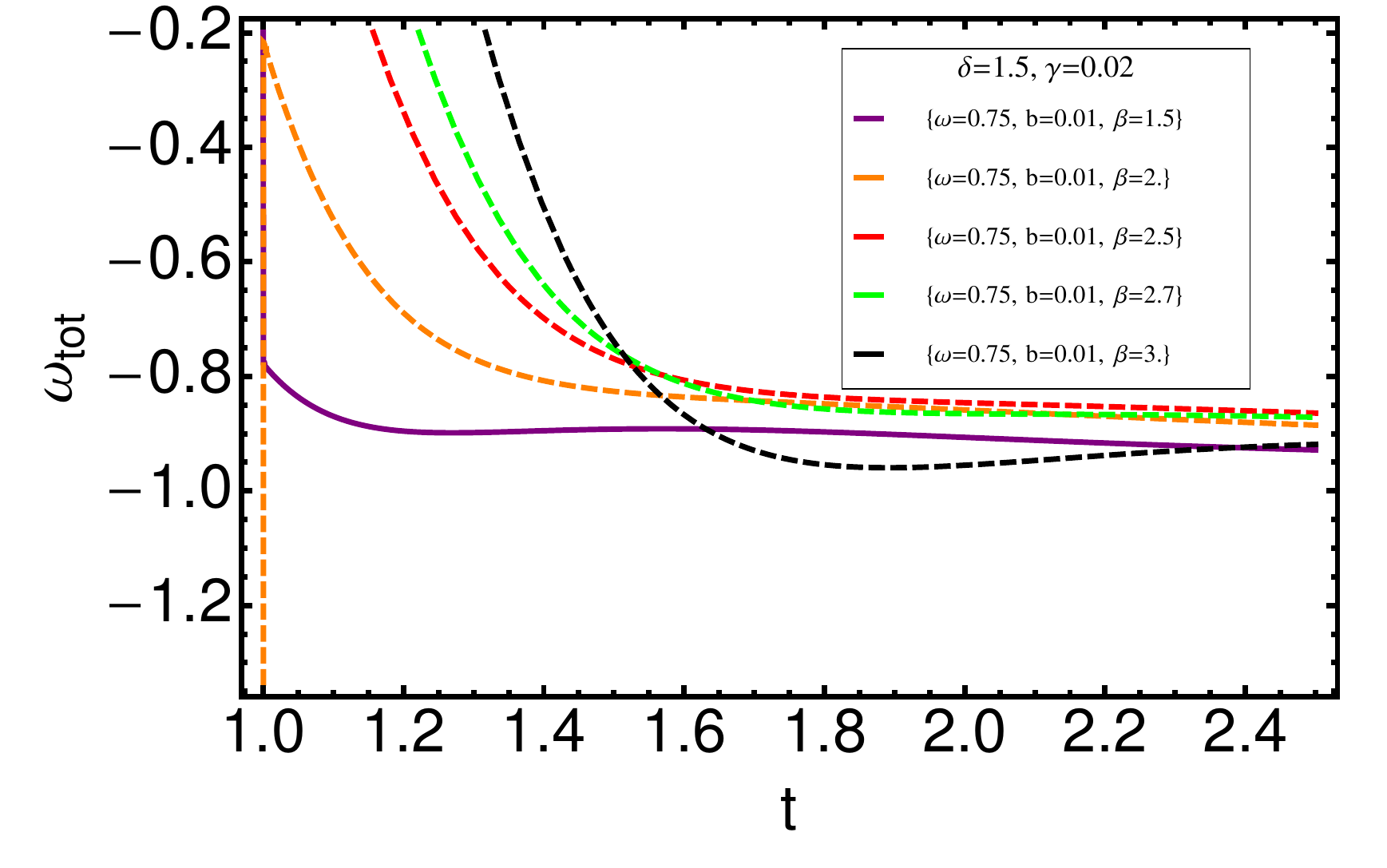}\\
\includegraphics[width=50 mm]{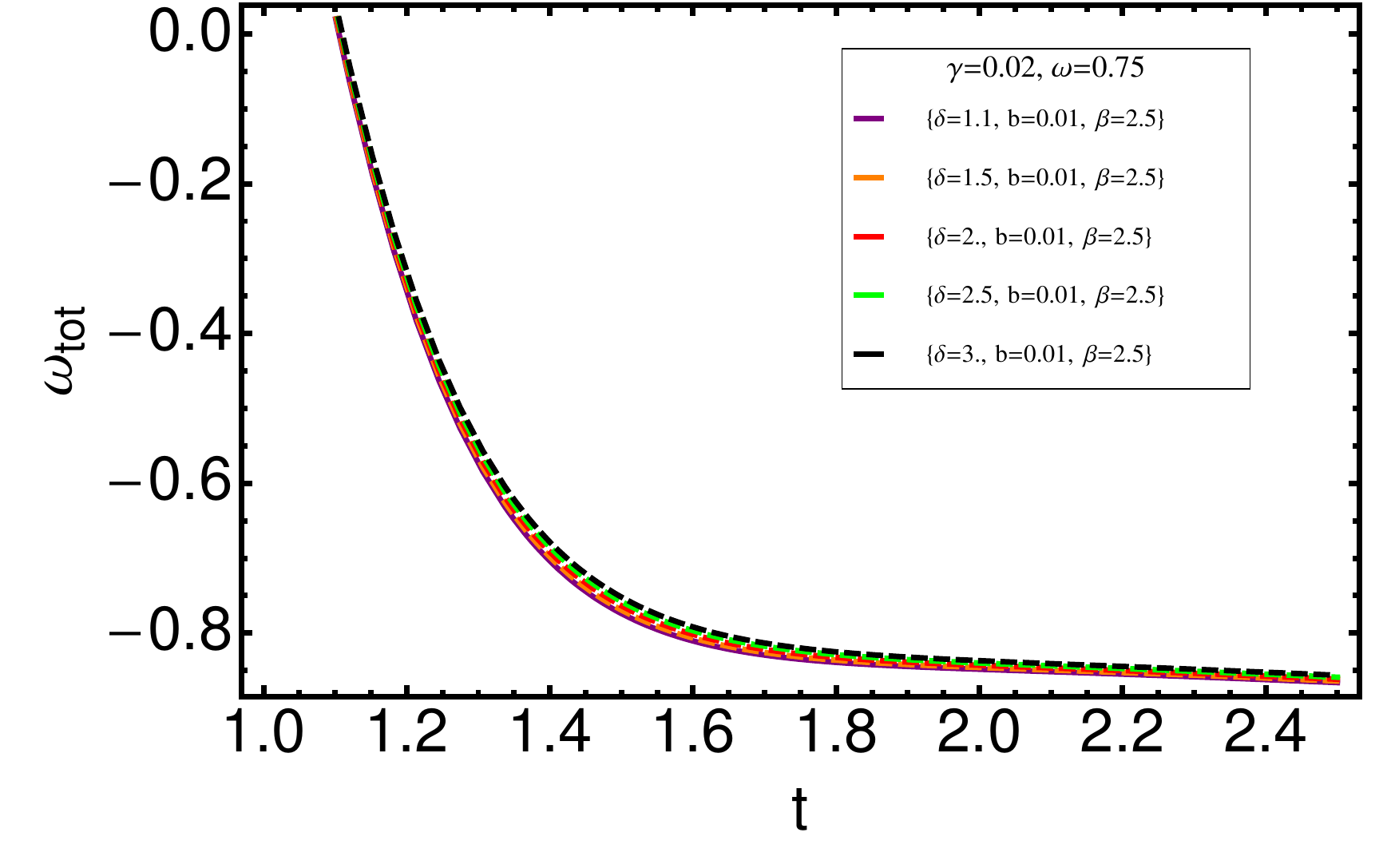}&
\includegraphics[width=50 mm]{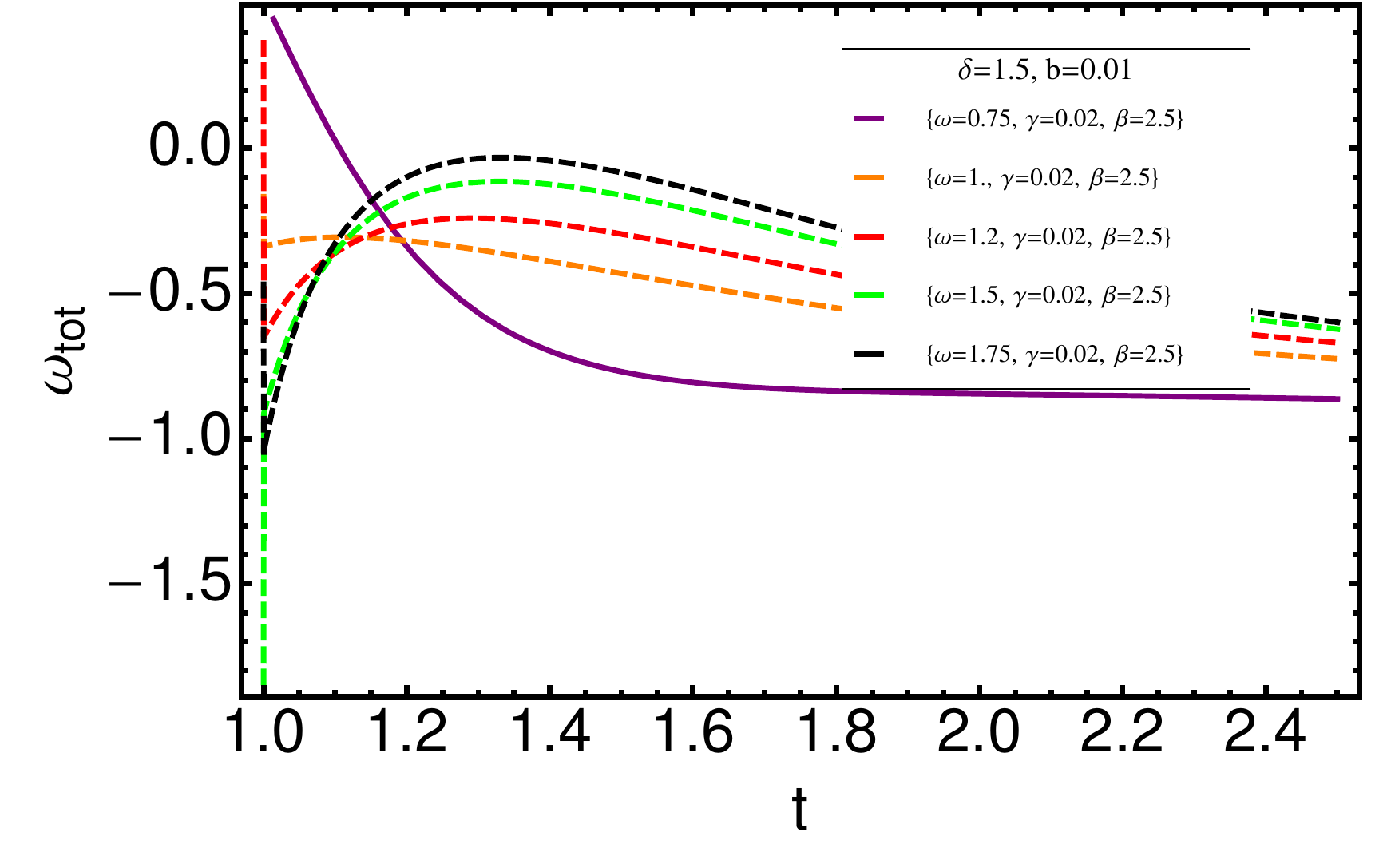}
 \end{array}$
 \end{center}
\caption{Behavior of $\omega_{tot}$ against $t$. Model 3}
 \label{fig:9}
\end{figure}

Now we will pay attention to the last model of our paper. For this model we assume that $\Lambda(t)$ can be a function of deceleration parameter as well. For simplicity and as an example we assume that  
\begin{equation}\label{eq:M3}
\Lambda(t)=q(\phi^{2}+\phi^{-2})-\delta e^{-\beta \phi}.
\end{equation}
We will establish mathematical formulas to describe dynamics of $G$ and $H$. Then, we will analyse obtained behaviors of cosmological parameters. Dynamics of $G$ with the Eq. (\ref{eq:M3}) can be writen as
\begin{equation}\label{eq:G3}
\frac{\dot{G}(t)}{G(t)}+\frac{\dot{\Lambda}(t) }{3H^{2} -q(\phi^{2}+\phi^{-2})+\delta e^{-\beta \phi}}=0,
\end{equation}
where 
\begin{equation}
\dot{\Lambda}(t)=\beta \delta \dot{\phi} e^{-\beta \phi}-\frac{(-\frac{2\dot{\phi}}{\phi^{3}}+2\phi\dot{\phi})}{H}\frac{\ddot{a}}{a}+\frac{(\phi^{-2}+\phi^{2})}{H^{2}}\left (\frac{\ddot{a}}{a} \right )^{2}-\frac{(\phi^{-2}+\phi^{2})}{H}\frac{a^{(3)}}{a},
\end{equation}
and
\begin{equation}
\frac{\ddot{a}}{a}=-\frac{4\pi G}{3}( ( 1+3\omega)\rho_{b} + 2\dot{\phi}^{2}+4e^{-\beta \phi} )+\frac{(-1-\frac{\dot{H}}{H^{2}})(\phi^{2}+\phi^{-2})-\delta e^{-\beta \phi}}{3}.
\end{equation}
After some mathematical manipulations with $\frac{\ddot{a}}{a}$, $a^{(3)}$ can be found. For the Hubble parameter we will have the following differential equation
\begin{equation}\label{eq:H3}
\dot{H}-\left [\frac{8 \pi G}{(\phi^{-2}+\phi^{2})} \left ( \rho_{b}+\frac{1}{2}\dot{\phi}^{2} \right )+ \frac{(8 \pi G -\delta )}{(\phi^{-2}+\phi^{2})}e^{-\beta \phi} - 1 \right ]H^{2}+\frac{3}{(\phi^{-2}+\phi^{2})}H^{4}=0.
\end{equation}
Starting the analysis from the graphical behavior of $G$ we will finish this section with analysis of $\omega_{tot}$. In this case $G$ is also increasing function over the evolution of the history of the Universe (top-panel and bottom-left plots present behavior of $G$ as a function of $b$, $\gamma$, $\beta$ and $\delta$ respectively). For all cases $\omega<1$. The bottom-right plot shows that with increasing $\omega$, $G$ can become from increasing function to decreasing-increasing function. This transition observed when $\omega \geq 1$ 
(see Fig. \ref{fig:7}). Figs. \ref{fig:8}-\ref{fig:9} showed graphical behaviors of $q$ and $\omega_{tot}$. For the early stages of evolution this model can contain transition from $q>0$ to $q<0$. Impact of interaction is investigated at top-left plot. $q$ as a function of $\beta$ over time is investigated on the top-right plot of Fig. \ref{fig:8}. 
We see when $\beta<2$ (particular example is a blue line with $\beta=1.5$, which gives $q=-0.58$ for later stages of evolution) we have ever accelerated model. With $\beta>2$ we will get transition in $q$. Behavior of $q$ as a function of $\omega$ is given on the bottom-right plot. Blue line with $\omega=0.75$ with $\delta=1.5$, $\beta=2.5$, $\gamma-0.02$ and $b=0.01$ corresponds to the decreasing-increasing behavior of $q$ indicating transition from $q>0$ to $q<0$ phase. $\omega \geq 1$ we have universe which is accelerated at early stages, then has transition to $q>0$. Finally, at later stages of evolution transition to the accelerated expansion phase is guaranteed. $\omega_{tot}$ indicates decreasing behavior when $\omega<1$, while it is decrasing and increasin function for $\omega\geq 1$ (
see Fig. \ref{fig:9}). In this case  also, as it was in case of the first and second models, we have good agreement with believe and obtained results, particular believe is that that  acceleretated expansion can be induced by the dark energy with $\omega_{DE}<0$. 

\newpage
\section*{\large{Discussions}}
In this article we consider two-component fluid Universe with a varying gravitational "constant" $G$ and varying cosmological "constant" $\Lambda$. One of the components of the effective fluid is thought to be a barotropic fluid, while the second component is associated with the quintessence model of a DE. An interaction of the phenomenological origin as a function of energy densities of components and scalar field $\phi$, derivative of the field $phi$, is taken to be. We proposed 3 different models for $\Lambda(t)$ and for each case we have analysed graphical behaviors of cosmological parameters ($H$, $q$, $G$, $\omega_{tot}$, $\omega_{Q}$, $\phi$ and $V$). To be able to solve the system of differential equations, which will describe dynamics of our Universe, we suppose that the form of the potential of the field is given. And as an a particular example we consider a potential $V(\phi)=e^{-\beta \phi}$ of the form, where $\beta$ is a constant. According to the $1\sigma$ level from $H(z)$ data $q\approx -0.3$ and $H_{0}=68.43\pm 2.8 \frac{Km}{sMpc}$ \cite{Kumar}. On the other hand from data of $SNe Ia$ we have $q \approx -0.43$ and $H_{0}=69.18 \pm 0.55\frac{Km}{sMpc}$ \cite{Kumar}. Also joint test using $H(z)$ and $SNe Ia$ give $-0.39 \leq q \leq -0.29$ and $H_{0}=68.93 \pm 0.53\frac{Km}{sMpc}$ \cite{Kumar}. Recent astronomical data based on anew infrared camera on the $HST$ gives $H_{0}=73.8 \pm 2.4\frac{Km}{sMpc}$ \cite{Riess1}. The other prob using galactic clusters data suggest $H_{0} = 67 \pm 3.2\frac{Km}{sMpc}$ \cite{Beutler}. Finally, $\Lambda CDM$ model suggests $q \rightarrow -1$ and the best fitted parameters of the Ref. \cite{Visser} say that $q \approx -0.64$. Conclusion of the presented facts is that that  generally $q \geq-1$.
Performing analysis of our models we see clearly, that $q \geq-1$ condition is can be satisfied, moreover we see that the carefull choose of values of the model parametres observational facts can be recovered. We also would like to mention that the analysis of our toy models indicate a possibility to realise all regimes with $\omega_{tot} \approx -1$ including de Sitter phase, as well as. In case of the first model $G$ is a decreasing function independent on the value $\omega$, while for second and third models we had an opposite picture i.e $G$ is an increasing function when $\omega<1$. Despite, to the first model, second and thirds models indicate that when $\omega \geq1$ $G$ will decrease and then will start to increase for later stages of evolution. EoS parameter $\omega_{tot}$ of the effective fluid is bigger than $-1$ when $\omega<1$ (latter stages of evolution). While either $\omega<1$ or $\omega \neq 1$ $\omega_{tot} \rightarrow -1$ at latter stages of evolution independent of the form of $\Lambda(t)$ under the consideration. Future possible works can include consideration of different forms for interaction $Q$, is it also possible to investigate the problem in presence of varying viscosity. Performing stafinder and dynamical analysis could be an other interesting work to complete the full investigation of the models.

\newpage{}

\newpage
\section*{Appendix}
In this section we added analysis of the Hubble parameter $H$, EoS parameter $\omega_{Q}$ of the dark energy. The graphical behavior and evolution of the scalar field $\phi$ and the potential $V$ with respect to cosmic time $t$ is given also. Behavior of the cosmological parameter discussed according to the model parameters and model parameters are taken to satisfy the well know fact that $V(t) \leftarrow 0$, when $t\leftarrow \infty$. Graphical analysis of $\omega_{Q}$ reveals an interesting fact about its evolution. We see that it is a DE model only for the later stages of the evolution, while for early epoch of the evolution it can be interpreted as a fluid with a positive EoS. It is an interesting fact and additional analysis should be performed on this question. This behavior also we observed during the analysis of $\omega_{tot}$, when a transition from a positive EoS a transition to a fluid with a negative EoS corresponding to the DE is in face. The question of the behavior of the potential we have discussed already and we would like to mention that the field is an increasing function.   
\begin{figure}[!ht]
 \begin{center}$
 \begin{array}{cccc}
\includegraphics[width=50 mm]{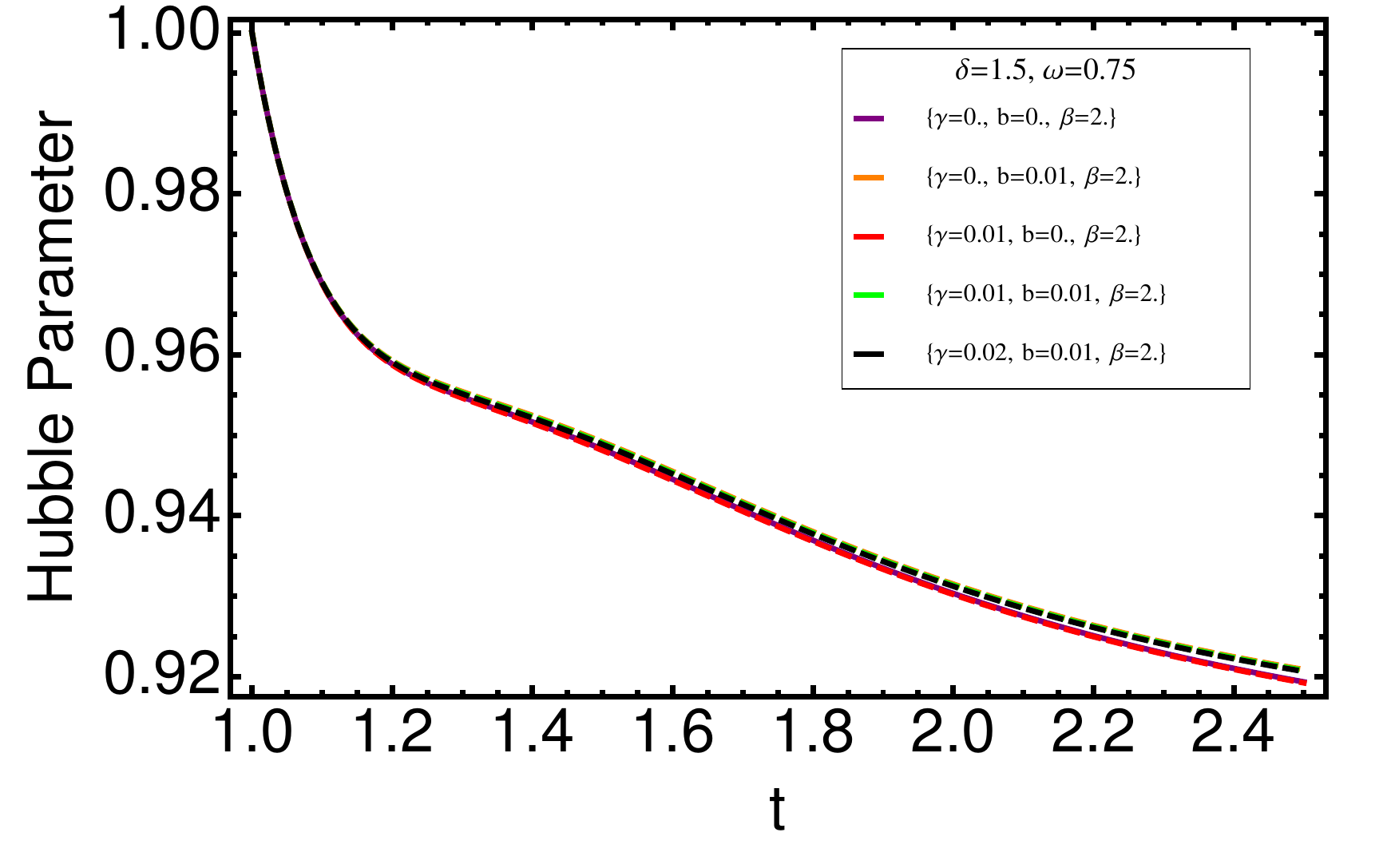} &
\includegraphics[width=50 mm]{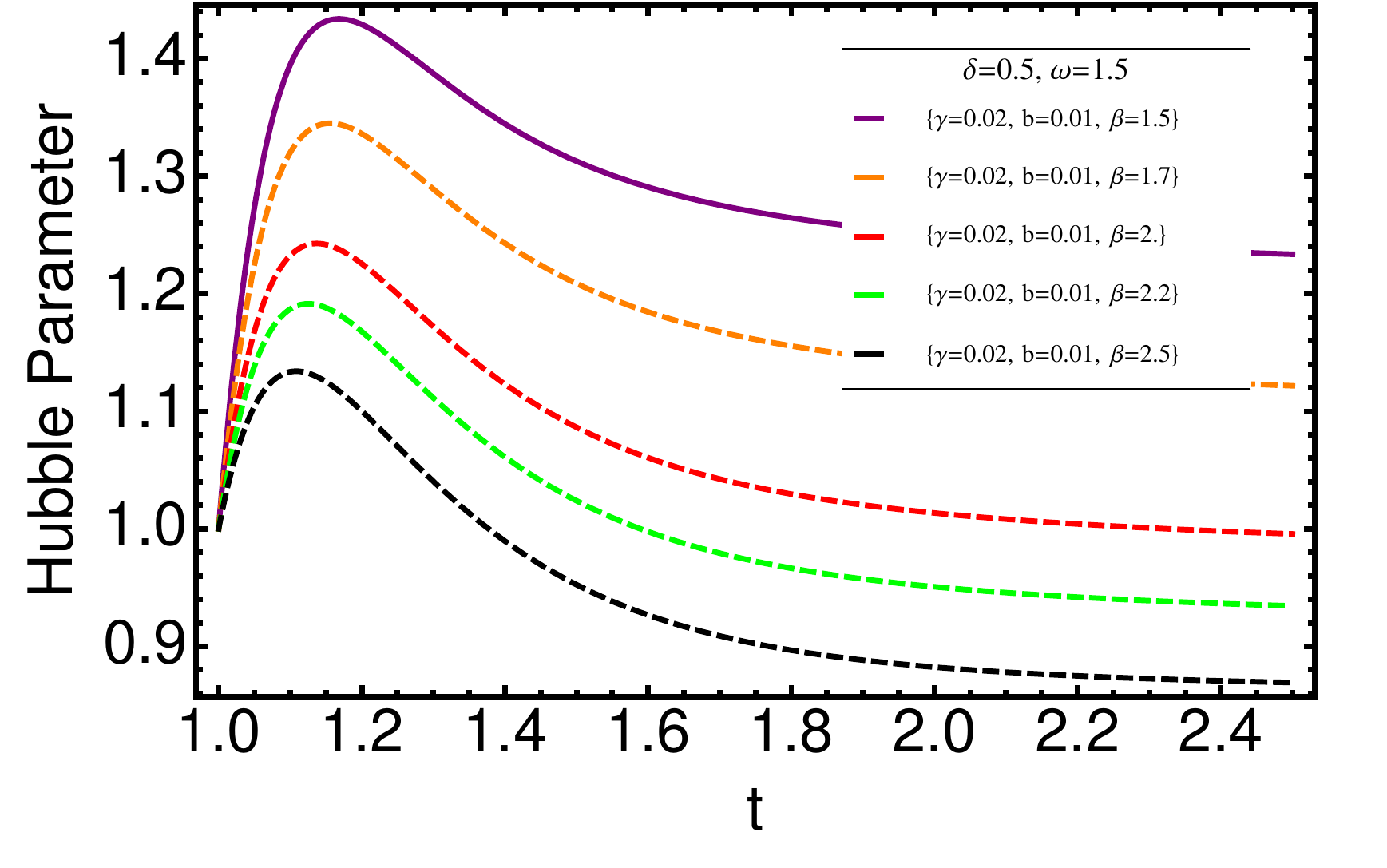}\\
\includegraphics[width=50 mm]{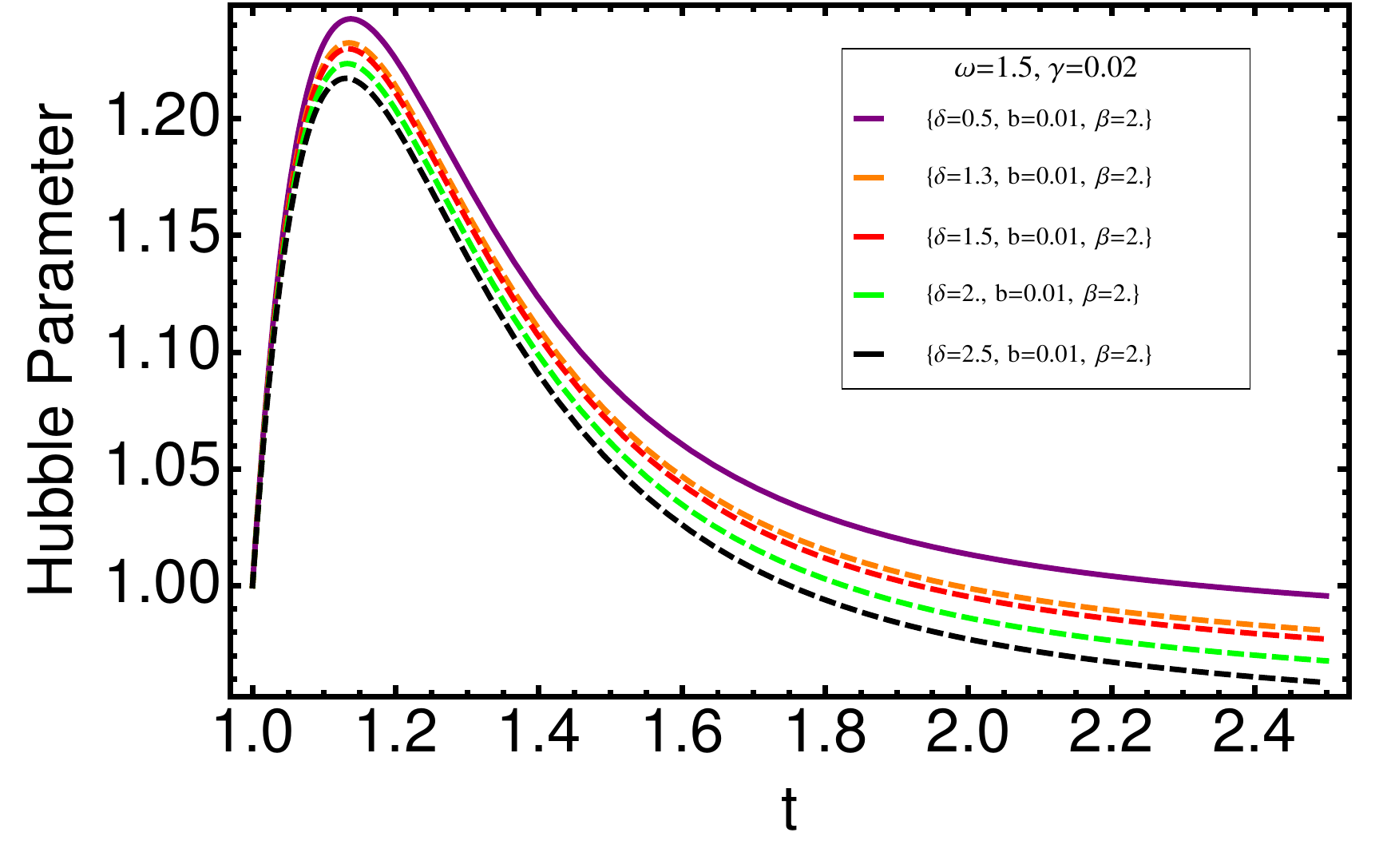}&
\includegraphics[width=50 mm]{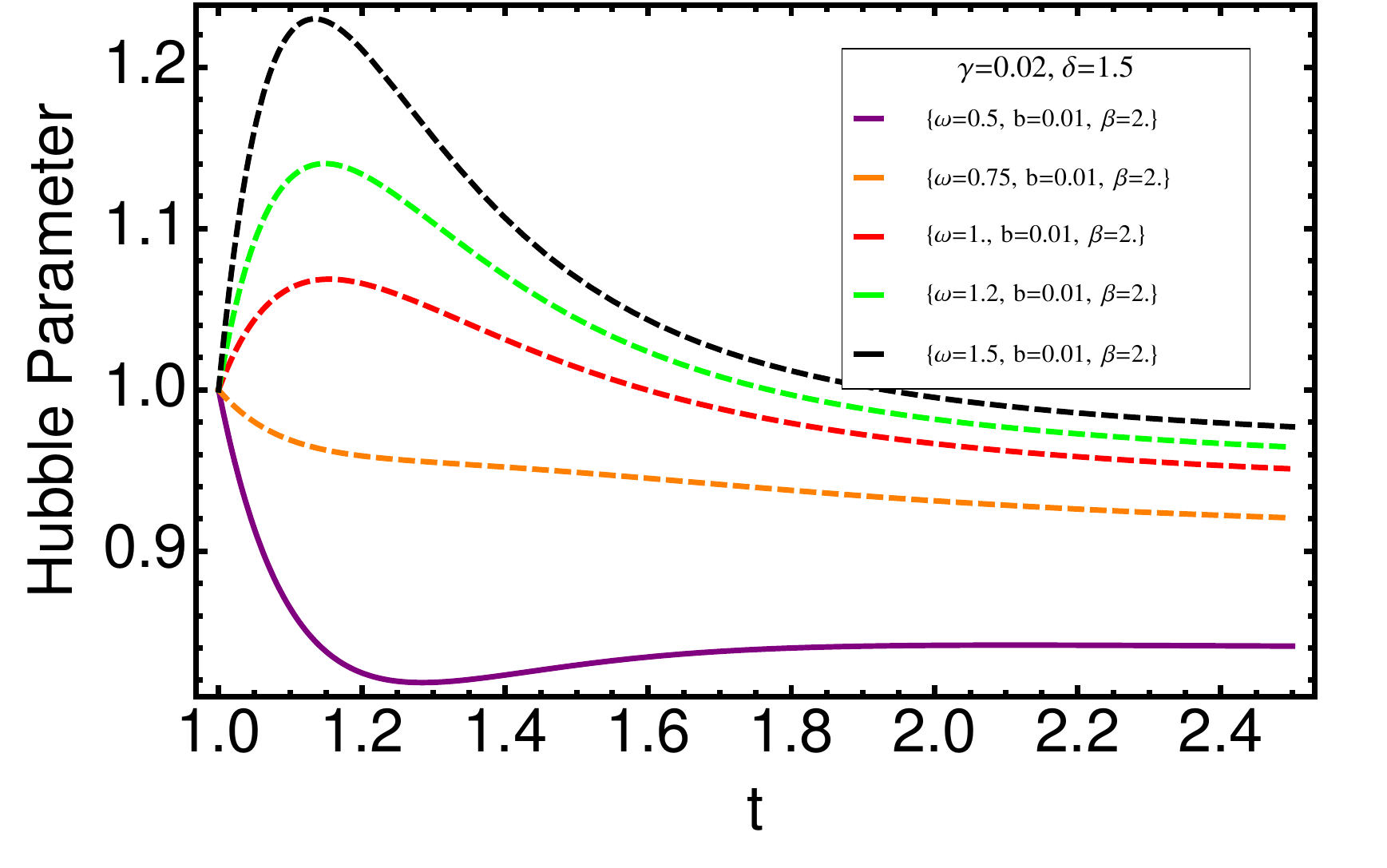}
 \end{array}$
 \end{center}
\caption{Behavior of $H(t)$ against $t$. Model 1}
 \label{fig:10}
\end{figure}

\begin{figure}[!ht]
 \begin{center}$
 \begin{array}{cccc}
\includegraphics[width=50 mm]{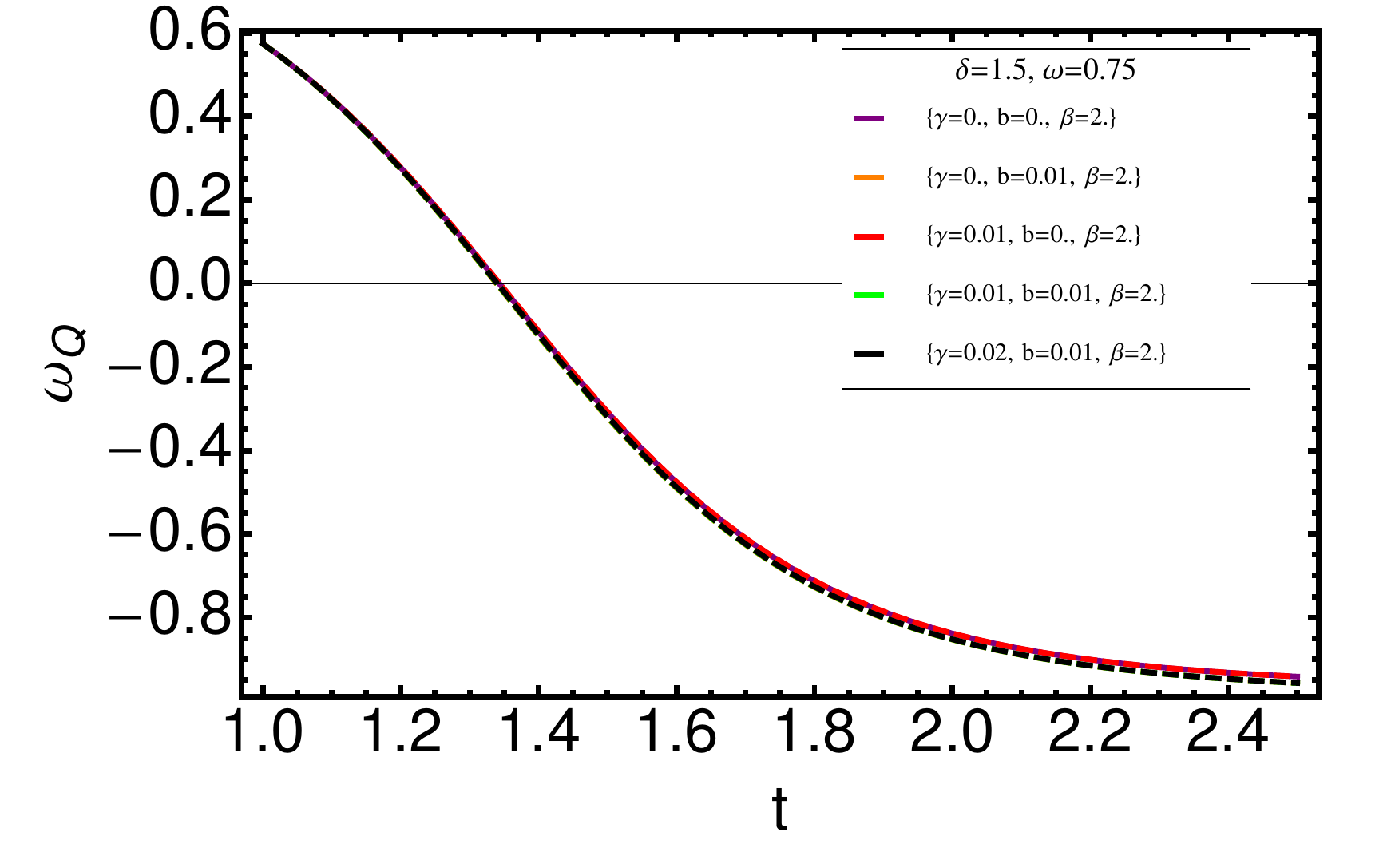} &
\includegraphics[width=50 mm]{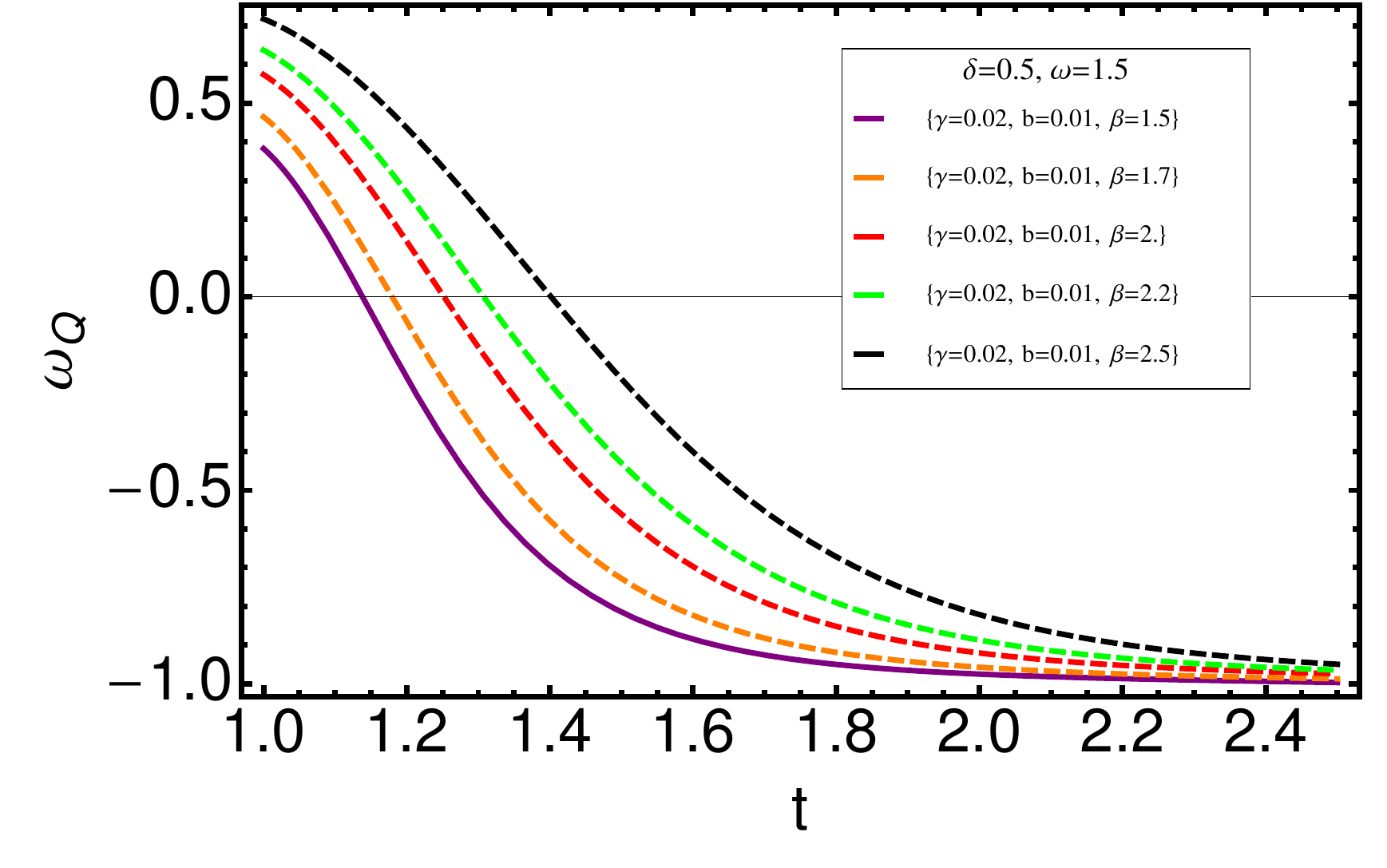}\\
\includegraphics[width=50 mm]{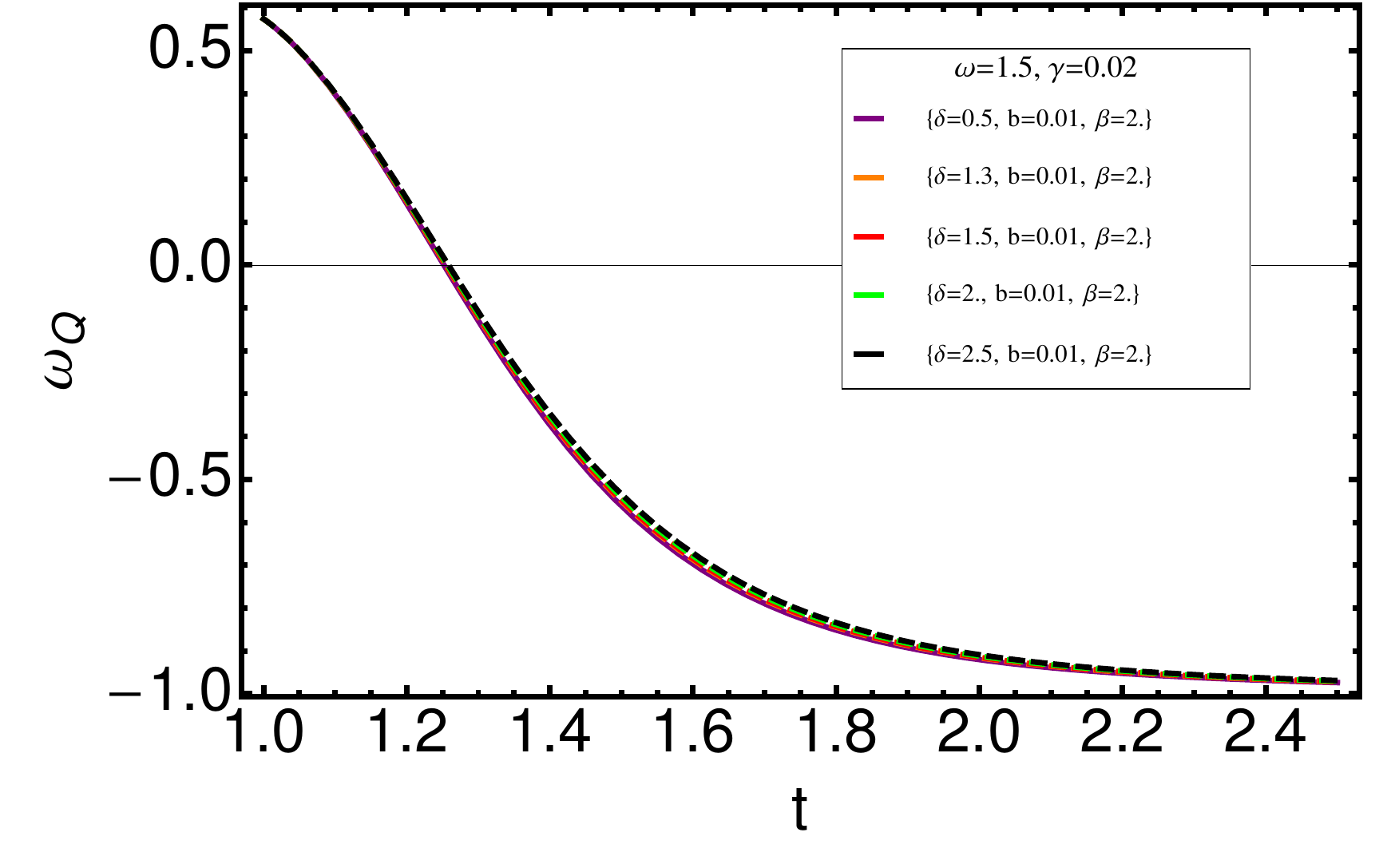}&
\includegraphics[width=50 mm]{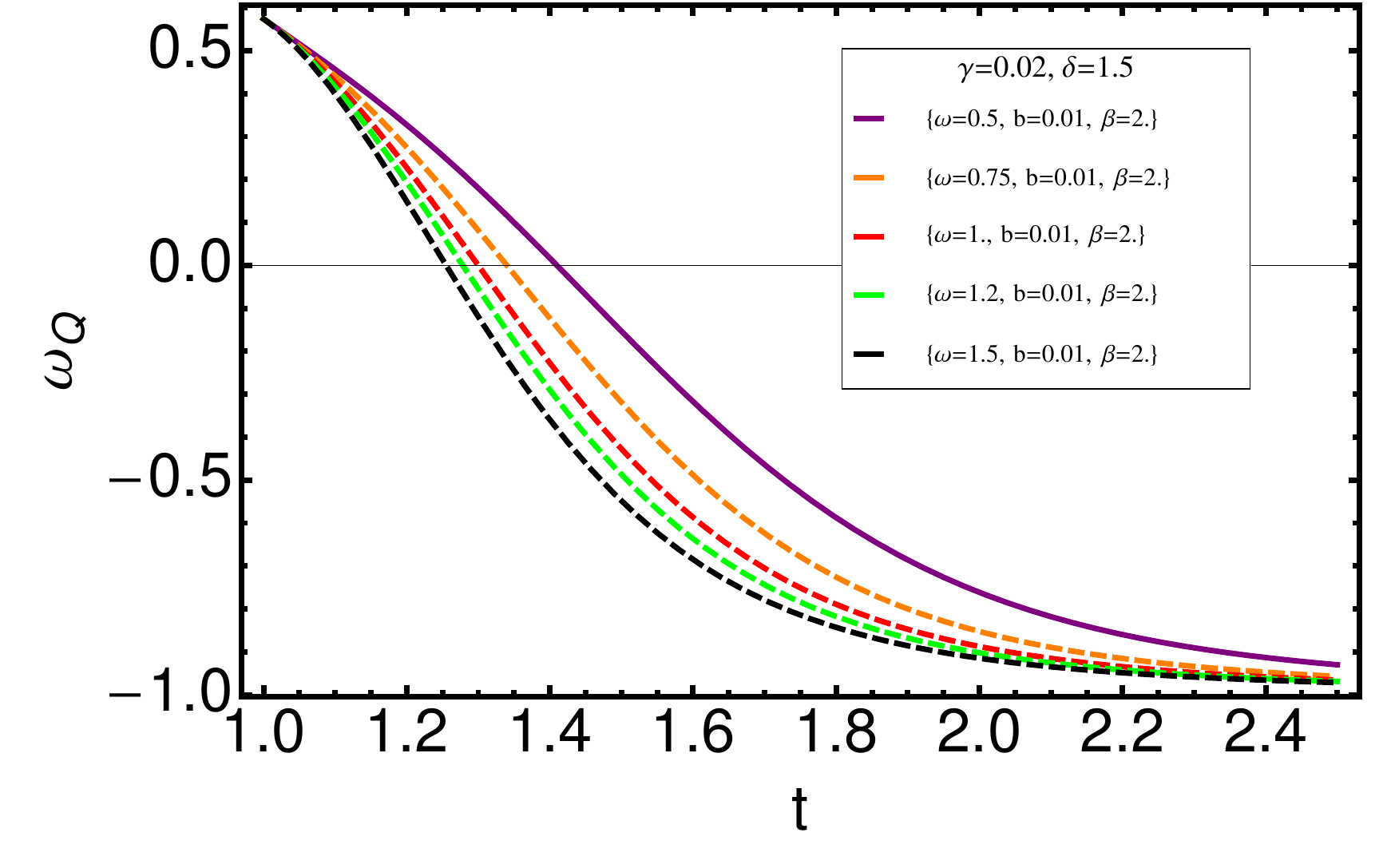}
 \end{array}$
 \end{center}
\caption{Behavior of $\omega_{Q}$ against $t$. Model 1}
 \label{fig:11}
\end{figure}

\begin{figure}[htb]
 \begin{center}$
 \begin{array}{cccc}
\includegraphics[width=50 mm]{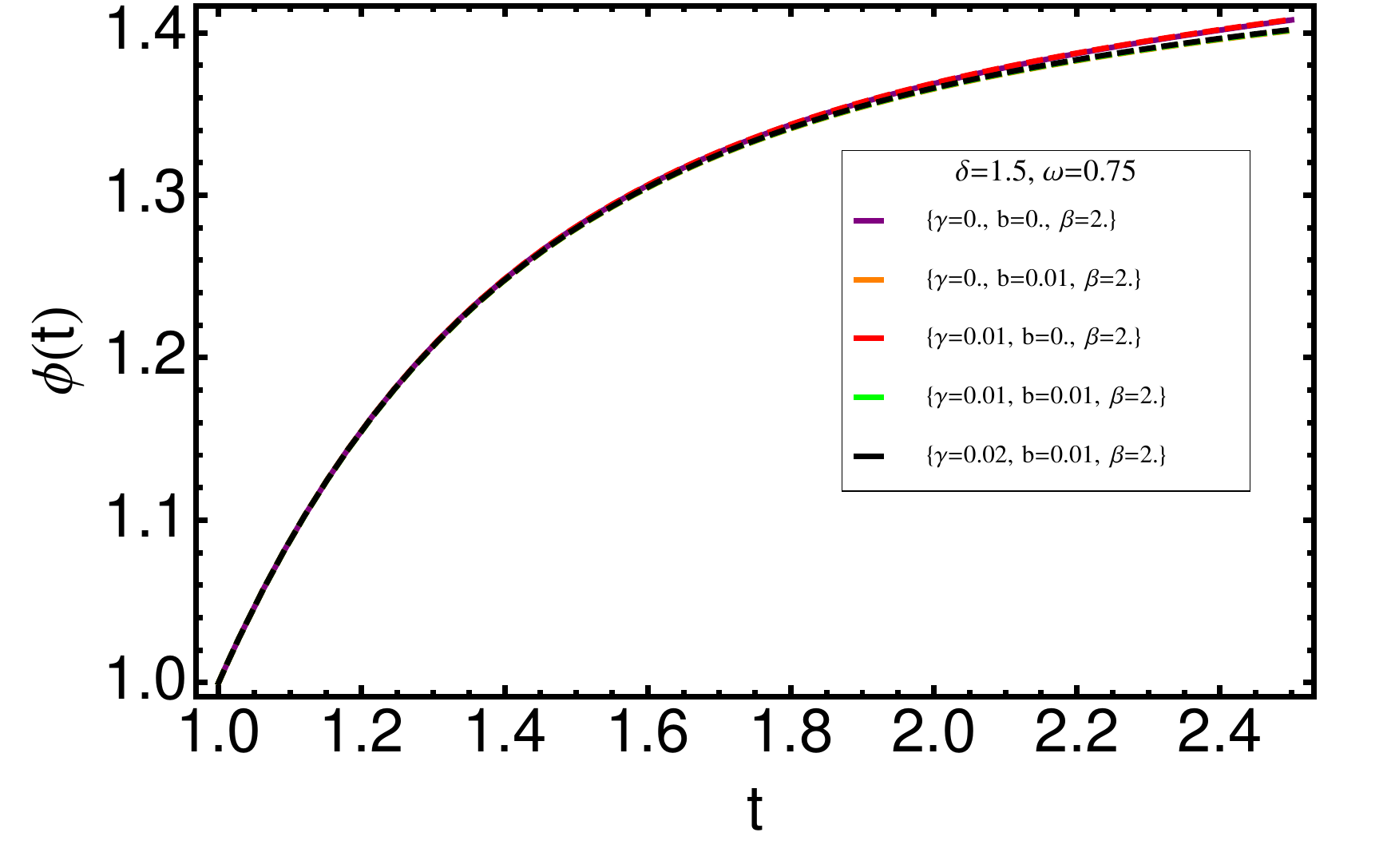} &
\includegraphics[width=50 mm]{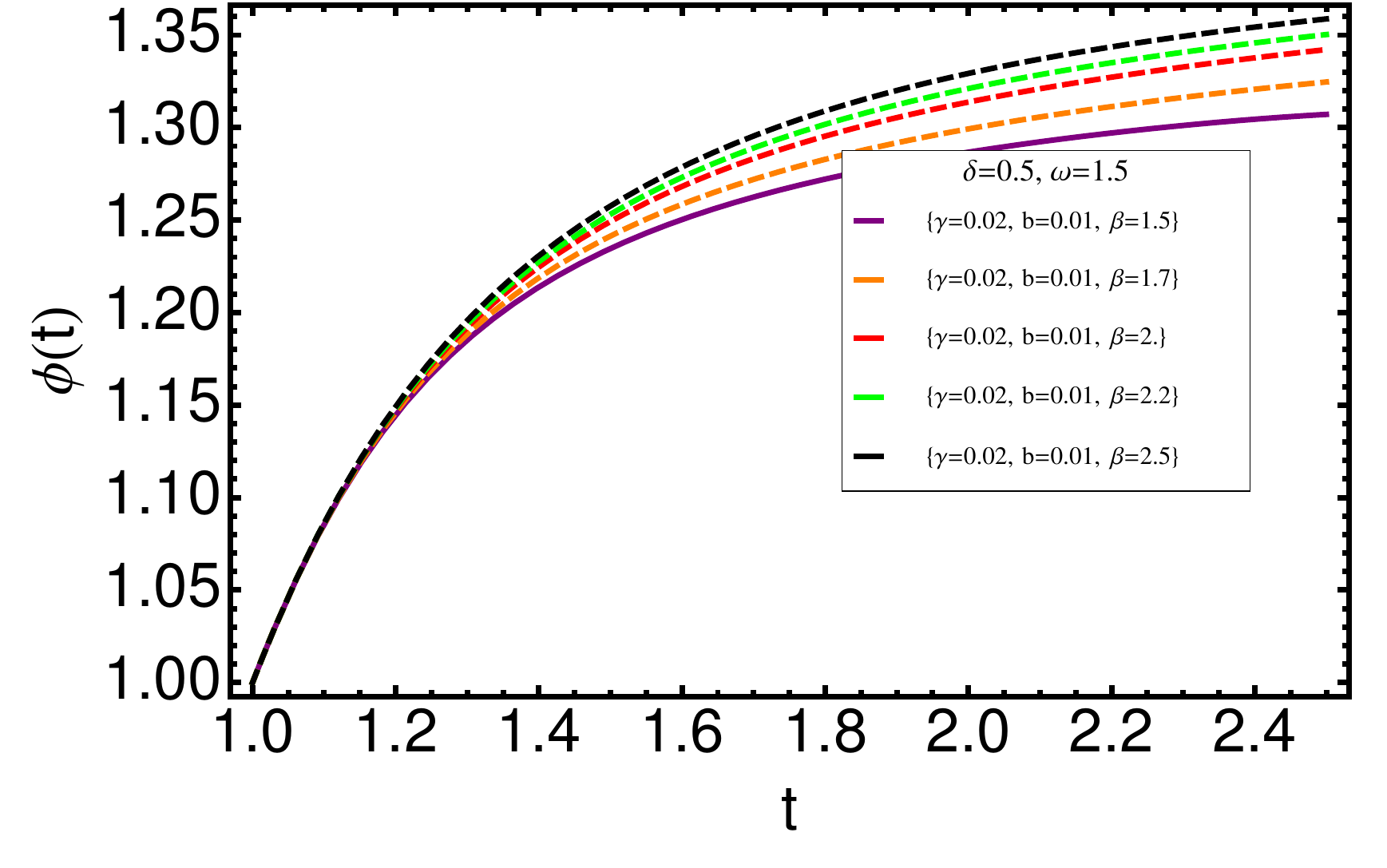}\\
\includegraphics[width=50 mm]{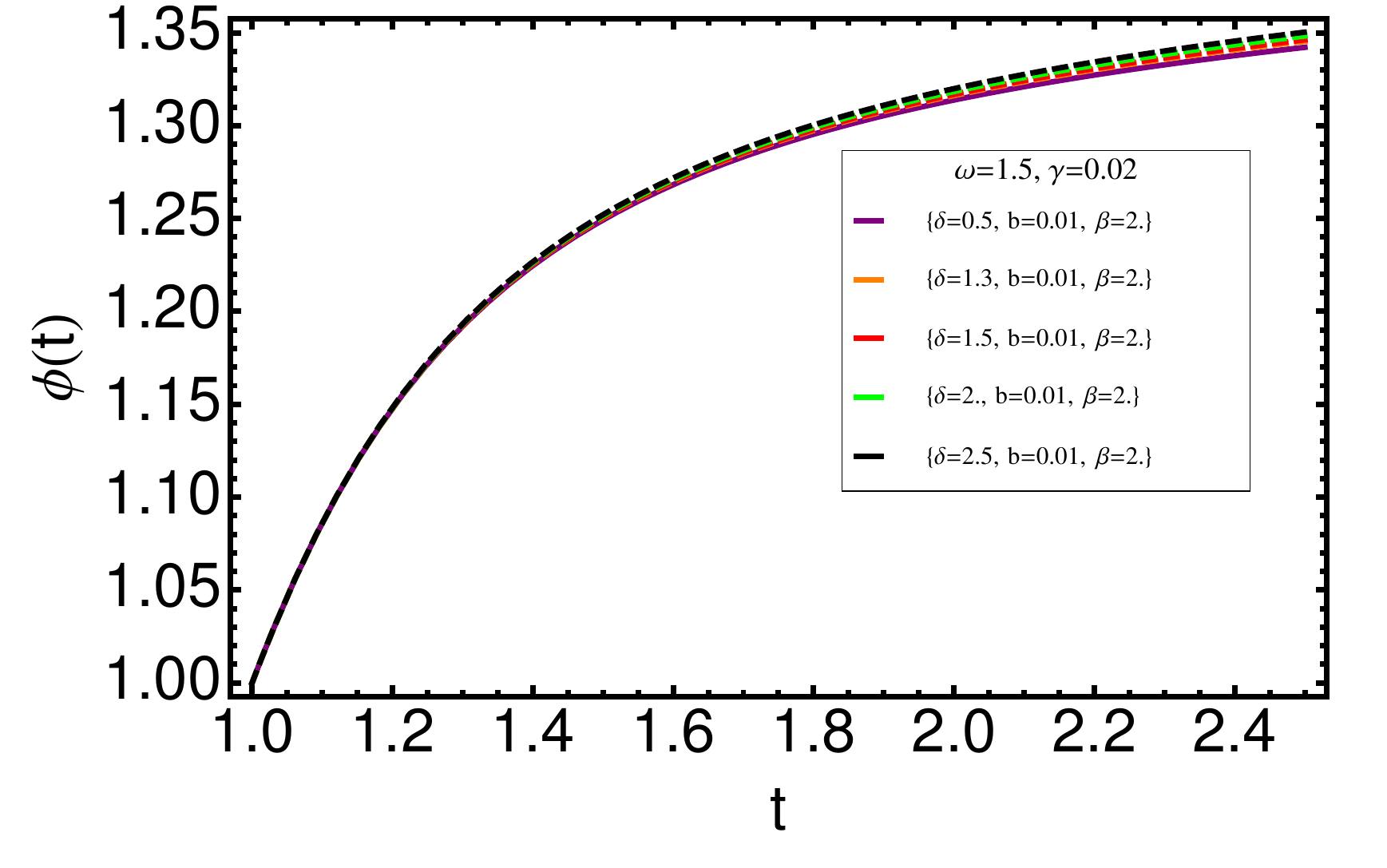}&
\includegraphics[width=50 mm]{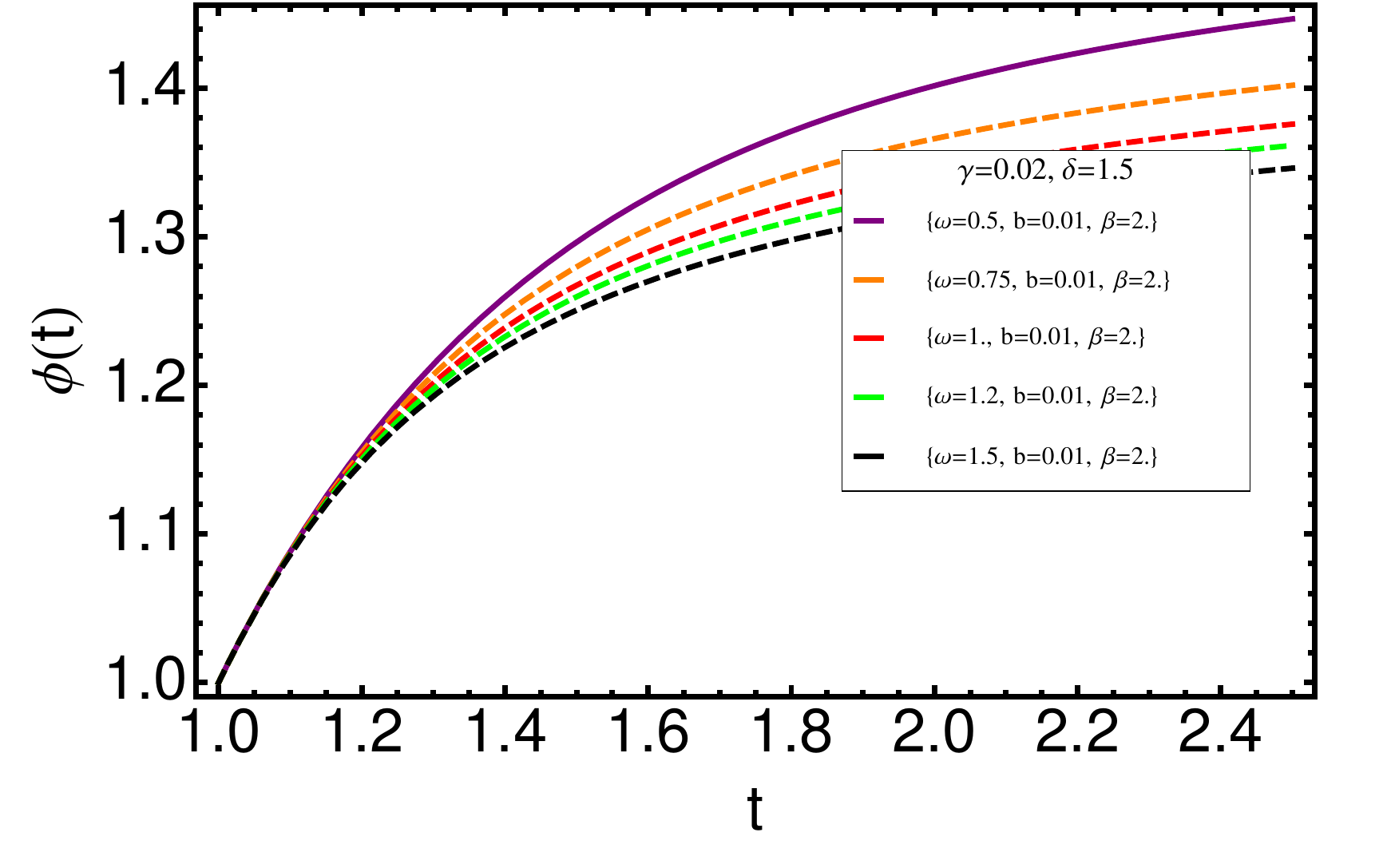}
 \end{array}$
 \end{center}
\caption{Behavior of $\phi$ against $t$. Model 1}
 \label{fig:12}
\end{figure}

\begin{figure}[h!]
 \begin{center}$
 \begin{array}{cccc}
\includegraphics[width=50 mm]{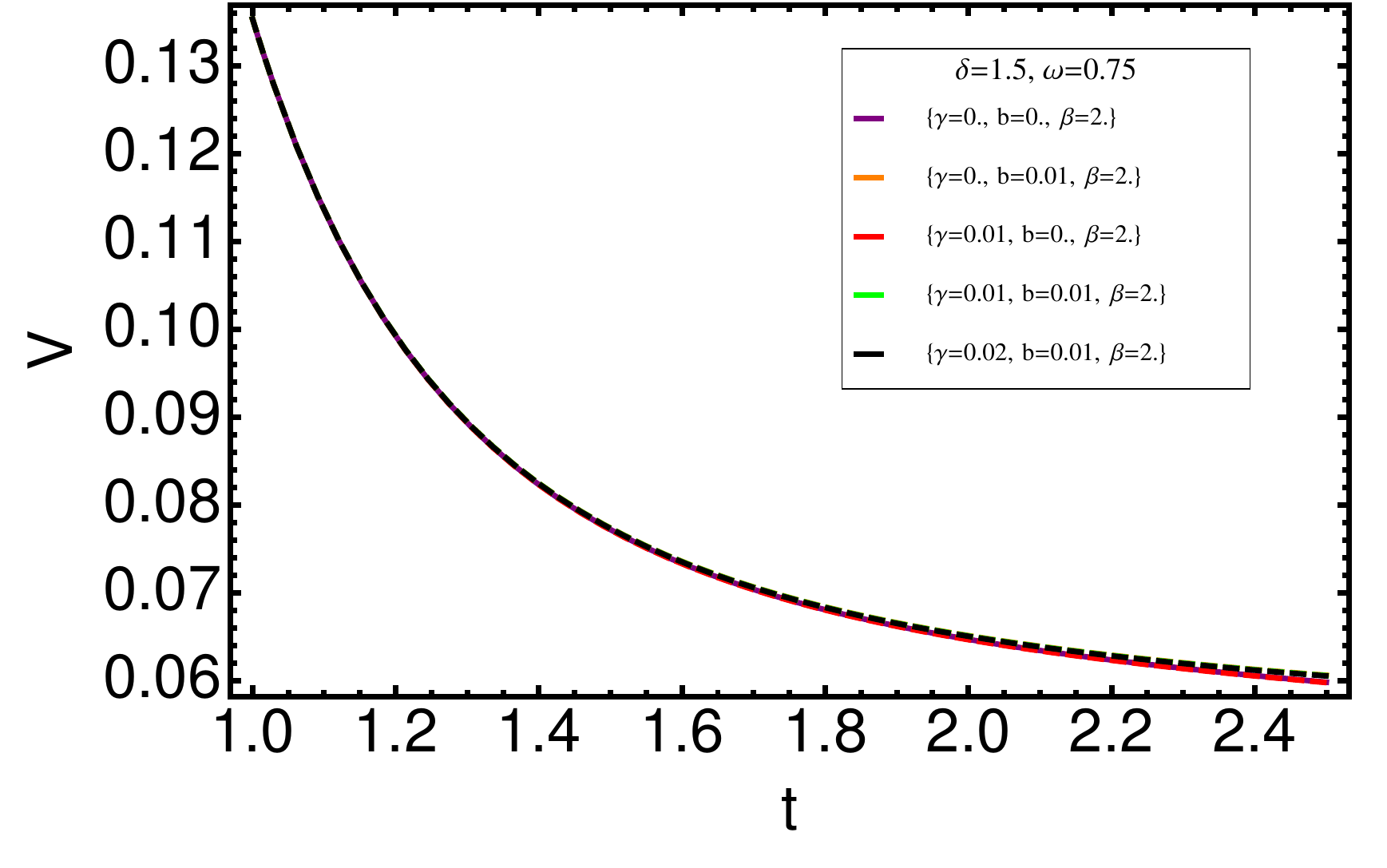} &
\includegraphics[width=50 mm]{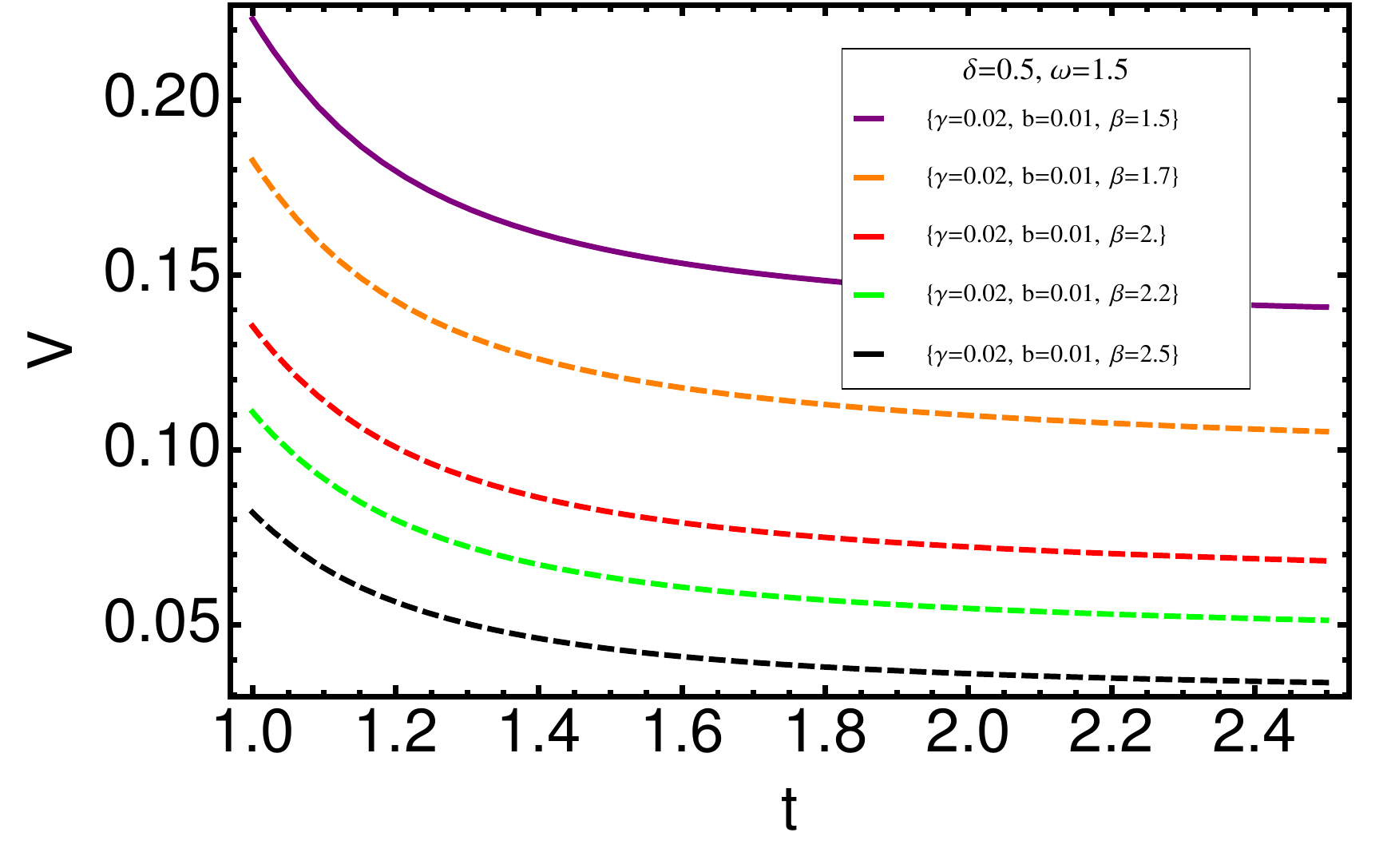}\\
\includegraphics[width=50 mm]{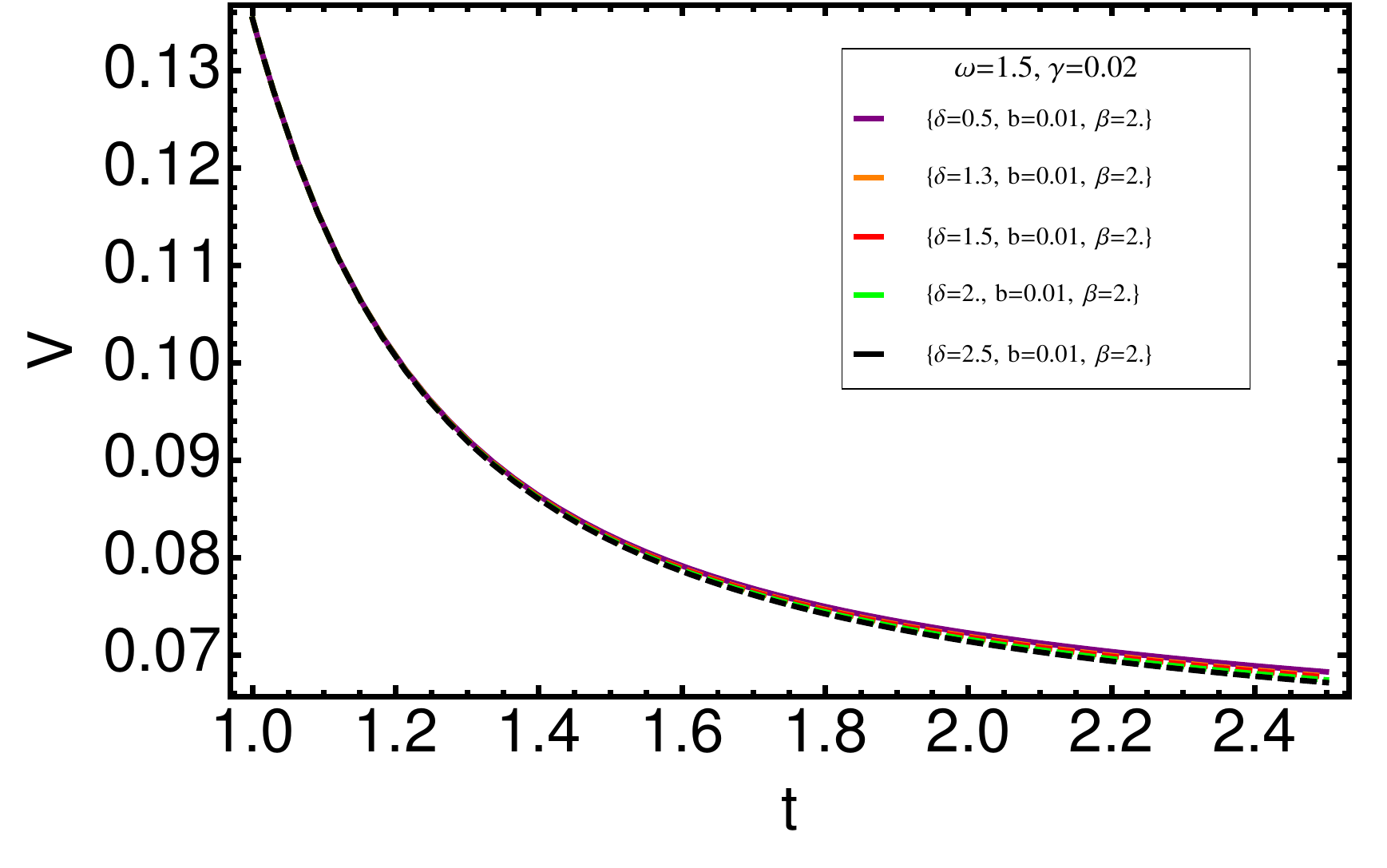}&
\includegraphics[width=50 mm]{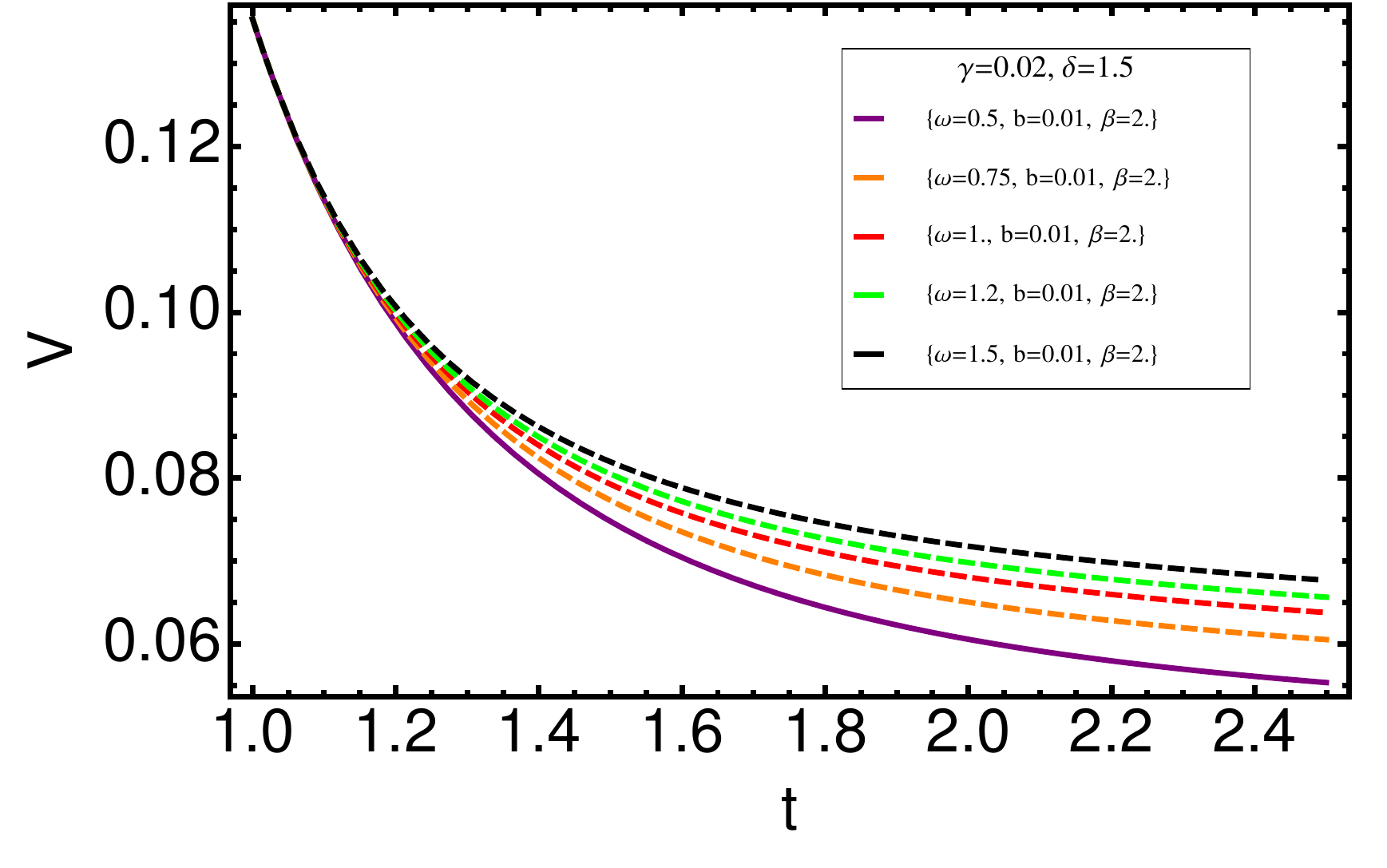}
 \end{array}$
 \end{center}
\caption{Behavior of $V(t)$ against $t$. Model 1}
 \label{fig:13}
\end{figure}

\begin{figure}[h!]
 \begin{center}$
 \begin{array}{cccc}
\includegraphics[width=50 mm]{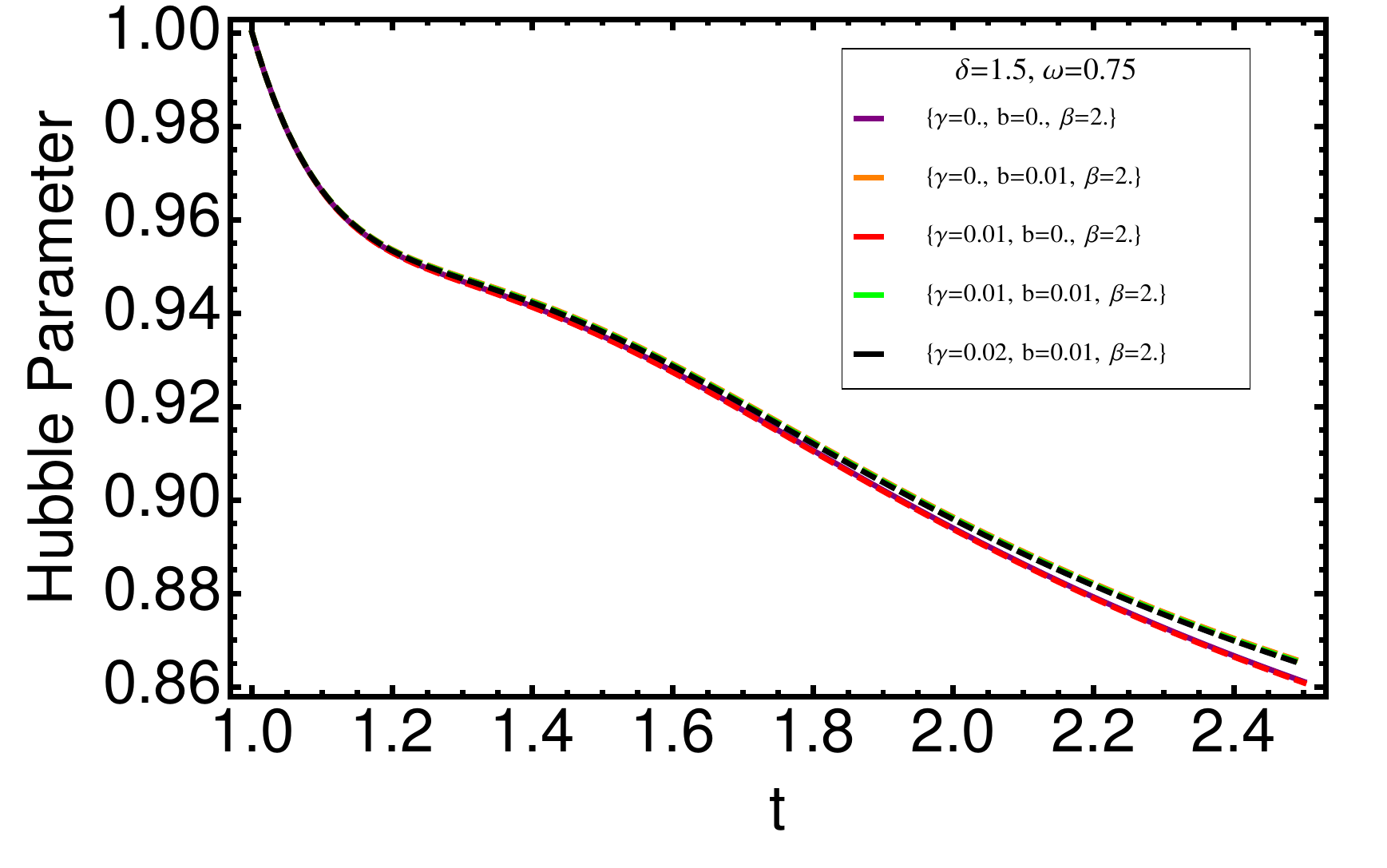} &
\includegraphics[width=50 mm]{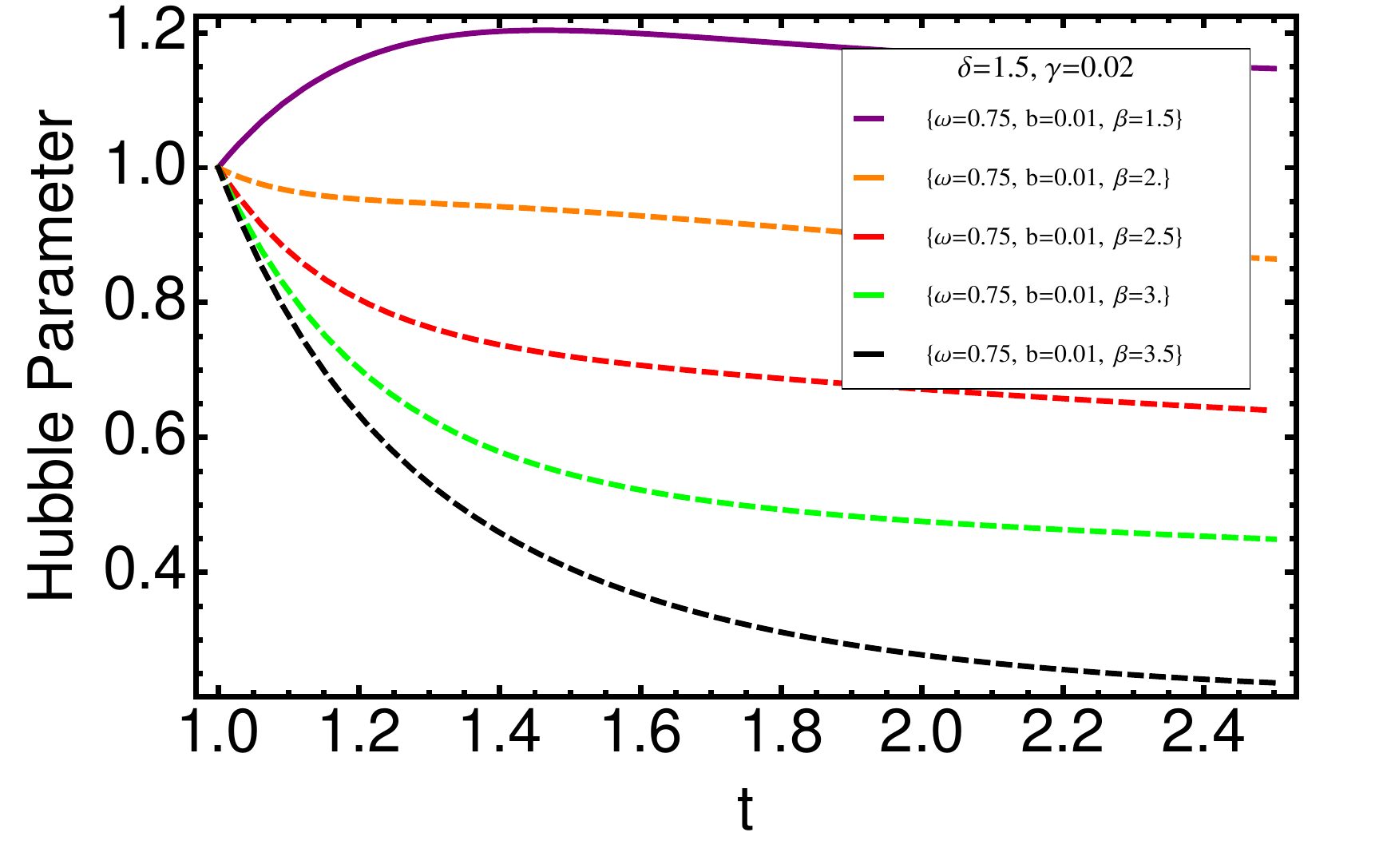}\\
\includegraphics[width=50 mm]{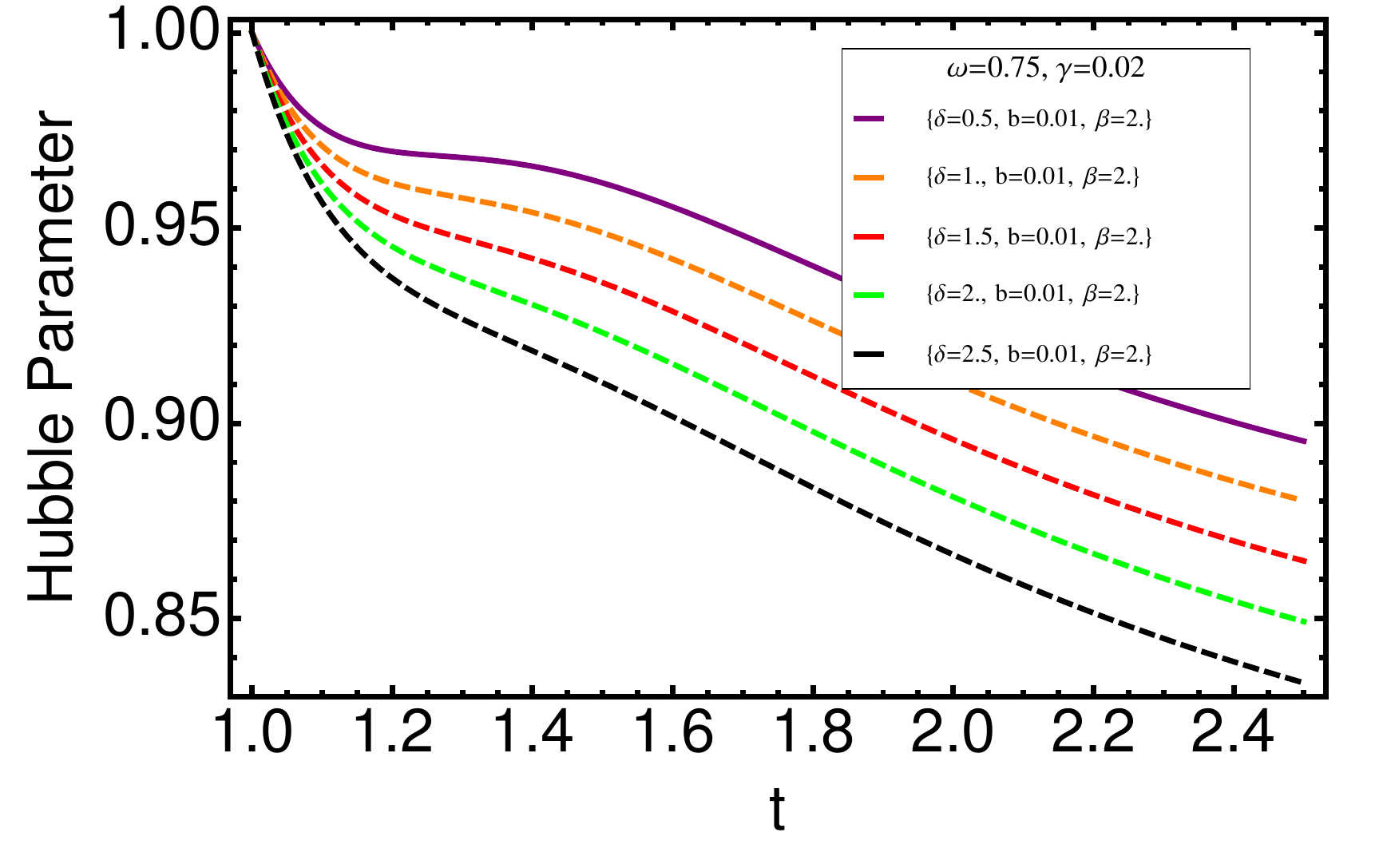}&
\includegraphics[width=50 mm]{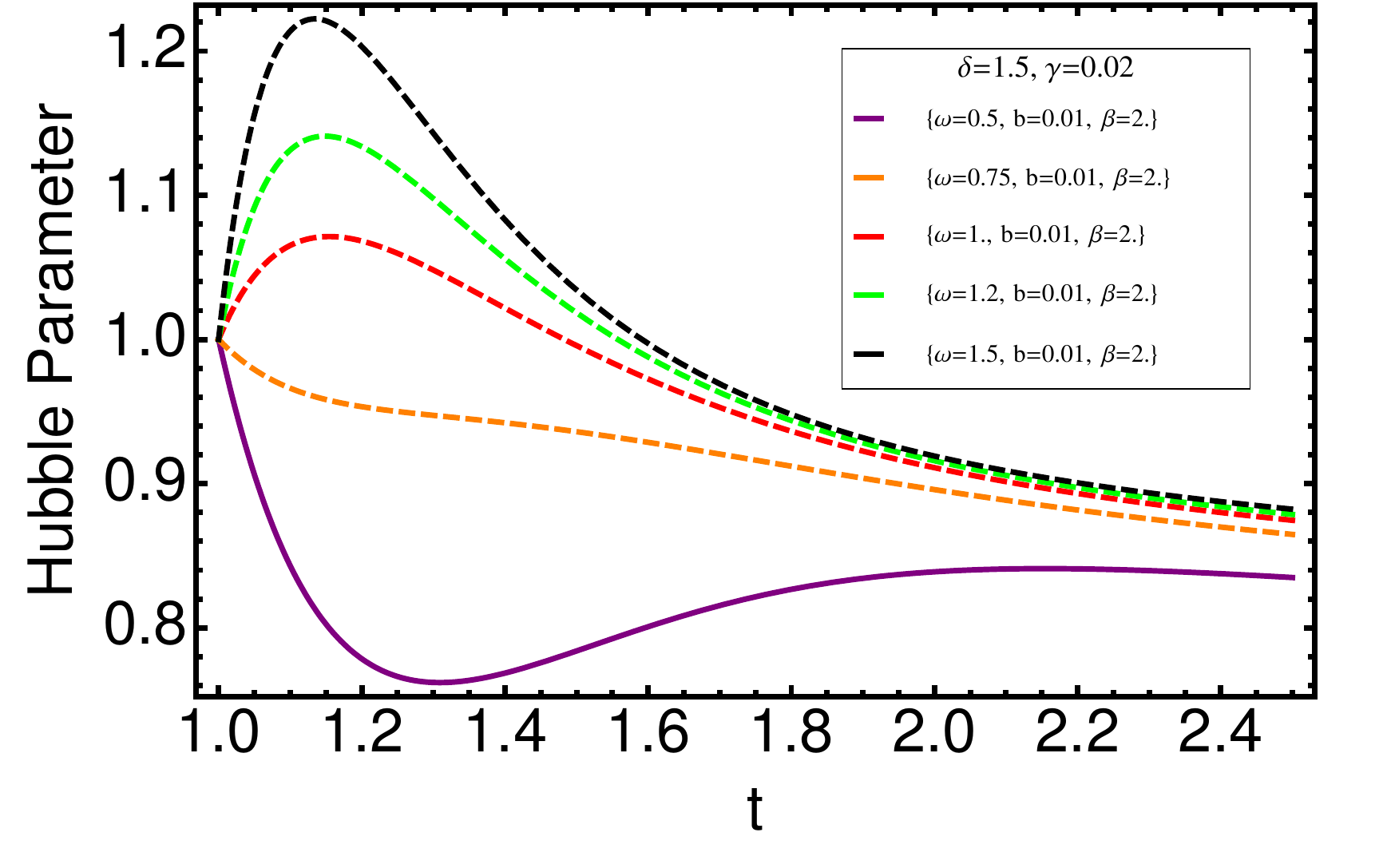}
 \end{array}$
 \end{center}
\caption{Behavior of $H(t)$ against $t$. Model 2}
 \label{fig:14}
\end{figure}

\begin{figure}[h!]
 \begin{center}$
 \begin{array}{cccc}
\includegraphics[width=50 mm]{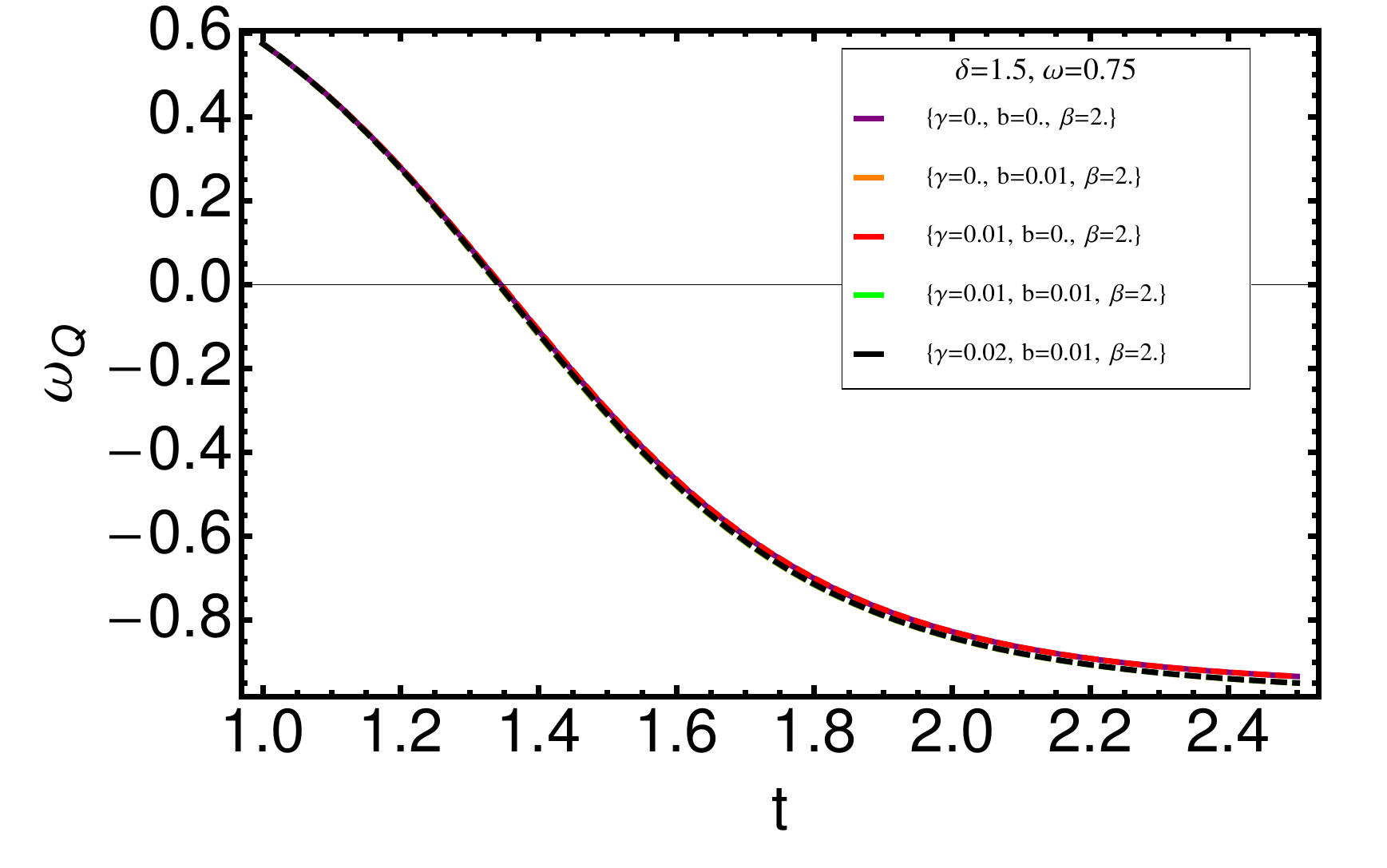} &
\includegraphics[width=50 mm]{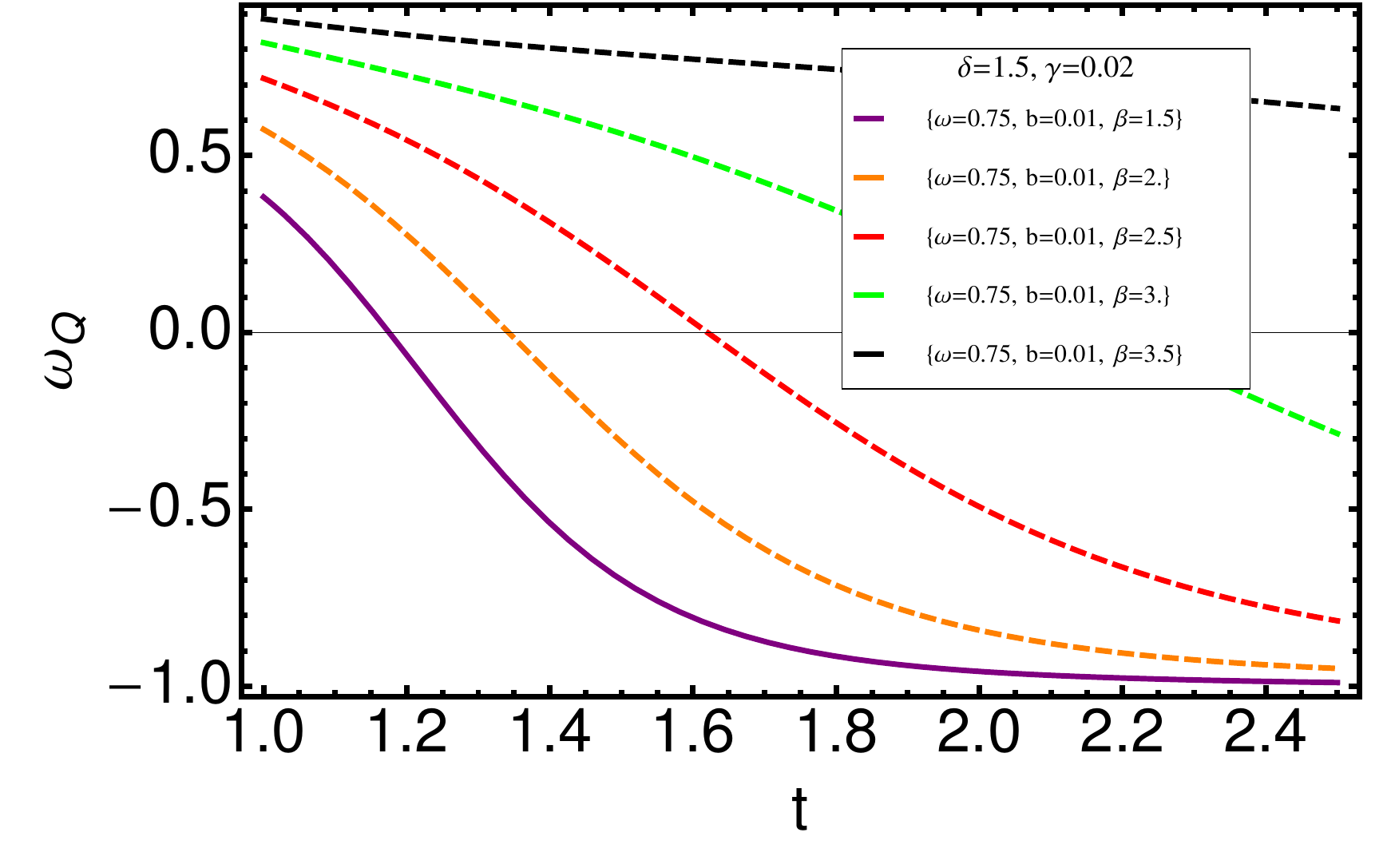}\\
\includegraphics[width=50 mm]{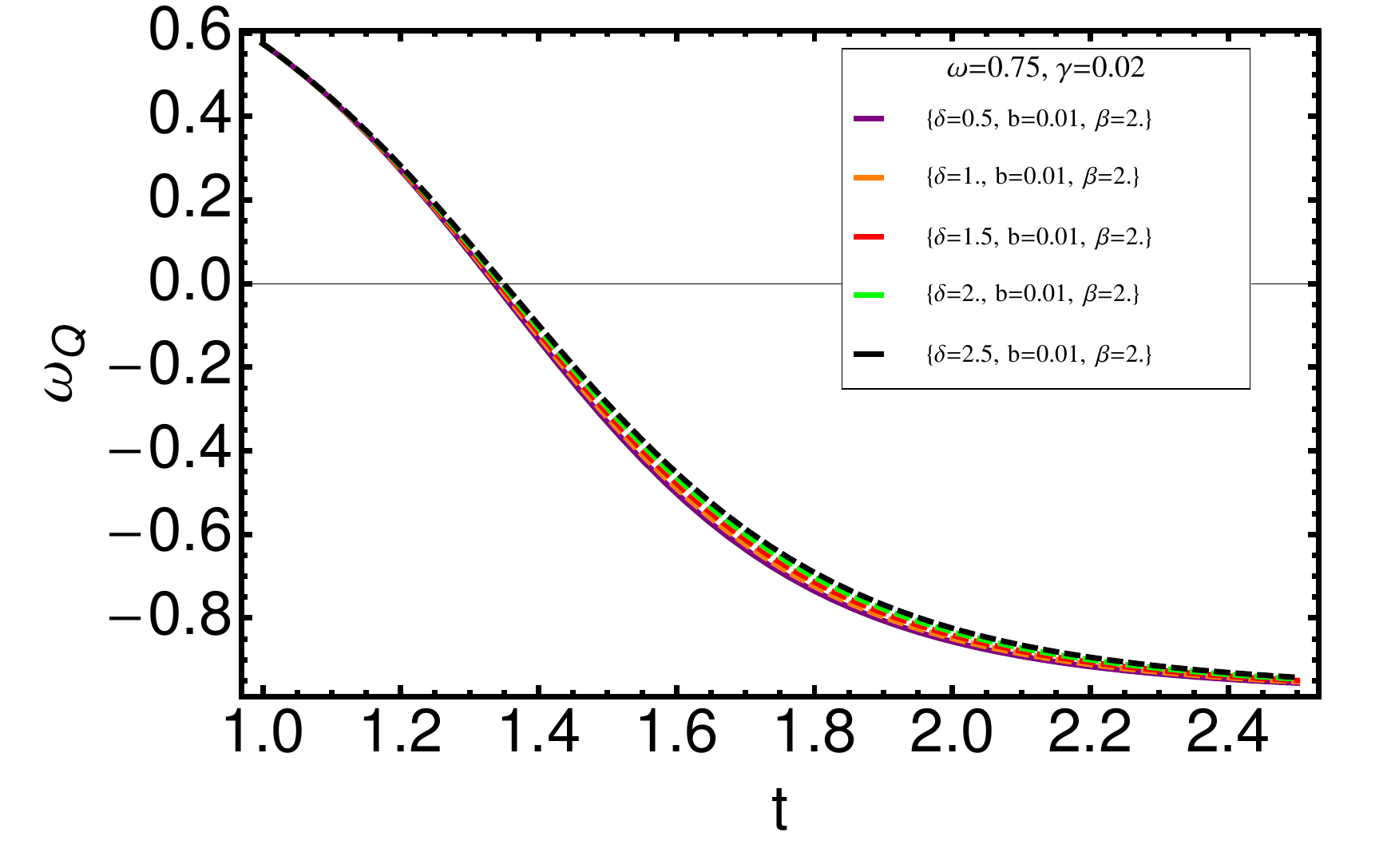}&
\includegraphics[width=50 mm]{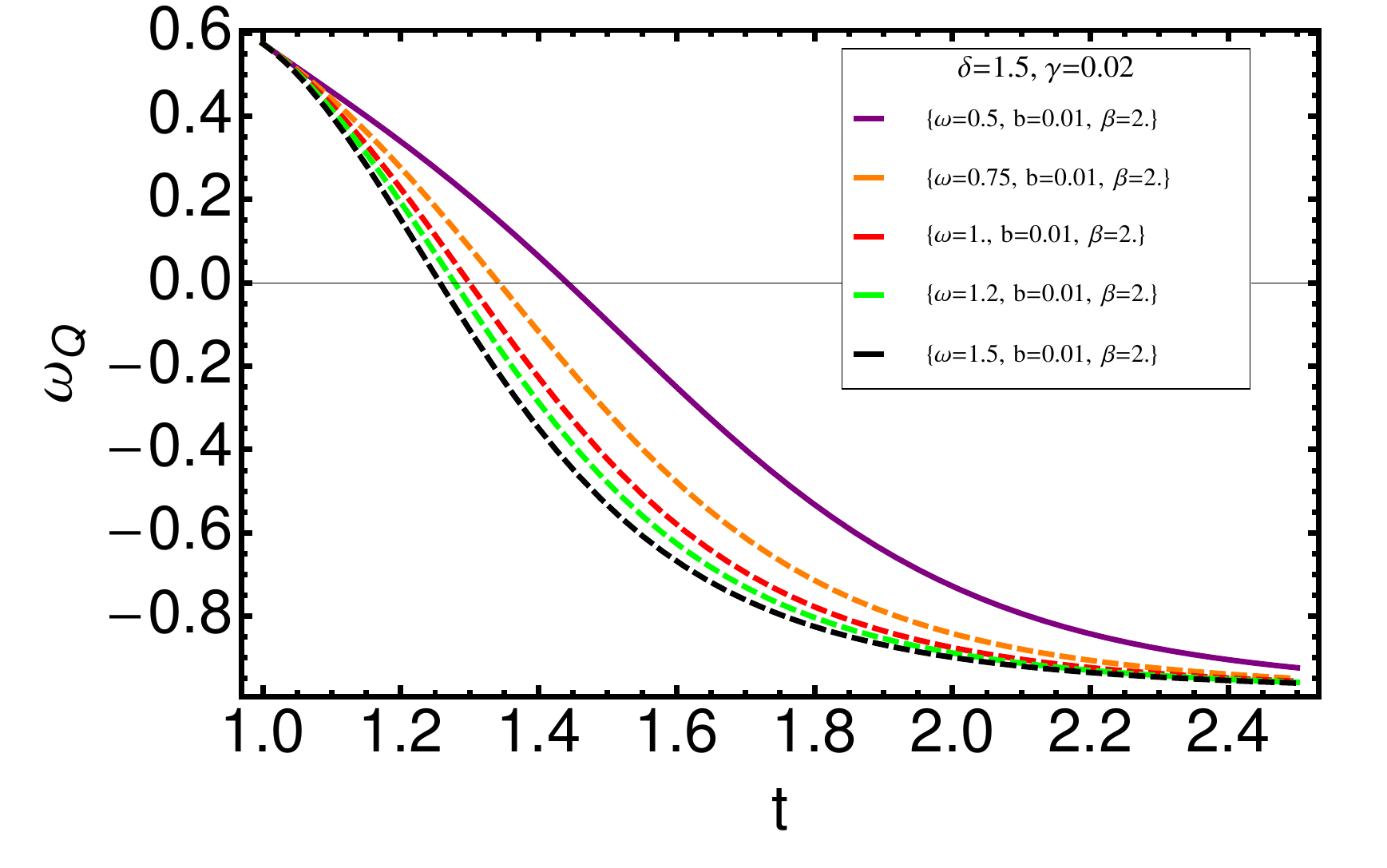}
 \end{array}$
 \end{center}
\caption{Behavior of $\omega_{Q}$ against $t$. Model 2}
 \label{fig:15}
\end{figure}

\begin{figure}[h!]
 \begin{center}$
 \begin{array}{cccc}
\includegraphics[width=50 mm]{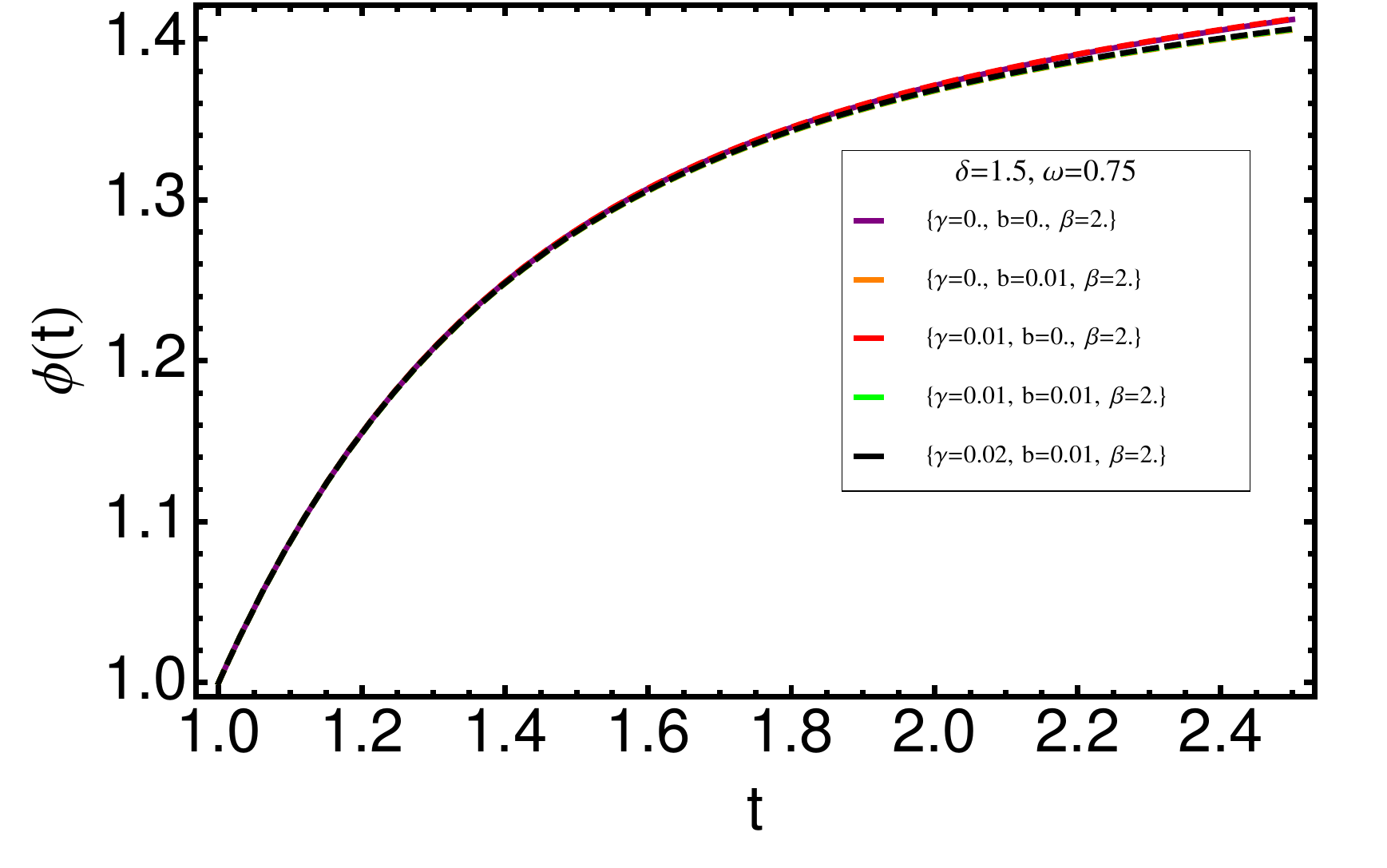} &
\includegraphics[width=50 mm]{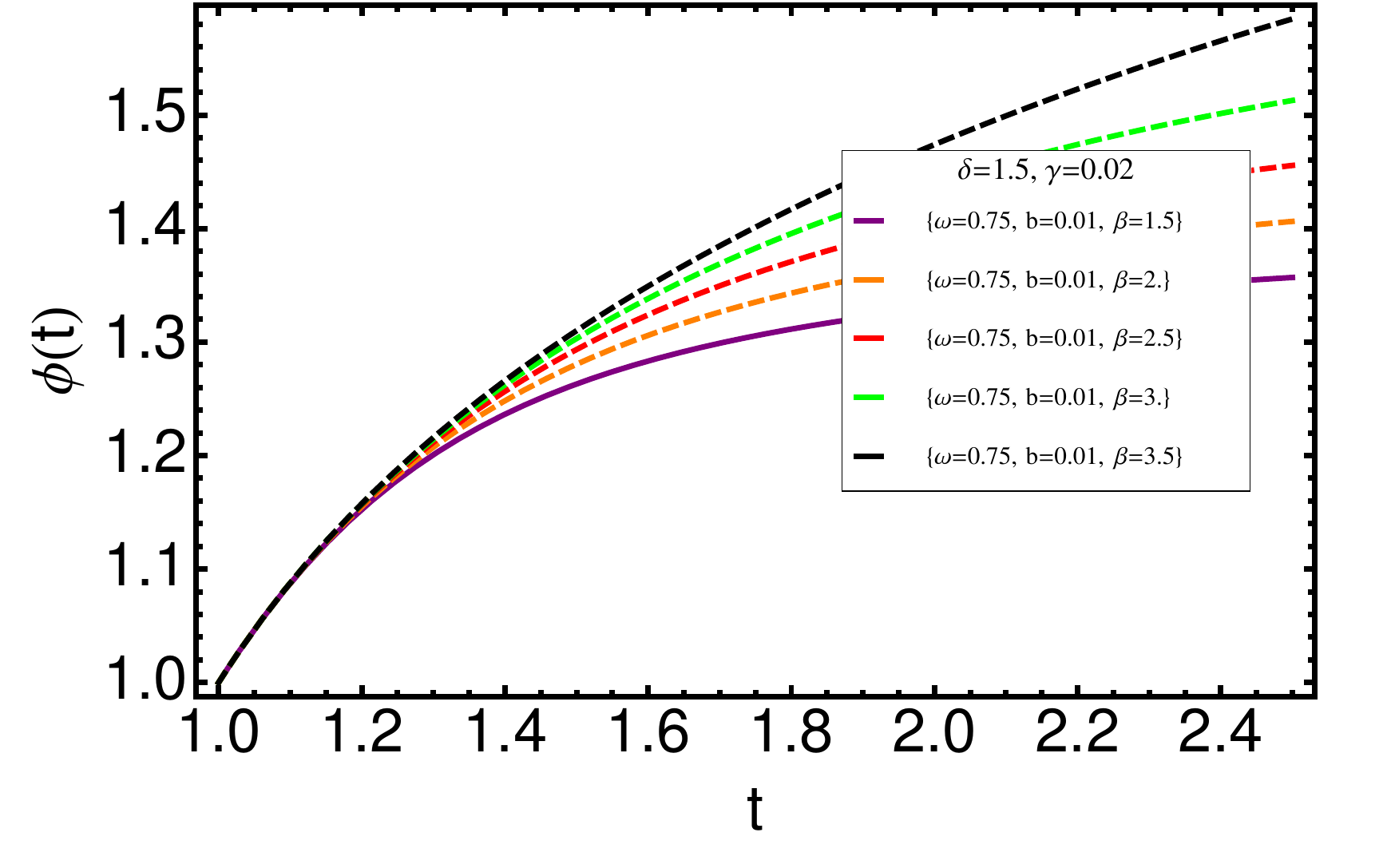}\\
\includegraphics[width=50 mm]{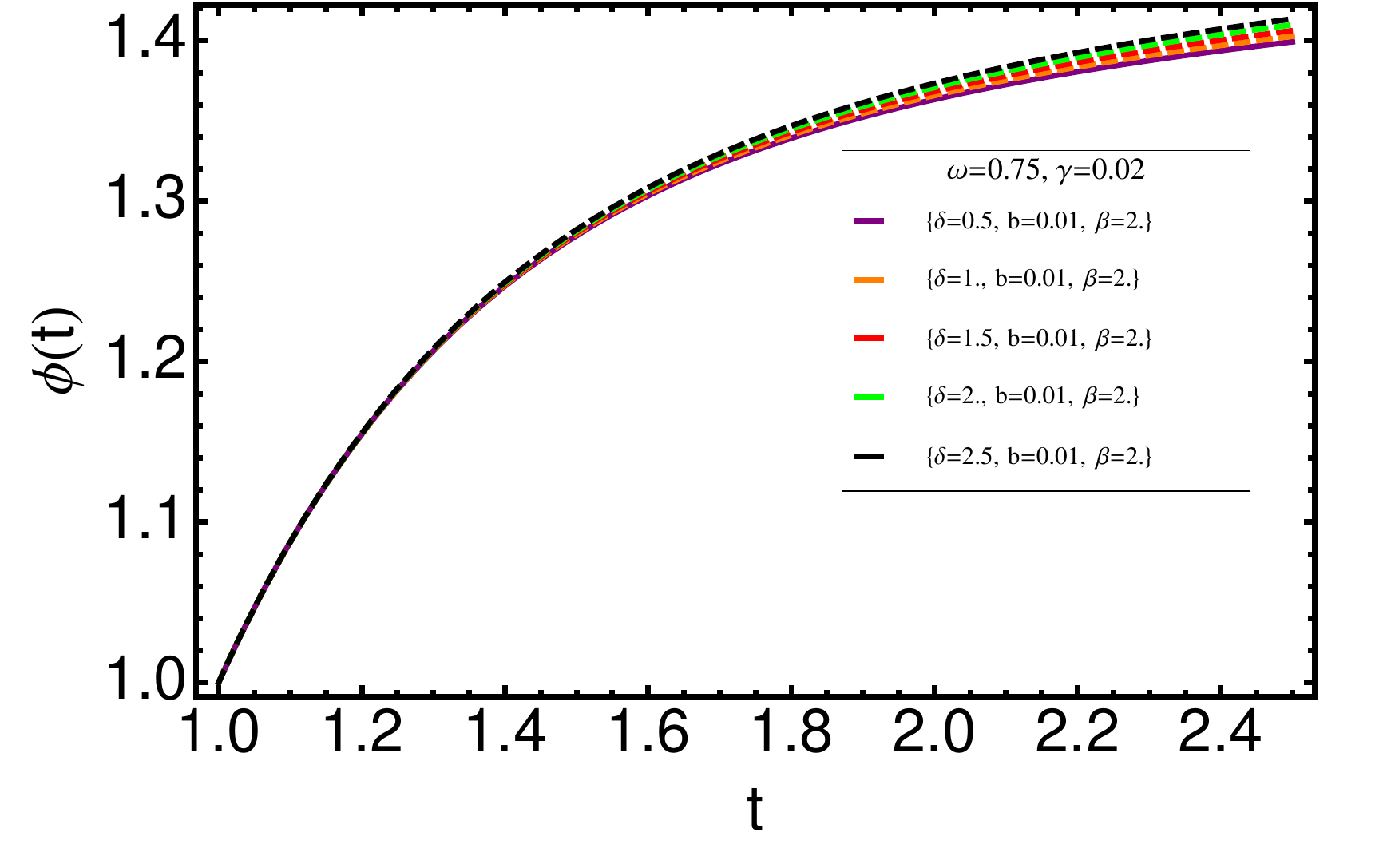}&
\includegraphics[width=50 mm]{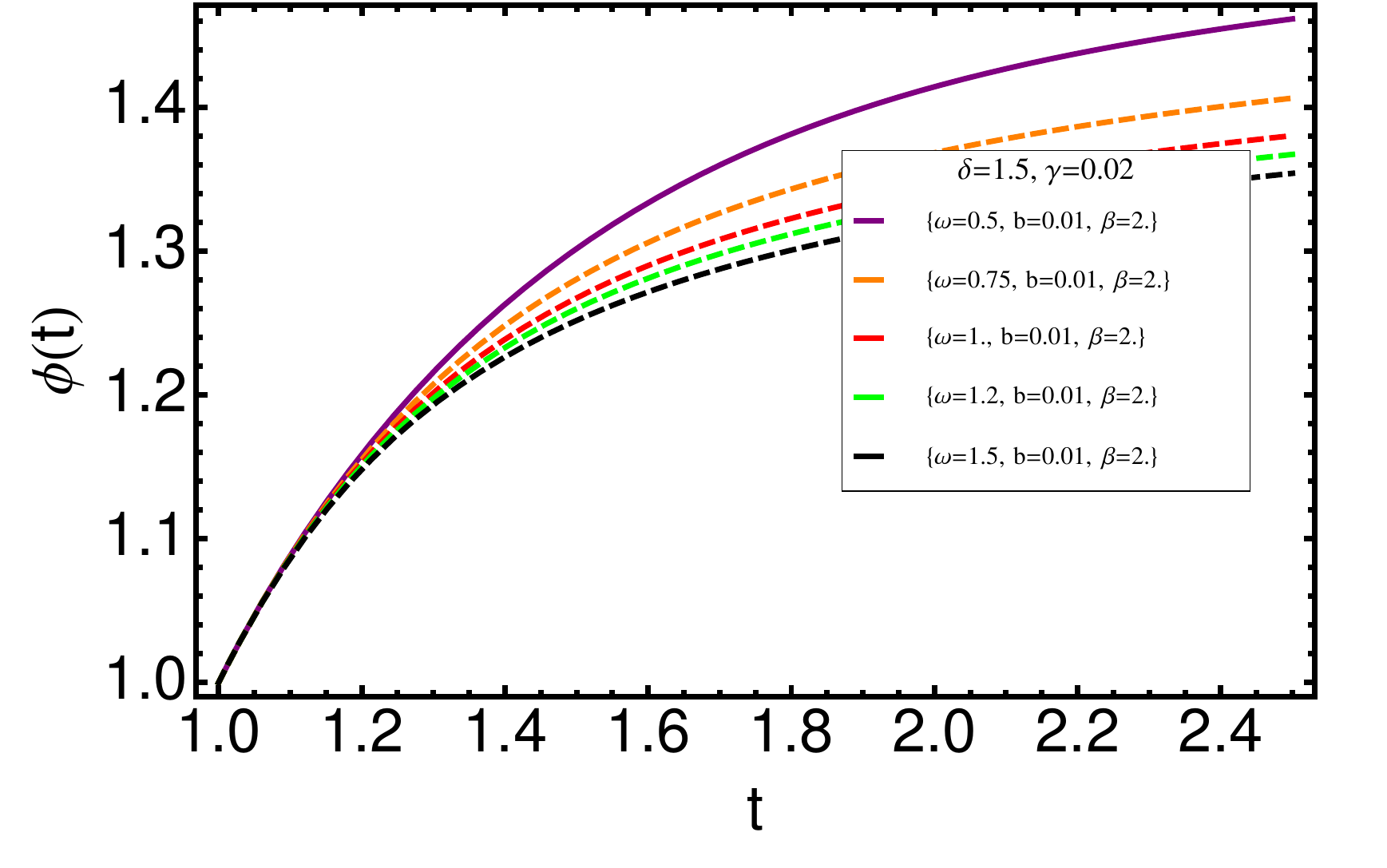}
 \end{array}$
 \end{center}
\caption{Behavior of $\phi$ against $t$. Model 2}
 \label{fig:16}
\end{figure}

\begin{figure}[h!]
 \begin{center}$
 \begin{array}{cccc}
\includegraphics[width=50 mm]{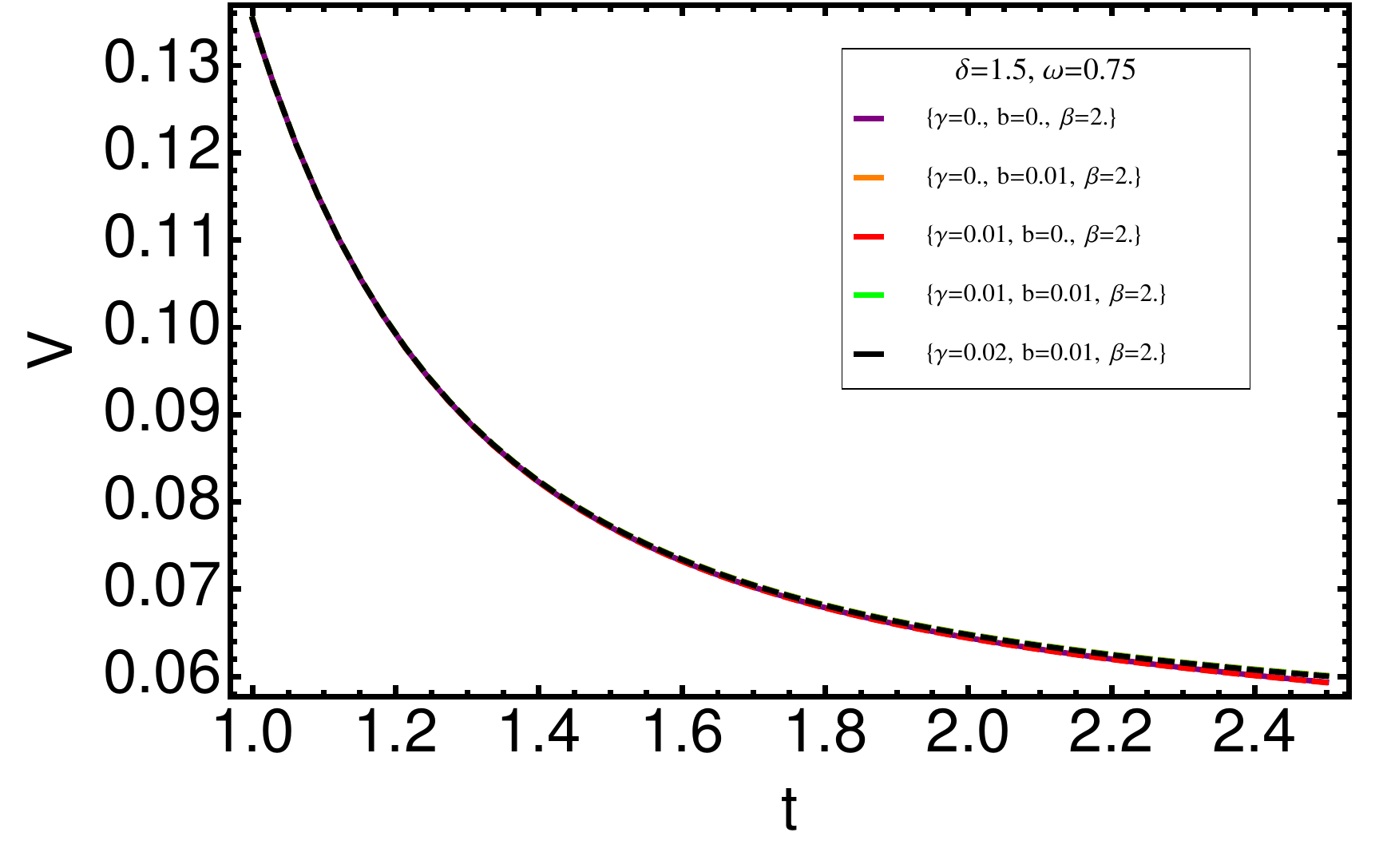} &
\includegraphics[width=50 mm]{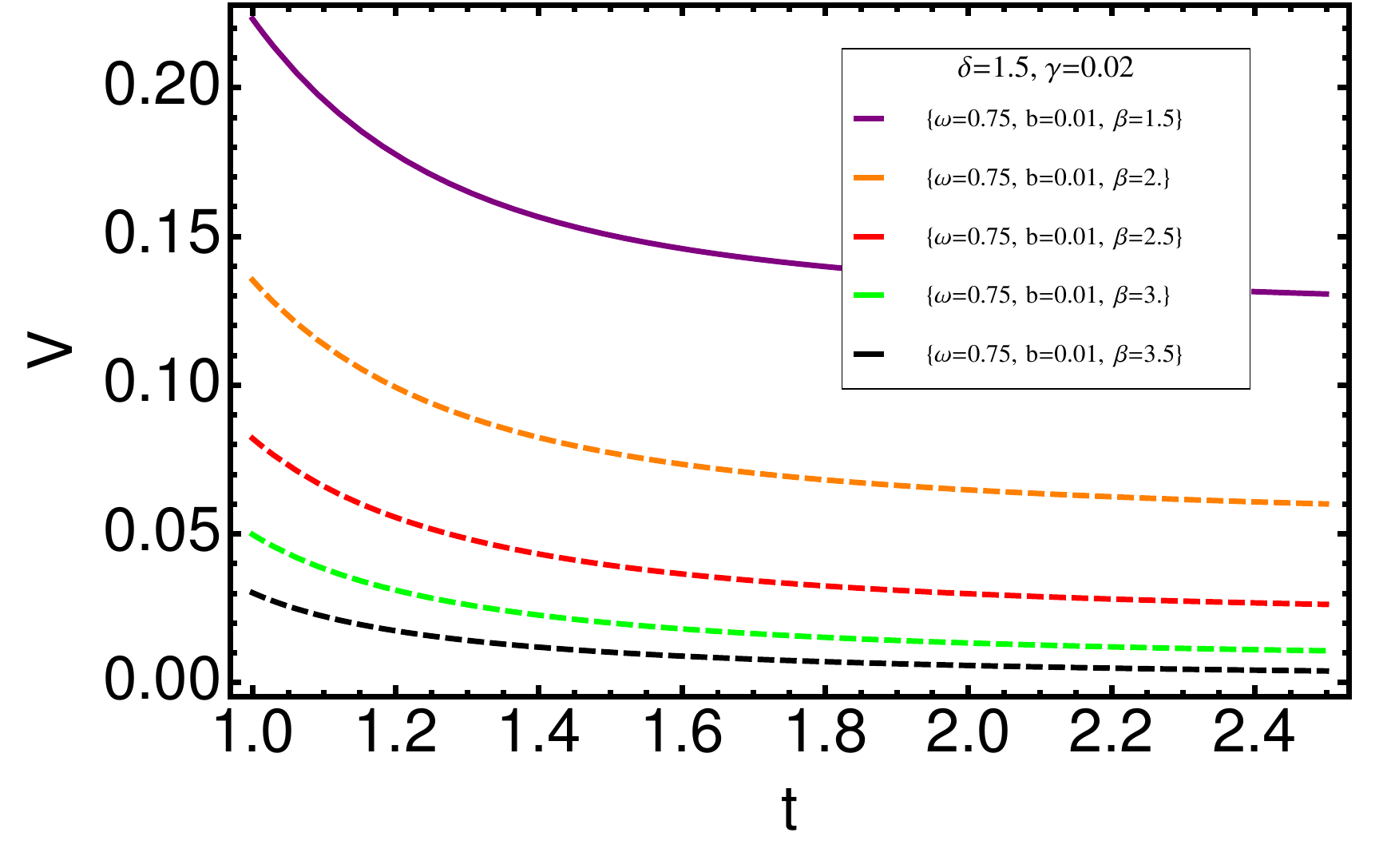}\\
\includegraphics[width=50 mm]{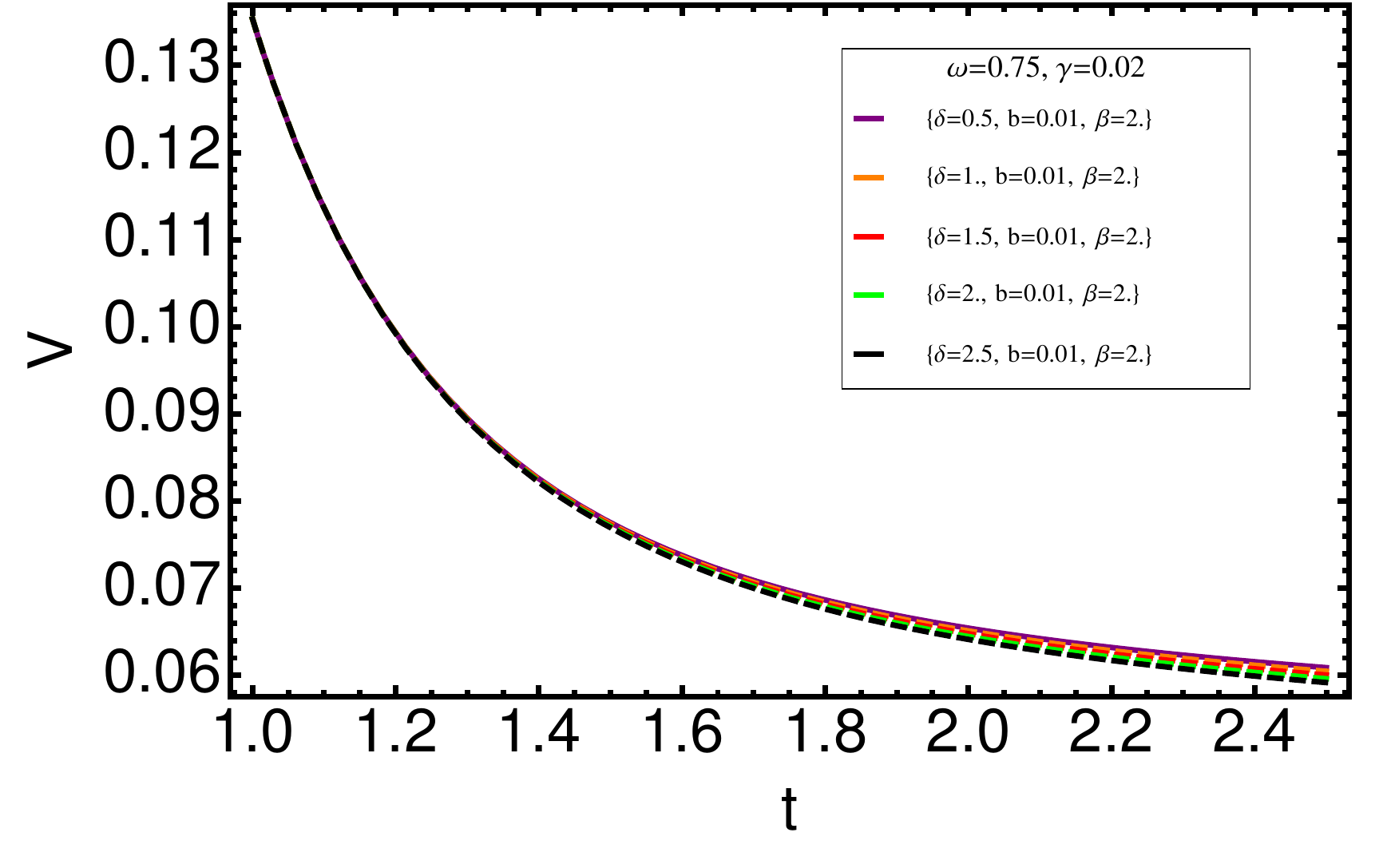}&
\includegraphics[width=50 mm]{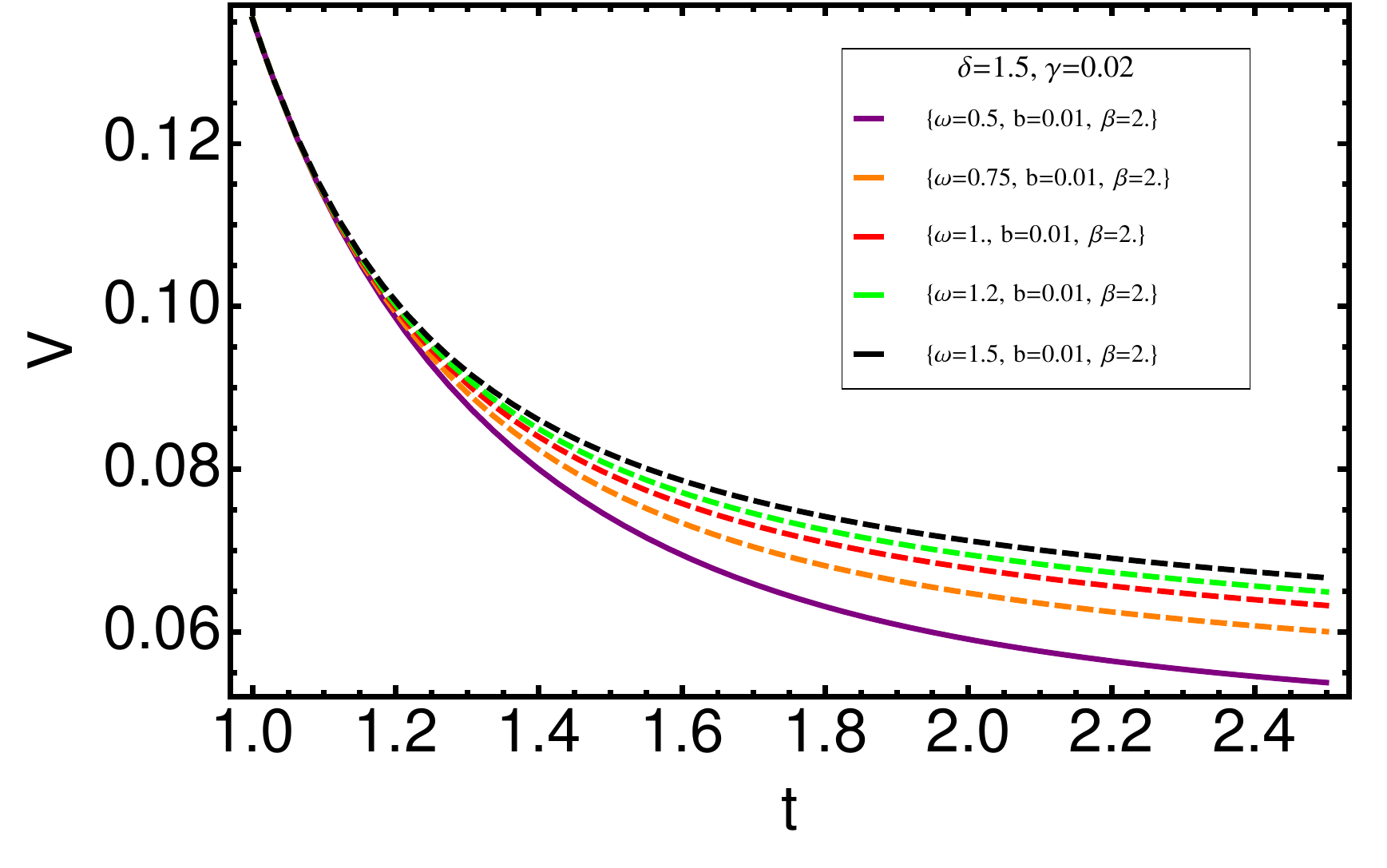}
 \end{array}$
 \end{center}
\caption{Behavior of $V(t)$ against $t$. Model 2}
 \label{fig:17}
\end{figure}

\begin{figure}[h!]
 \begin{center}$
 \begin{array}{cccc}
\includegraphics[width=50 mm]{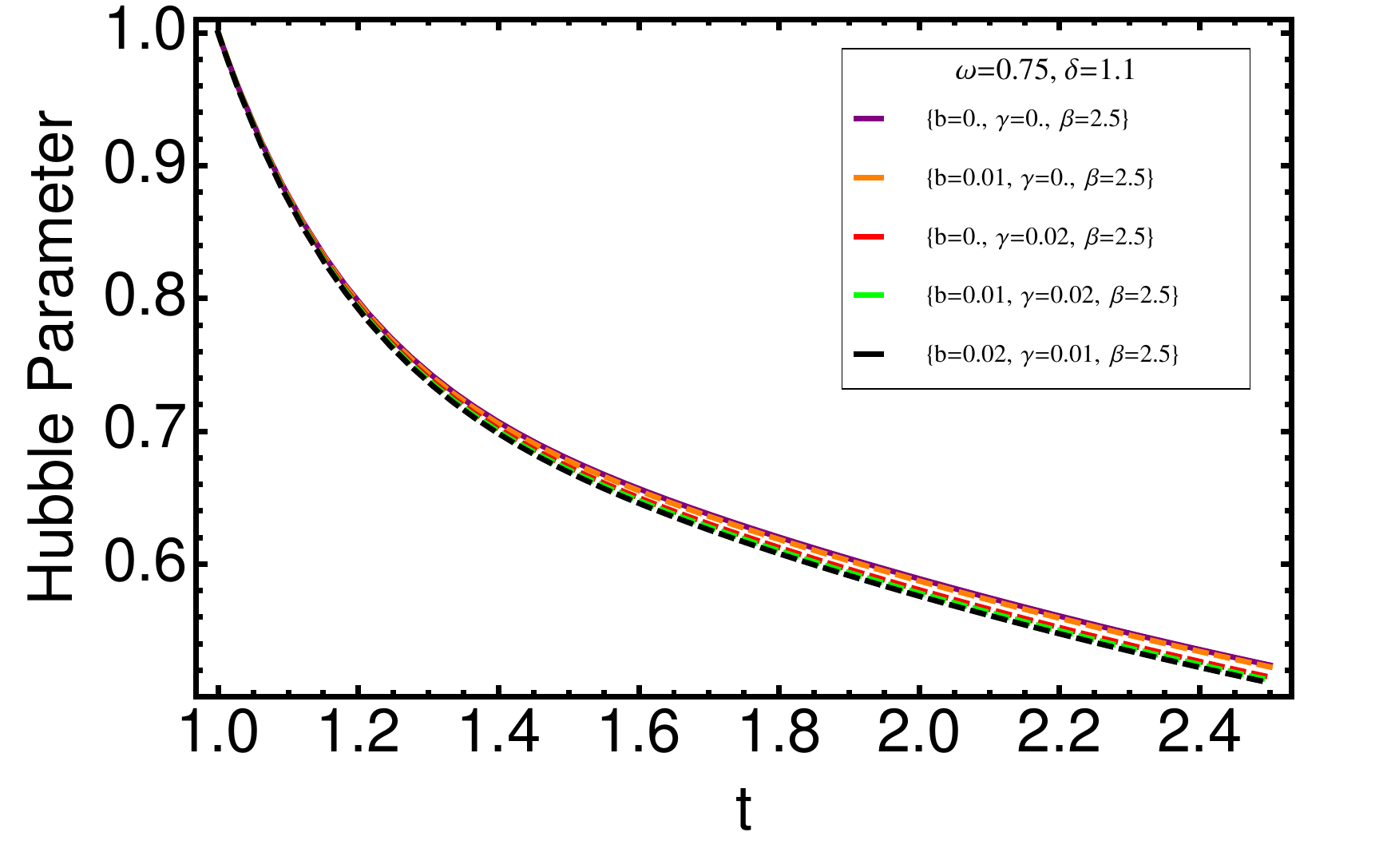} &
\includegraphics[width=50 mm]{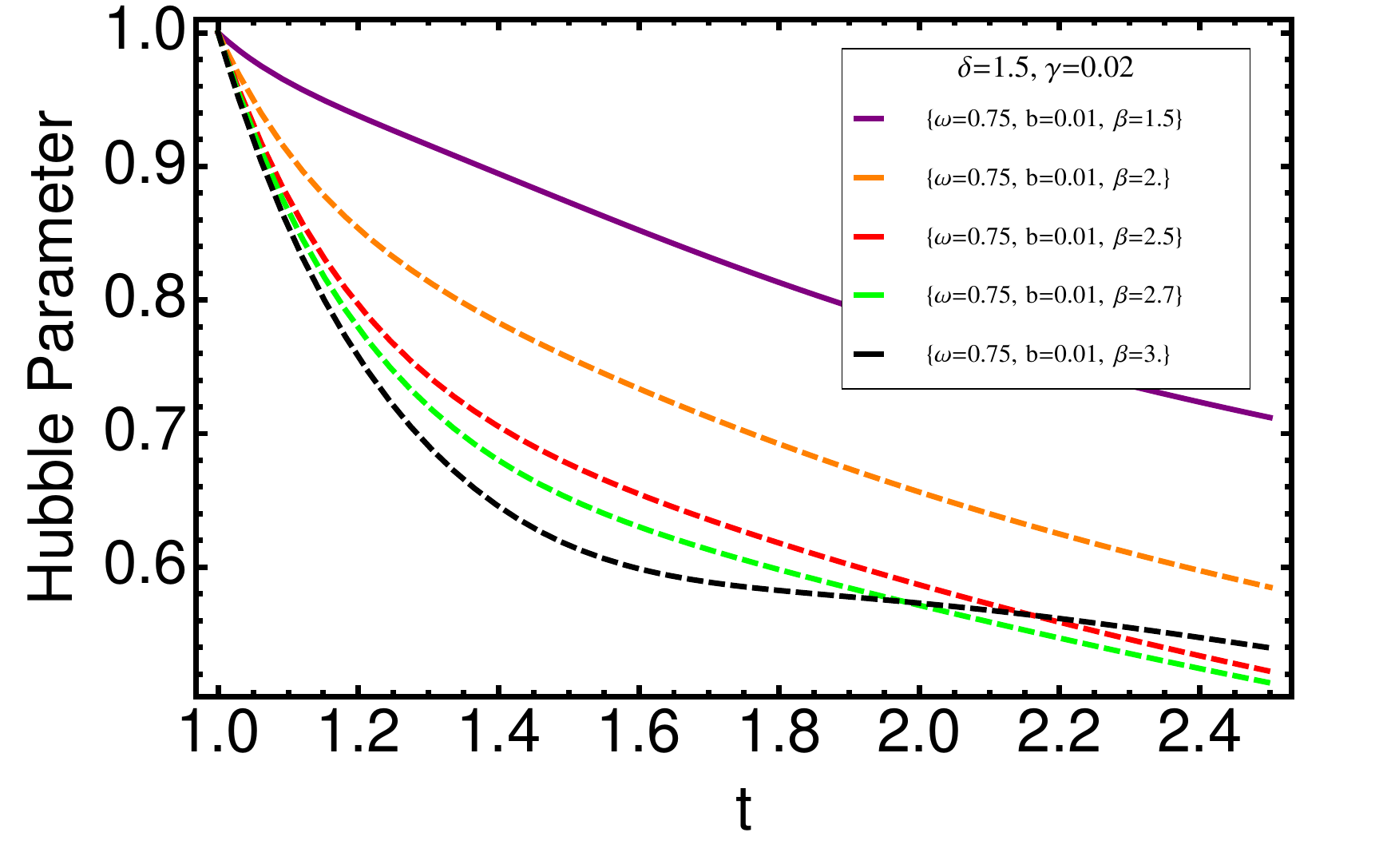}\\
\includegraphics[width=50 mm]{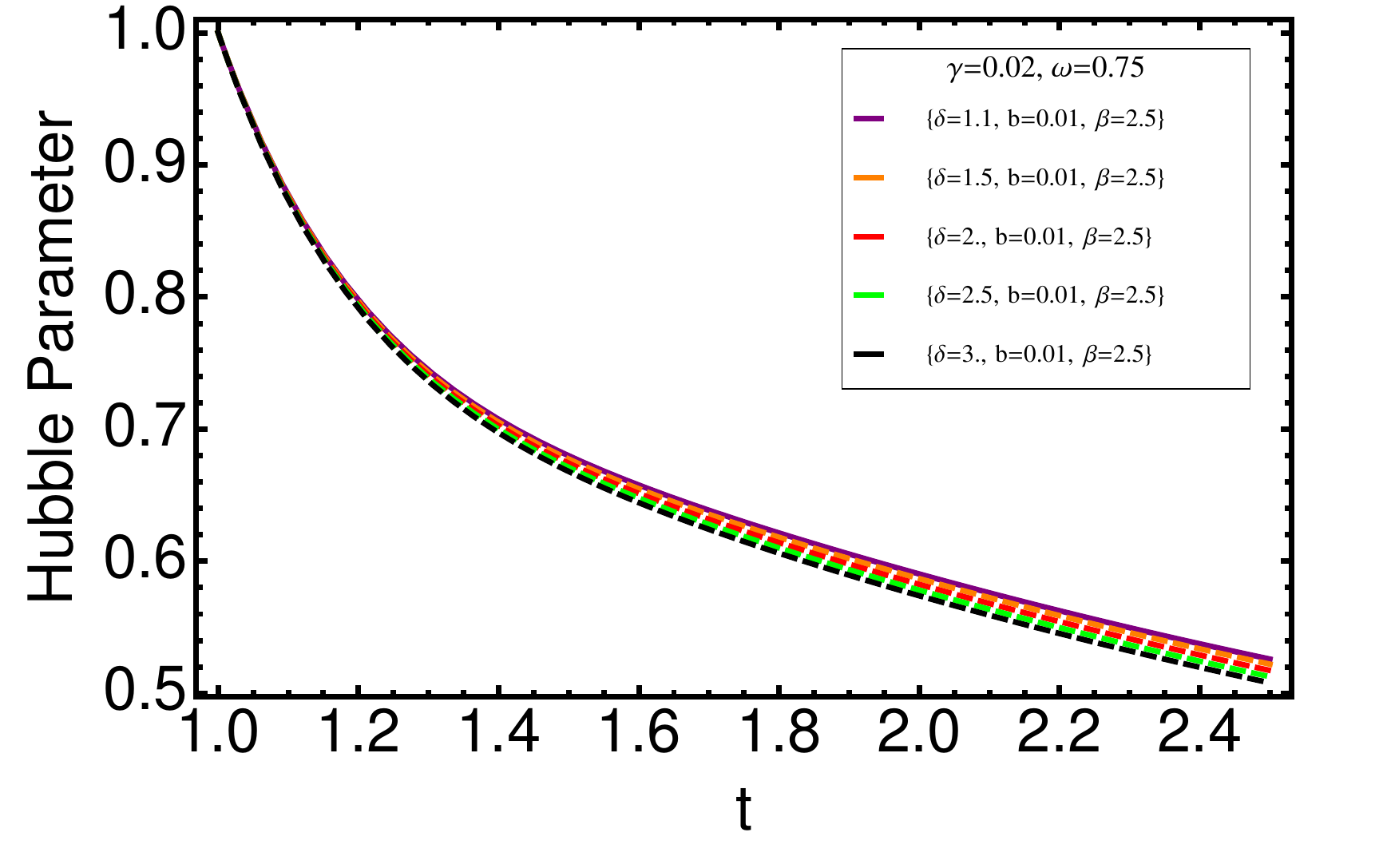}&
\includegraphics[width=50 mm]{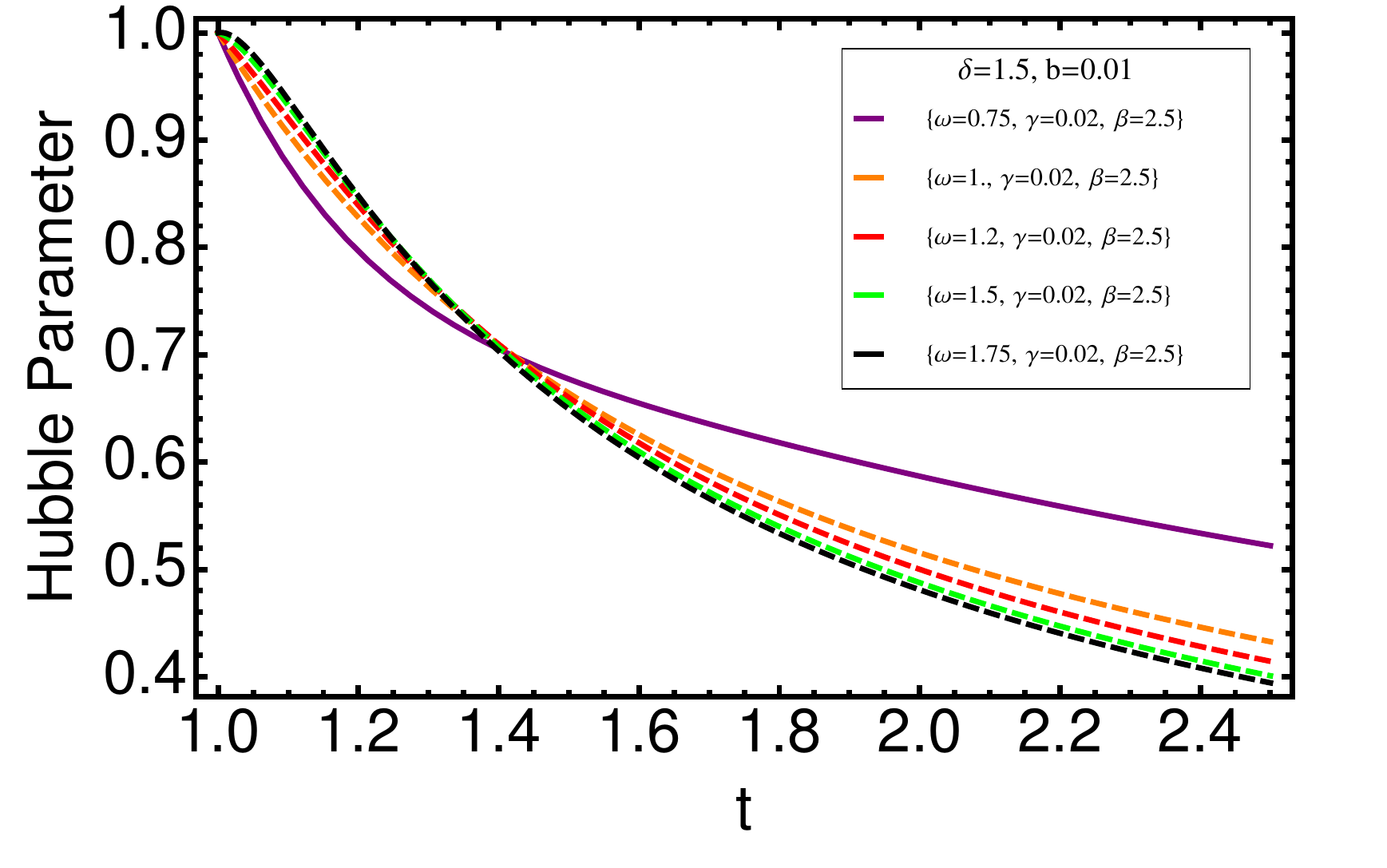}
 \end{array}$
 \end{center}
\caption{Behavior of $H(t)$ against $t$. Model 3}
 \label{fig:18}
\end{figure}

\begin{figure}[h!]
 \begin{center}$
 \begin{array}{cccc}
\includegraphics[width=50 mm]{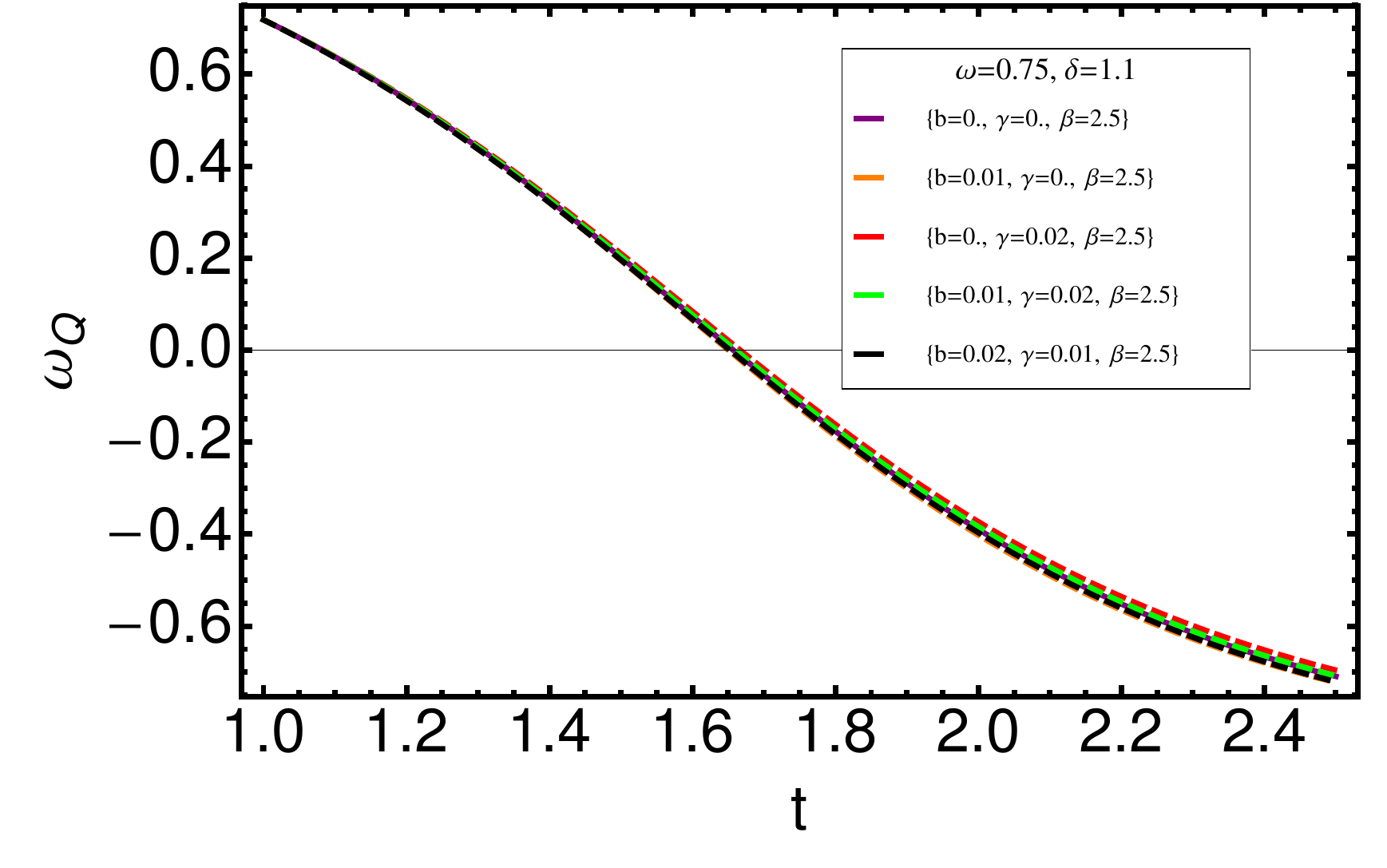} &
\includegraphics[width=50 mm]{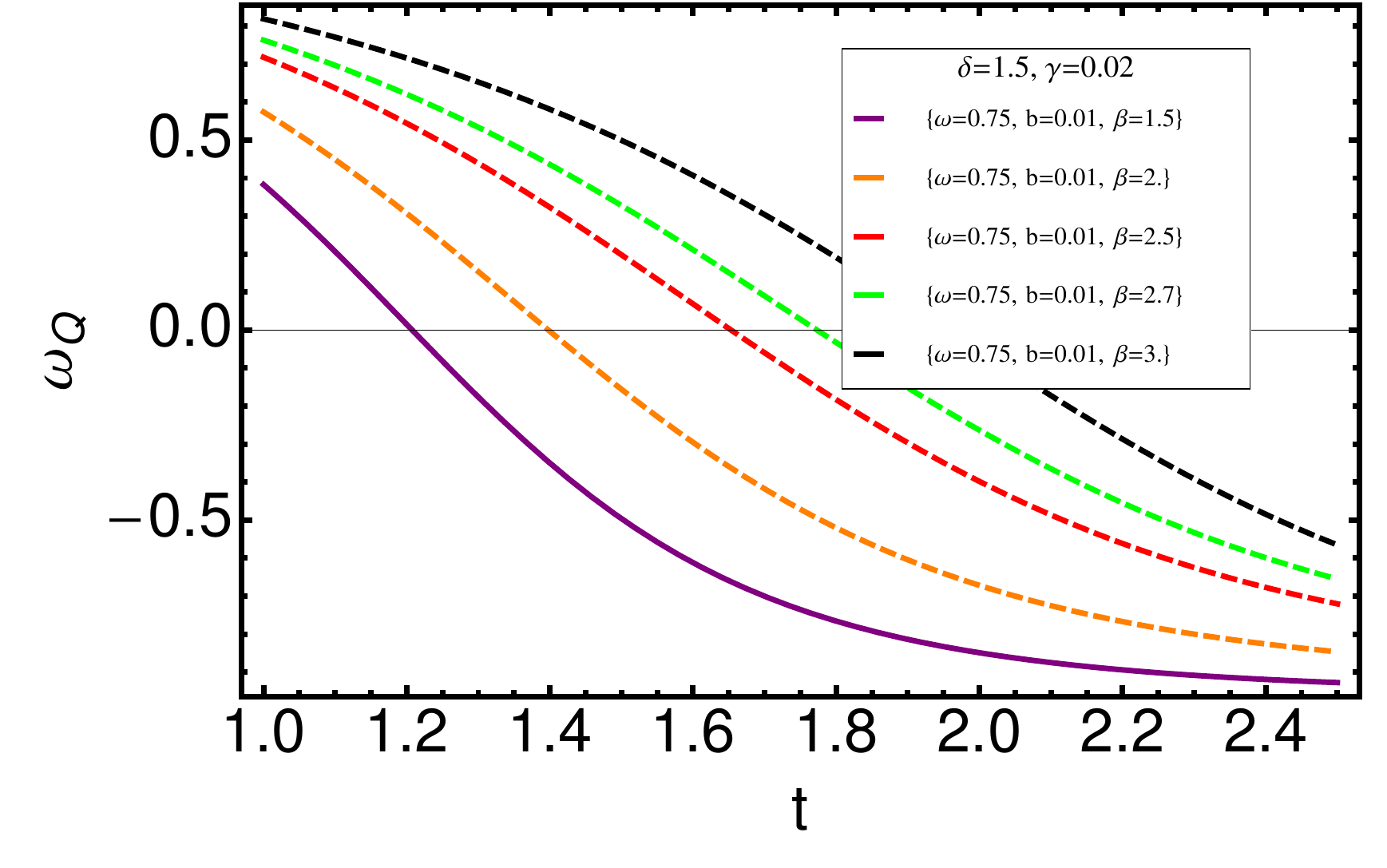}\\
\includegraphics[width=50 mm]{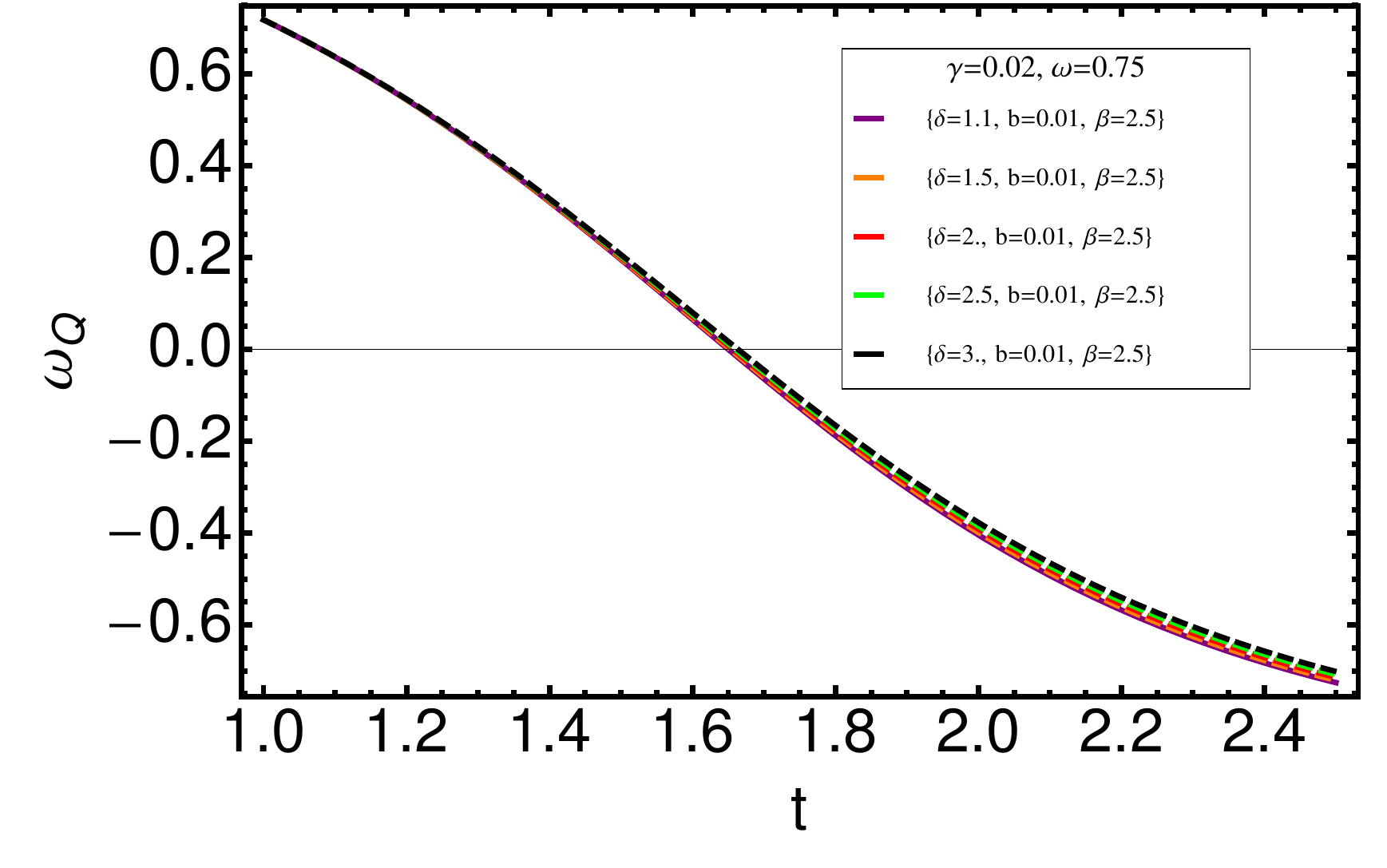}&
\includegraphics[width=50 mm]{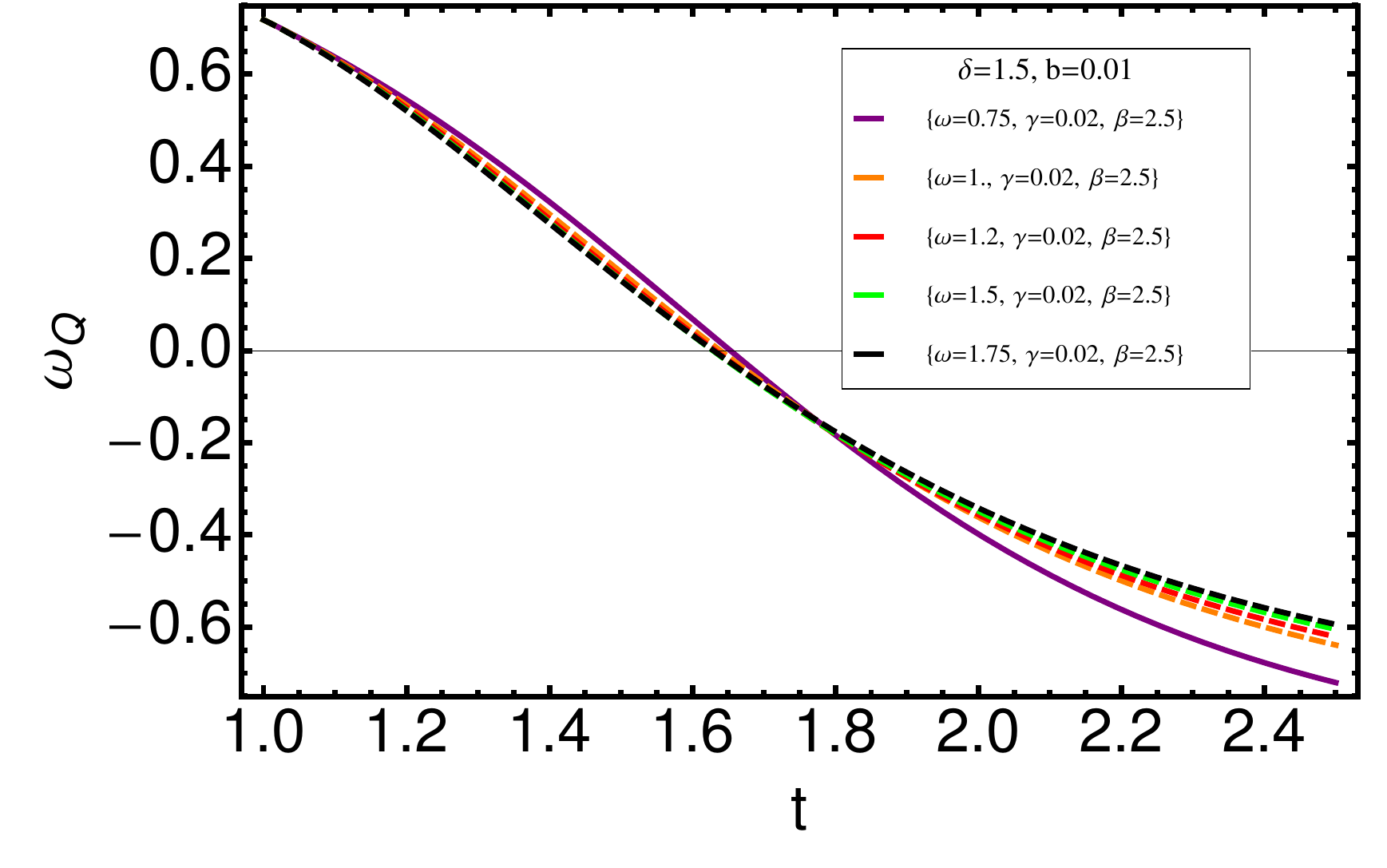}
 \end{array}$
 \end{center}
\caption{Behavior of $\omega_{Q}$ against $t$. Model 3}
 \label{fig:19}
\end{figure}

\begin{figure}[h!]
 \begin{center}$
 \begin{array}{cccc}
\includegraphics[width=50 mm]{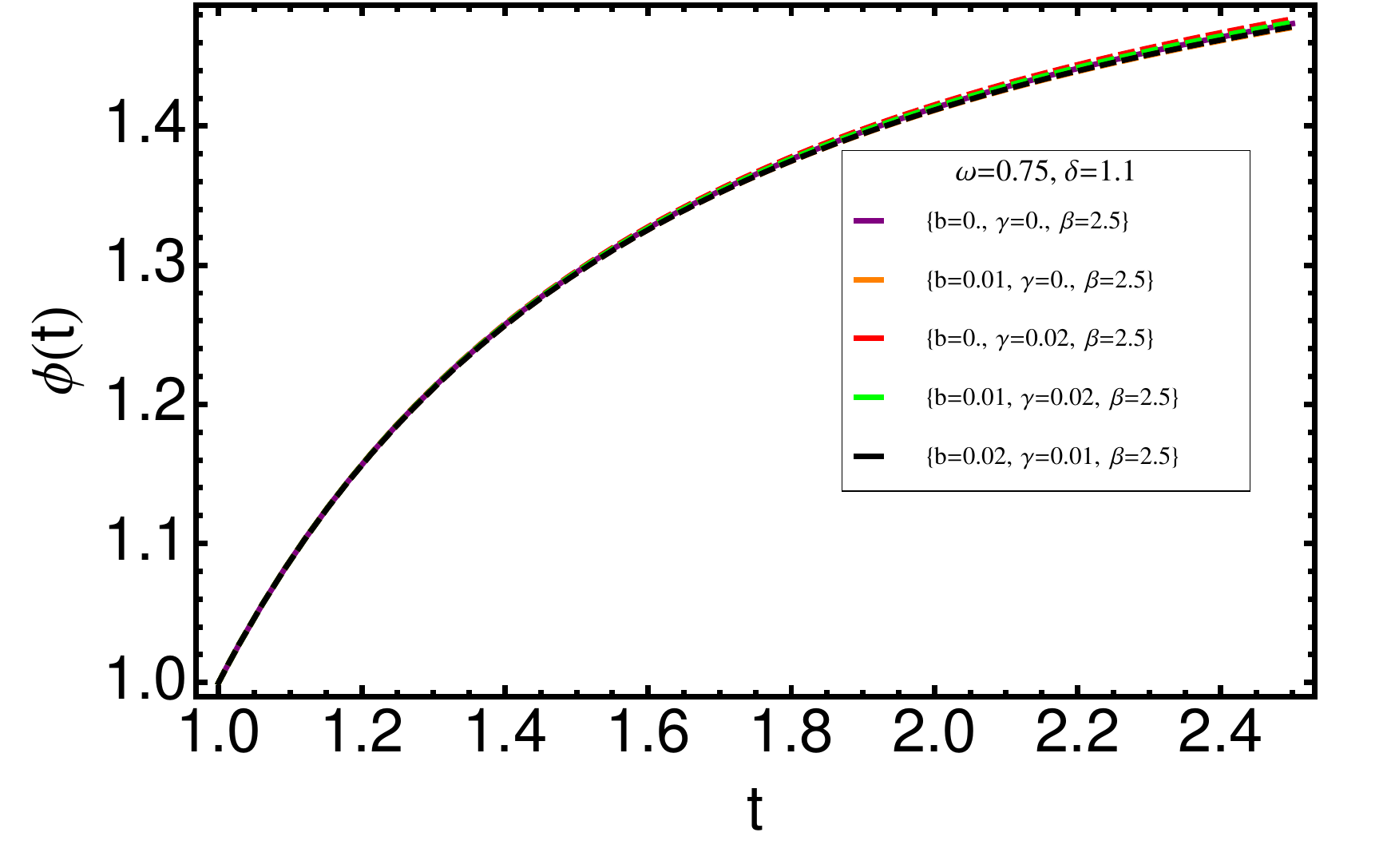} &
\includegraphics[width=50 mm]{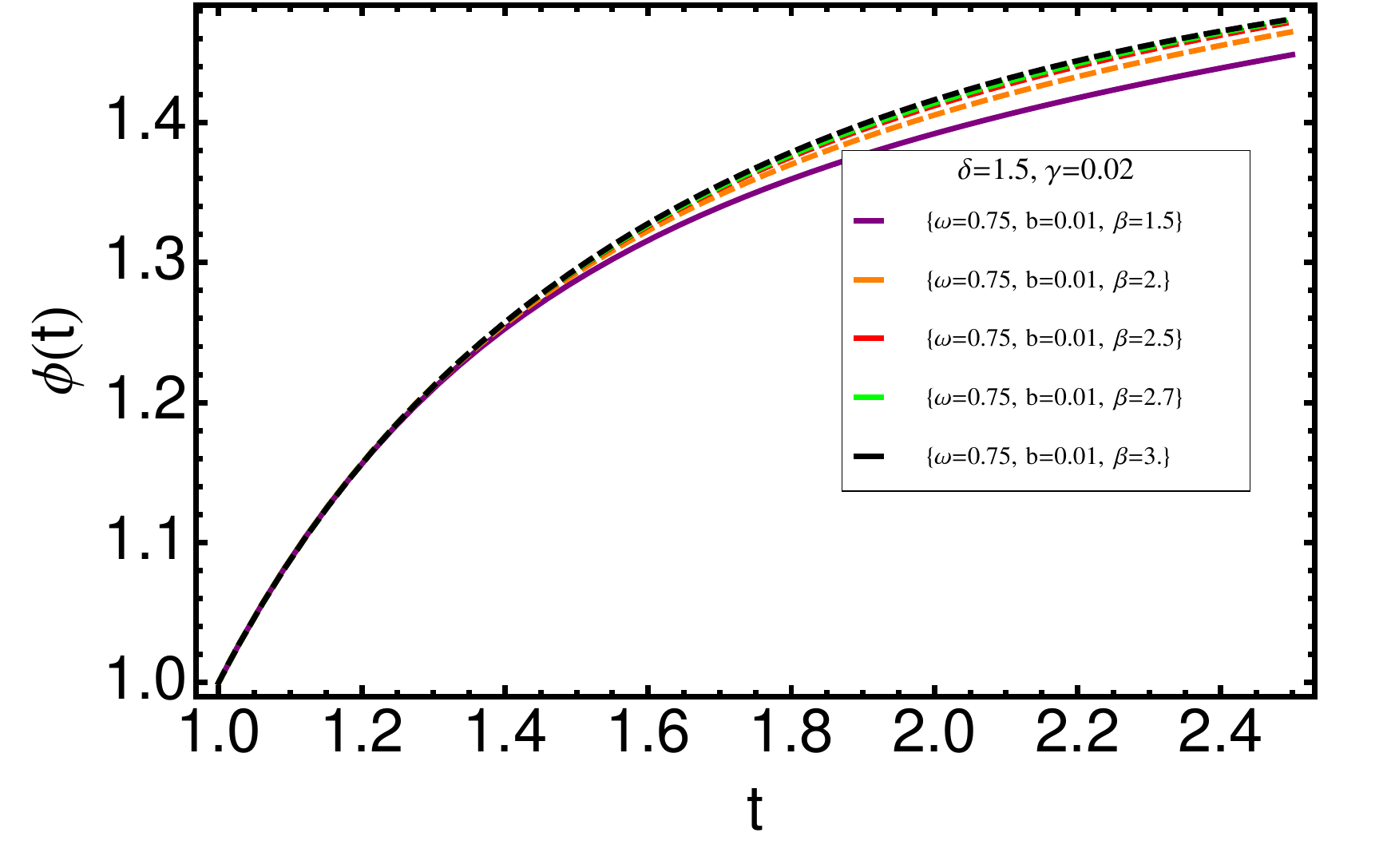}\\
\includegraphics[width=50 mm]{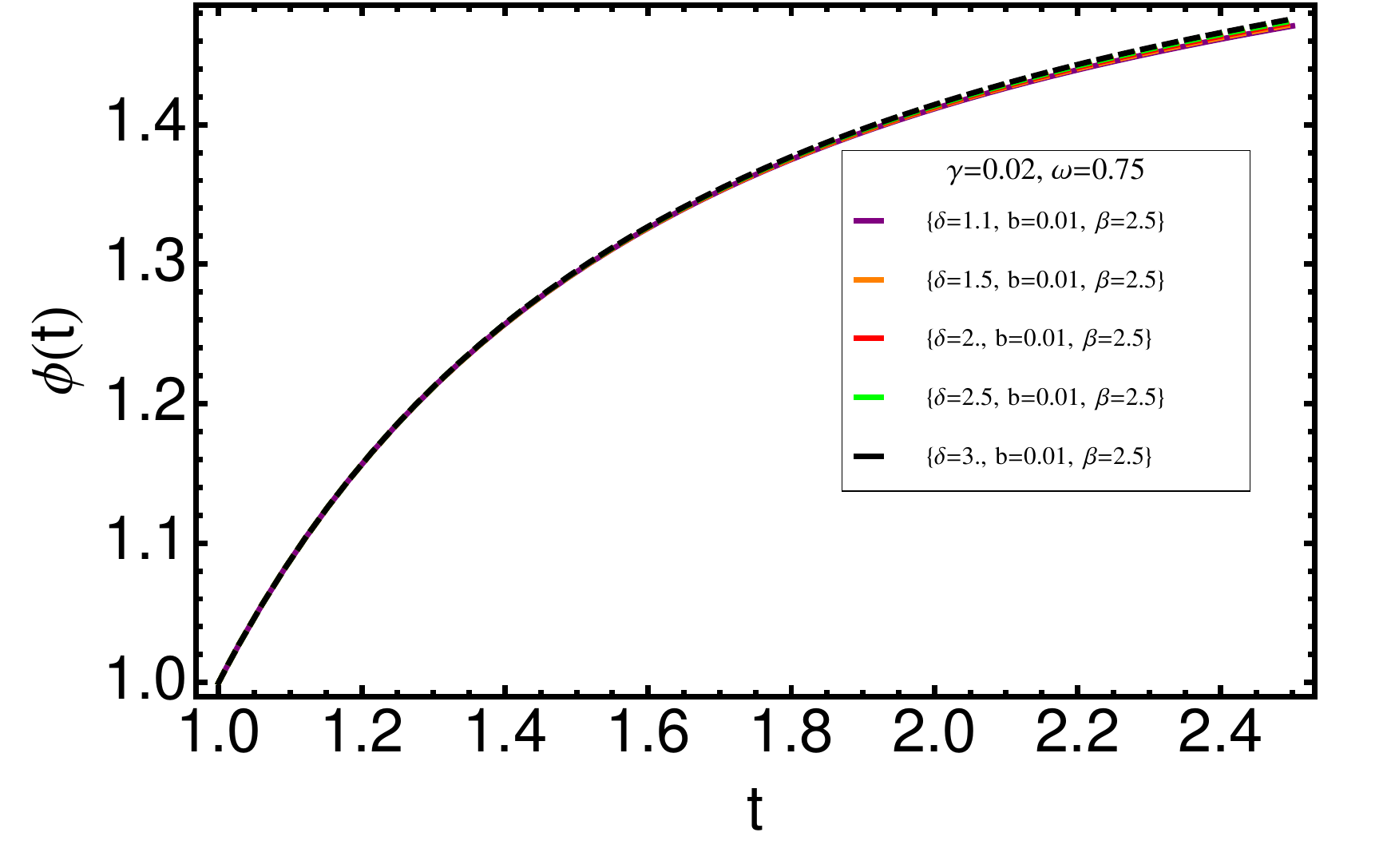}&
\includegraphics[width=50 mm]{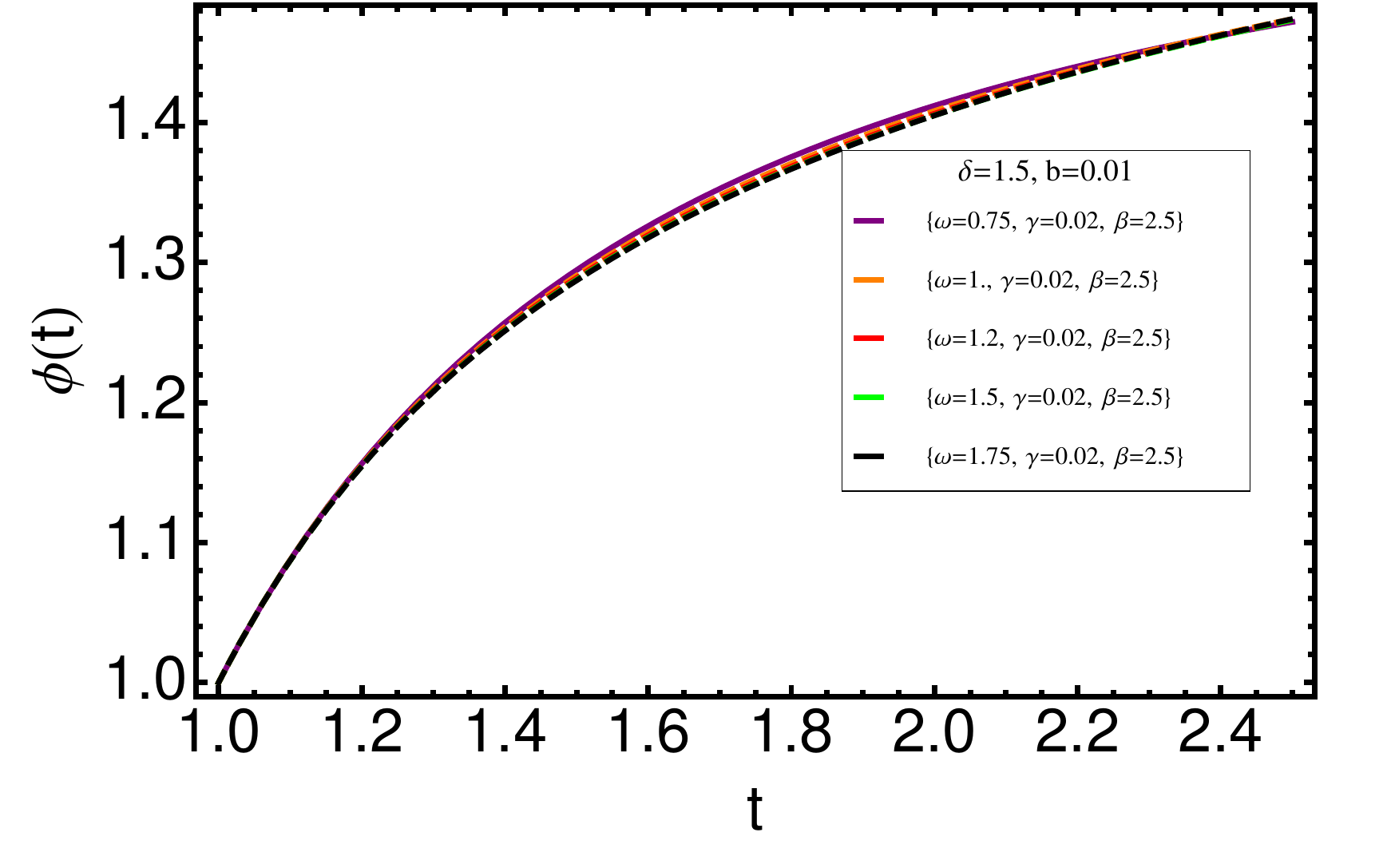}
 \end{array}$
 \end{center}
\caption{Behavior of $\phi$ against $t$. Model 3}
 \label{fig:20}
\end{figure}

\end{document}